\documentclass[prx,superscriptaddress,aps, bibliography, longbibliography, twocolumn, reprint, showpacs, footnoteinbib]{revtex4-1}
\usepackage[breaklinks=true,colorlinks=true,linkcolor=blue,urlcolor=blue,citecolor=blue]{hyperref}
\usepackage{bbm}

\usepackage[T1]{fontenc}
\usepackage[utf8]{inputenc}
\usepackage{color}
\definecolor{note_fontcolor}{rgb}{1, 0, 1}
\usepackage{amsmath}
\usepackage{amssymb}
\usepackage{graphicx}

\makeatletter

\usepackage{graphicx}

\newcommand{\be}{\begin{equation}}
\newcommand{\ee}{\end{equation}}
\def\ba{\begin{aligned}}
\def\ea{\end{aligned}}
\newcommand{\bea}{\begin{eqnarray}}
\newcommand{\eea}{\end{eqnarray}}
\newcommand{\bes}{\begin{subequations}}
\newcommand{\ees}{\end{subequations}}

\newcommand{\la}{\left\langle}
\newcommand{\ra}{\right\rangle}

\newcommand{\lp}{\left(}
\newcommand{\rp}{\right)}

\renewcommand{\hat}[1]{{\widehat #1}}

\begin{document}
\title{ Localization transition on the Random Regular Graph as an unstable tricritical point
in a log-normal Rosenzweig-Porter random matrix ensemble. }

\author{V.~E.~Kravtsov}
\address{Abdus Salam International Center for Theoretical Physics - Strada Costiera~11, 34151 Trieste, Italy}
\address{L. D. Landau Institute for Theoretical Physics - Chernogolovka, Russia}

\author{I.~M.~Khaymovich}
 \address{Max-Planck-Institut f\"ur Physik komplexer Systeme, N\"othnitzer Stra{\ss}e~38, 01187-Dresden, Germany }

\author{B. L. Altshuler}
\address{Department of Physics, Columbia University, New York, NY 10027, USA}
\address{Russian Quantum Center, Skolkovo, Moscow Region 143025, Russia}

\author{L. B. Ioffe}
\address{Department of Physics, University of Wisconsin – Madison, Madison, WI 53706 USA}
\address{Google Inc., Venice, CA 90291 USA}

\begin{abstract}
Gaussian Rosenzweig-Porter (GRP) random matrix ensemble is the only one in
which the robust
multifractal phase and ergodic transition have a status of a mathematical theorem.
  Yet, this phase in GRP model is oversimplified:
the spectrum of fractal dimensions is degenerate and the mini-band in the local
spectrum is not multifractal. In this paper we suggest an extension of the GRP model by
adopting a logarithmically-normal (LN) distribution of off-diagonal matrix elements. A family of
such  LN-RP models is parametrized by a symmetry parameter $p$ and
it interpolates between the GRP at $p\rightarrow 0$ and Levy ensembles at
$p\rightarrow\infty$. A special point $p=1$ is shown to be the simplest approximation
to the Anderson localization model on a random regular graph.
 We study in detail the phase diagram of LN-RP model and
show that $p=1$ is a tricritical point where the multifractal phase first collapses.
This collapse is shown to be
unstable with respect to the truncation of the log-normal distribution. We   suggest
a new criteria of stability of the non-ergodic phases  and prove that
the Anderson transition in LN-RP model is discontinuous at all $p>0$.

\end{abstract}

\maketitle

\section{Introduction}
The development of Quantum Computing algorithms and the problem of
Many Body Localization (MBL)~\cite{BAA} in interacting systems (e.g. in spin chains)
ignited
an interest to the single-particle (Anderson) localization on random graphs as a proxy
for MBL.
The analogy
comes from the mapping in which bit strings of spins-$1/2$ (describing many-body configurations) correspond to sites on a graph and
interaction provides transitions between these bit strings represented by a link between sites. The bit strings
directly accessible
from a given one are uniquely determined by the interaction Hamiltonian, so does
the structure and topology of the corresponding graph.

\begin{figure}[tb]
\center{
\includegraphics[width=0.8 \linewidth,angle=0]{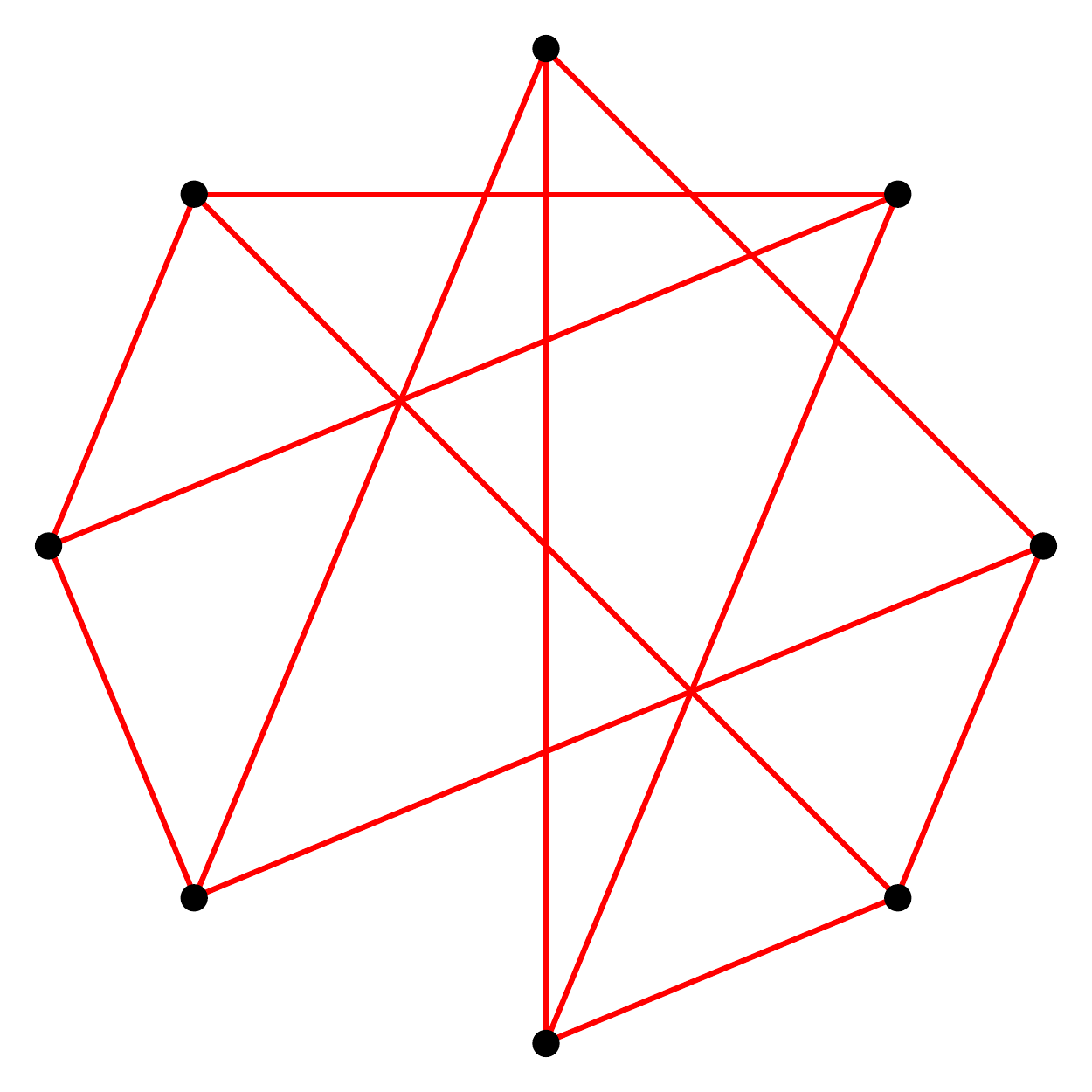}}
\caption{(Color online) \textbf{Random regular graph} with the branching number $K+1=3$
and $N=8$. }
\label{Fig:RRG}
\end{figure}

Since the seminal work~\cite{AGKL} there is a mounting evidence that a representative class of
interacting systems can be modeled by a graph with a local
tree structure but without a boundary. The simplest graph of such kind is
 a random regular graph (RRG), Fig.~\ref{Fig:RRG}, which was
suggested in Refs.~\cite{DeLuca2014,AnnalRRG} as a toy model for many-body localization.

In particular, it was conjectured in these works~\cite{DeLuca2014,AnnalRRG} that a special Non-Ergodic Extended (NEE)
phase is realized on RRG which random eigenfunctions are multifractal. In view of the
above analogy with the system of interacting qubits this statement, if correct, is of
extreme importance. Recently such NEE states were suggested~\cite{ASm1, ASm2} as mediators for
implementing efficient population transfer in the Grover's Quantum Search Algorithms
with potential application to Machine Learning.

While the existence of NEE phase on RRG is still under debate~\cite{AnnalRRG, Tikh-Mir1, Tikh-Mir2, Scard-Par,RRG_R(t),deTomasi2019subdiffusion,Lemarie2017,Lemarie2020_2loc_lengths} it is rigorously proven to exist in
a much simpler Rosenzweig-Porter (RP) random matrix model~\cite{RP, gRP, Warzel,Biroli_RP,Ossipov_EPL2016_H+V,Amini2017,Monthus}.
On the face of it, the RP model seems to have little to do with RRG. However,
in this paper we
show that the modification of RP model, the {\it logarithmically normal}
RP model (LN-RP), offers
the simplest approximation to RRG which accounts for a key difference between an RRG and
a finite Cayley tree.

This LN-RP model is important on its own, as it interpolates between the Gaussian RP
model and the L\'evy random matrix models (see, e.g.,~\cite{Bouchard_Levy_Mat,Biroli_Levy_Mat} and references therein) with power-law distribution of
the off-diagonal matrix elements. We introduce a symmetry parameter $p$ that
controls this interpolation with $p=0$ corresponding to the Gaussian RP model
and $p=\infty$ corresponding to the L\'evy RP model.

A particular case of the
Anderson model on RRG corresponds to $p=1$ due to the hidden $\beta$-symmetry
(see Eqs.~(6.5)-(6.8) in Ref.~\cite{AbouChacra}, Eqs.~(D.2),~(D.17) in
Ref.~\cite{AnnalRRG} and Appendix~\ref{App_sec:P(U)})
on the local Cayley tree.
We show that $p=1$ is the {\it tricritical}
point such that for $p<1$ the LN-RP model supports the NEE phase, while at $p\geq 1$ a
direct transition from the localized to the Ergodic Extended (EE) phase occurs.
However, this EE phase is unstable, as it results from large off-diagonal matrix
elements from the tail of the logarithmically-normal distribution. Any truncation
of this tail (as well as breaking the $\beta$-symmetry) is shown to lead to the reappearance of the NEE phase separating localized
and EE phases.

The analytical theory of the Ergodic (ET) and Localization transitions (AT) developed
in this paper is verified
by extensive numerics based on the Kullback-Leibler divergence~\cite{KLdiv, KLdiv_book}
of certain correlation functions of wave function coefficients~\cite{KLPino}. We also
present
the theory of this new measure of eigenfunction statistics.

The rest of the paper is organized as follows.
In Sec.~\ref{sec:RRG_motivation} we provide the motivation of the model with
long-tailed distribution of off-diagonal elements and its relevance to the
Anderson problem on RRG.
In Sec.~\ref{sec:LN-RP_model} we describe LN-RP model and introduce
the symmetry parameter $p$.
In Sec.~\ref{sec:Phase_diagram} we present the phase diagram of LN-RP model
based on the   Anderson localization and Mott ergodicity criteria.
Section~\ref{sec:KL} shows the numerical data confirming the  analytical
predictions of Sec.~\ref{sec:Phase_diagram} by making use of the
two types of Kullback-Leibler divergence. The critical values  of both ergodic
and localization transitions with the corresponding critical exponents are
extracted from the exact diagonalization with finite-size scaling analysis.
In Sec.~\ref{sec:Truncation} we show that
for  large symmetry parameter $p\geq 1$
the system is unstable with respect to the emergence of the multifractal phase
by pushing
the ergodic transition to smaller disorder values.
In Sec.~\ref{sec:MF-hybridization} we develop a new theory of stability of
non-ergodic states with respect to their
hybridization. The analytical theory of the wave function support set fractal
dimension is presented in Sec.~\ref{sec:D_1}.
Conclusions and discussion of the results 
are given in Sec.~\ref{sec:Conclusion}.
The Appendices~\ref{App_sec:RRG-LN-RP_correspondence}~--~\ref{App_sec:Stability} give the details of the derivation of the
above results and in-depth discussion of the new methods and ideas.

\section{Definitions and emergence of long-range models
on tight-binding hierarchical graphs}\label{sec:RRG_motivation}
\begin{figure}[t!]
\center{
\includegraphics[width=1.0 \linewidth,angle=0]{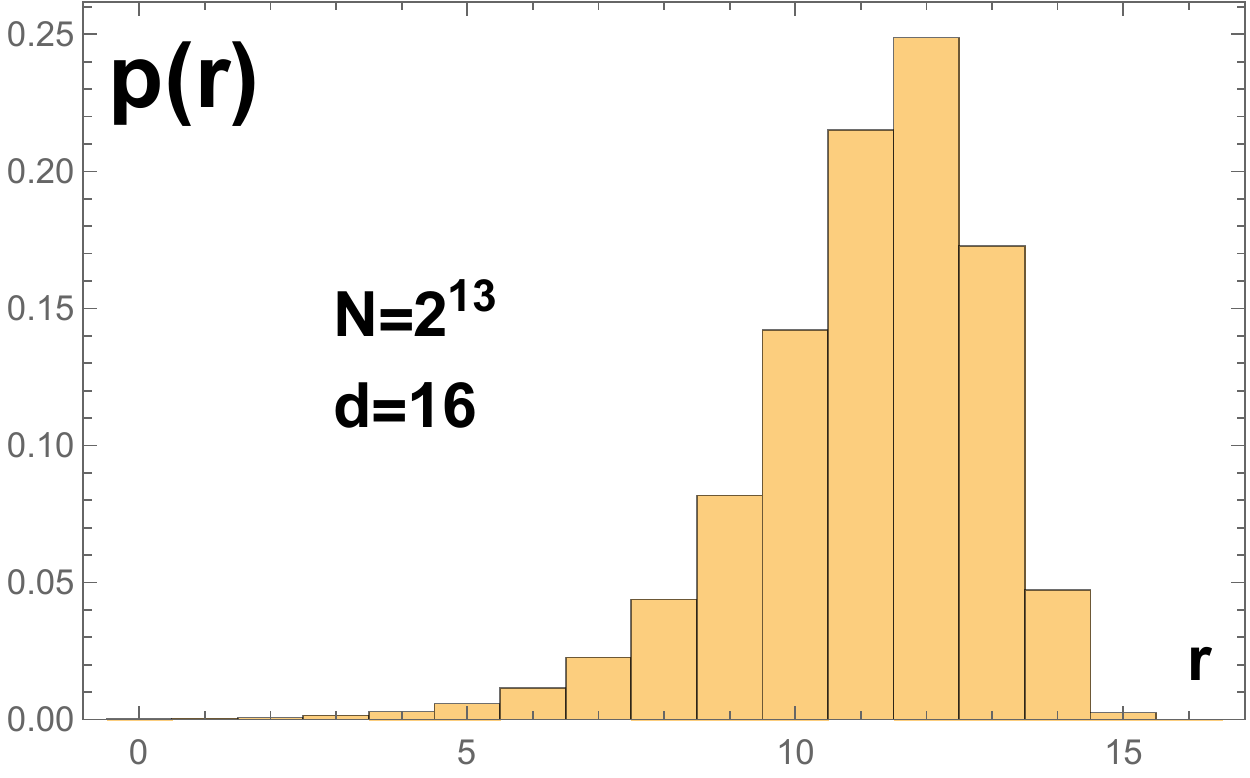}}
\caption{(Color online) \textbf{Distribution of distances $r$ between two points on RRG} of
$N=2^{13}=8192$ sites and branching number $K=2$. The diameter of the graph $d=16$
is only by 4 larger than the most probable distance $r_{*}=12$, both being
approximately equal to $\ln N/\ln K$ in the limit $N\rightarrow\infty$. }
\label{Fig:distances}
\end{figure}

Let us first sketch the mapping of the Anderson localization model with {\it nearest
neighbor}
hopping on RRG to the log-normal Rosenzweig-Porter random matrix model with {\it infinite-range} hopping.
The reader interested
in the properties of the LN-RP model itself can go directly to the next section.

The random $K$-regular graph (RRG) of $N$ sites is a graph in which any
site is connected to $K+1$ other sites in a random way (see Fig.~\ref{Fig:RRG}).
The hopping amplitude  $V=1$ between nearest neighbor sites  is fixed for all links.
The Anderson localization model on RRG adds a
random potential $\varepsilon_{n}$ fluctuating independently  at any site $n$ with the
site-independent distribution $F(\varepsilon)$ characterized by the variance
$\la \varepsilon_n^2\ra \sim W^{2}$.

RRG is known to have locally
a Cayley-tree structure with the branching number $K$. However, in contrast to a
finite Cayley tree which has a strict hierarchical structure and grows from a special
point (the root)
onwards, any point of RRG can be considered as a root of a local Cayley tree.
This is because RRG has loops (of which overwhelming majority are long loops
with the length of
the order of the diameter of the graph)
which connect one local tree with the other thus making all points of the graph
statistically equivalent.

The local tree structure and the predominance of long loops on RRG
lead to the exponential growth of the distribution $p_N(r)\sim K^{r}$
of distances $r$ between pairs of points on this graph.
This growth lasts nearly up to the maximal distance on RRG (the diameter $d$), followed by an abrupt drop to zero, Fig.~\ref{Fig:distances} (see also Fig.~12 in~\cite{AnnalRRG}).
The most probable distance, $r=r_*$, differs from the graph diameter $d$ only by few extra links
and for large graphs of $N$ points they are approximately equal $r_*\simeq d\simeq\ln N/\ln K$.
Moreover, due to the exponential growth of $p_{N}(r)$ with $r$,
in the thermodynamic limit $N\to\infty$ a finite fraction of pairs of sites
on RRG is at the most probable distance, $p_{N\to\infty}(r_*) \to f>0$.	
This ``condensation of large distances'' is the crucial point for the analogy between RRG and RP models.

Indeed, let's consider the set of equally spaced sites on RRG with the most
abundant distance $r_*\approx d-4$ (see Fig.~\ref{Fig:distances}).
An effective random matrix model involving only such (``marked'') sites which
constitute a finite fraction ($\approx 25\%$ at $K=2$) of all sites on RRG can be
written using the
Anderson impurity model (see  Appendix~\ref{App_sec:RRG-LN-RP_correspondence} for
more details). Those marked sites interact with each other through the remaining
{\it tree sites} similar to the {\it indirect interaction} between Anderson impurities
in a metal.  This indirect interaction is {\it long-range},  as well as
the RKKY interaction \cite{RKKY} of magnetic impurities mediated by electrons
in a metal. Since all the marked
sites are at  a distance $r_*\simeq d$ from each other, the effective
hopping matrix elements between them $ H_{{\rm eff},n\neq m}$ can be all taken
{\it independent and identically distributed} (i.i.d.). Thus we arrive at an
{\it infinite-range}
random matrix model of the Rosenzweig-Porter type.
The diagonal matrix elements
$H_{{\rm eff}, nn}=\varepsilon_{n}$ have the same statistics $F(\varepsilon)$ as the
on-site
energies on RRG.

In the same way as RKKY interaction
depends on the details of the Fermi surface in a metal, in our case the
 effective hopping matrix elements $ H_{{\rm eff},n\neq m}$ between marked sites
encode the
hierarchical structure of
the tree sites in their distribution function.
In Appendices~\ref{App_sec:Greens_funct_BL}~--~\ref{App_sec:P(U)}
we show that in contrast to the Gaussian Rosenzweig-Porter (RP) random matrix
theory (RMT)~\cite{RP,gRP},
the distributions of diagonal $F(\varepsilon)$ and off-diagonal $P(U)$
 matrix elements for a RP model associated with RRG
are drastically different.
Unlike  $F(\varepsilon)$ which   is {\it compact},
$\langle \varepsilon^{2n}\rangle \propto \langle\varepsilon^{2}\rangle^{n}$,
the  distribution of $P(U)$ of $ H_{{\rm eff},n\neq m}$ has a {\it fat tail}
which makes any moment
$\langle |U|^{q}\rangle \sim N^{-\gamma_{q}}$ of $U$
characterized by its own exponent
$\gamma_{q}$, the averages with large enough $q$ being divergent.

For not very large branching number $K$ the distribution
of $P(U)$ associated with RRG in its delocalized phase is
{\it logarithmically normal}.
The log-normal distribution of effective hopping reflects the hierarchical
structure of the tree sites and
follows from the representation of $ H_{{\rm eff},n\neq m}$ as a product
of one-point Green's functions on a tree along the path between the sites $n$ and $m$.
Due to the random graph structure the logarithm $ \ln H_{{\rm eff},n\neq m}$ is
represented by the sum of large number $r_*$ of nearly independent elements
and, thus, its distribution can be approximated by Gaussian (see
Appendices~\ref{App_sec:RRG-LN-RP_correspondence} and~\ref{App_sec:P(U)} for details of
derivation and limits of applicability).
This leads to the log-normal distribution of the hopping term itself.

We will see that the Rosenzweig-Porter RMT with the compact distribution of
diagonal elements and log-normal distribution of off-diagonal
elements (LN-RP)
is very rich with numerous potential applications which justify its detailed consideration
independently of the
analogy with the Anderson model on RRG.

\section{Log-normal Rosenzweig-Porter RMT}\label{sec:LN-RP_model}
As shown in Appendices~\ref{App_sec:Greens_funct_BL}~--~\ref{App_sec:P(U)}, the distribution function
$P(U)$ of off-diagonal elements of the RP ensemble associated with RRG is of the ``multifractal'' form of the
{\it large deviation ansatz}:
\be\label{MFform}
P(U) \sim U^{-1}\, {\rm exp}\left[-\ln N\,\mathfrak{G}\left(\frac{\ln U}
{\ln N}\right) \right],
\ee
where $ \mathfrak{G}(x)$ is a certain function and the large parameter $\ln N$ is proportional
to the diameter of RRG. This form is very special, as $\ln N$ appears both
in front of the function $ \mathfrak{G}(x)$ and in its argument. It emerges in many
different physical problems ranging from distribution of amplitudes of multifractal
wave functions to statistics of work in driven systems out-of-equilibrium
~\cite{NatComm}.

The simplest choice of $ \mathfrak{G}(x)$ is a linear function which gives rise to a power-law
distribution $P(U)$. However, in many relevant cases where the Central Limit Theorem
applies to $\ln U$, the distribution $P(U)$ is log-normal which
corresponds to a parabolic
 function $ \mathfrak{G}(x)$:
\be \label{LN-MF}
P(U)=\frac{A}{U}\,{\rm exp}\left[-\frac{\ln^{2}(U/U_{{\rm typ}})}
{2p\,\ln(U_{{\rm typ}}^{-1})}\right],
\;\; U_{{\rm typ}}\sim N^{-\gamma/2}.
\ee
This distribution is controlled by two parameters: the parameter $\gamma>0$
that governs the
typical value of the hopping matrix element in LN-RP model and the
symmetry parameter $p$.

The reason we refer to this parameter as the
{\it symmetry parameter} is related to the basic symmetry on the Cayley tree
(see Appendix~\ref{App_sec:P(U)}) which gives rise to the duality relation:
\be
P(U^{-1})=U^{4}\,P(U),\;\;\Leftrightarrow p=1.
\ee
When applied to Eq.~(\ref{LN-MF}) this relation requires $p=1$. However,
it is useful to keep this parameter free to interpolate between the LN-RP
with long-tail cut in $P(U)$ (the case $p\rightarrow 0$
which is equivalent to the Gaussian RP~\cite{RP, gRP})
and the case $p\rightarrow\infty$ when $P(U)$ approaches the L\'evy power-law
distribution.

Another model parameter $\gamma$ is related to the Lyapunov exponent $\lambda$ on the
disordered Cayley tree. It is defined~\cite{AnnalRRG} via the exponential decay of
the typical absolute value of the Green's function $|G_{r}|_{typ}$
with the distance $r$:
\be\label{def-lambda}
\lambda=-\lim_{r\rightarrow\infty}r^{-1}\,\ln |G_{r}|_{typ}.
\ee
By substituting $r=r_*\simeq d=\ln N/\ln K$ and
$|G_{r}|_{typ}\sim N^{-\gamma/2}$ in Eq.~(\ref{def-lambda}) one immediately obtains:
\be\label{gamma-lambda}
\gamma=\frac{2\lambda}{\ln K}.
\ee
As shown in Appendix~\ref{App_sec:P(U)}, the log-normal distribution of $P(U)$ is asymptotically
exact for RRG at small disorder.
It is also quantitatively accurate in the entire 
delocalized phase for not very large branching number $K$.

It is important to note that for the distribution Eq.~(\ref{LN-MF}) the scaling
of the typical value of $U$ with $N$ differs from that of the mean value.
The latter
exists only for $p<2$ and is given by $\langle U \rangle
\sim N^{-\gamma_{{\rm av}}/2}$, with
\be\label{typ-av}
\gamma_{{\rm av}}=\gamma\,(1-p/2),\;\;\;(0<p<2) \ .
\ee

\section{Phase diagram of LN-RP.  Collapse of
the multifractal phase at the tricritical point at $p=1$.}
\label{sec:Phase_diagram}

It was first shown in Ref.~\cite{gRP} that the Rosenzweig-Porter RMT with a Gaussian $P(U)$
has three phases: ergodic, $\gamma<\gamma_{ET}$, multifractal, $\gamma_{ET}<\gamma<\gamma_{AT}$, and localized, $\gamma>\gamma_{AT}$ and two transitions between them at $\gamma_{ET}=1$ and $\gamma_{AT}=2$. The same transition points are expected for
the LN-RP in the limit $p\rightarrow 0$. In this section we consider simple
``rule of thumb'' criteria formulated in Refs.~\cite{BogomolnyPLRBM2018,Nos} which show how the phase diagram of LN-RP is modified as
the symmetry parameter $p$ increases. The physical picture and details of these transition
and the corresponding phases will be considered in Sec.~\ref{sec:MF-hybridization}.

The first criterion, nicknamed as {\it Anderson localization criterion}, applies to
random matrices
with uncorrelated entries and states~\cite{BogomolnyPLRBM2018,Nos} that if the sum:
\be\label{Anderson}
S_{1}=\sum_{m=1}^{N}\langle |H_{n,m}|\rangle_{W} <\infty
\ee
converges in the limit $N\rightarrow\infty$ then the states are Anderson localized.
Here $\la ..\ra_W$ stands for the disorder averaging.

The physical meaning of this criterion is that the number is sites
in resonance with a given site $n$ is finite. Indeed, consider for simplicity the
box-shaped distribution $F(\varepsilon)$ of on-site energies. The probability that {\it two}
sites $n$
and $m$
are in resonance is:
\bea
P_{n\rightarrow m}=&& W^{-2}\int_{-W/2}^{W/2} d\varepsilon_{n}\int_{-W/2}^{W/2}
d\varepsilon_{m}\,
 \nonumber \\
&&\int_{|\varepsilon_{n}-\varepsilon_{m}|}^{\infty}\,P(H_{nm})\, d(H_{nm}).
\eea
Then simple integration over $(\varepsilon_{n}+\varepsilon_{m})/2$ and integration by parts
over $\varepsilon_{n}-\varepsilon_{m}$ gives:
\be\label{Pres}
P_{n\rightarrow m}=\int_{0}^{W}dU\,P(U)\,\left( \frac{2U}{W}-\frac{U^{2}}{W^{2}}\right)+
\int_{W}^{\infty} P(U)\,dU,
\ee
where $U=|H_{nm}|$.

One can easily see that at $U_{{\rm typ}}\sim N^{-\gamma/2}\ll O(1)$
the last integral in Eq.~(\ref{Pres}) is always small and the second term in the first integral is
at most of the same order as the first term. Thus with the accuracy up to a
constant of order 1 we obtain:
\be
P_{n\rightarrow m}\sim \frac{\langle |H_{nm}| \rangle_{W}}{W},
\ee
where the subscript
$W$ in $\langle ... \rangle_{W}$ implies that the distribution $P(U)$ should be truncated at
$U_{max}=W$.
The number of sites in resonance with the given site is
 the sum $\sum_{m}P_{n\rightarrow m}$ which coincides with
Eq.~(\ref{Anderson}) up to a pre-factor of order $O(1)$.

Importantly, the above derivation~\eqref{Pres} gives an elaboration  to Eq.~(\ref{Anderson}).
Indeed, in the case of the long-tailed distribution $P(U)$, one should cut it off
at $U_{{\rm max}}=W\sim O(1)$
in order to obtain a correct sufficient criterion of Anderson localization.
Note that such a cutoff of $P(U)$ is automatically embedded into the
RRG/LN-RP correspondence
(see Appendix~\ref{App_sec:Greens_funct_BL}).

The second criterion suggested in Refs.~\cite{BogomolnyPLRBM2018,Nos} and nicknamed in~\cite{Nos} {\it the Mott's criterion}
is a sufficient criterion of ergodicity. It states that if the sum
\be\label{Mott}
S_{2}=\sum_{m=1}^{N}\langle |H_{nm}|^{2}\rangle_{W} \rightarrow \infty
\ee
diverges in the limit $N\rightarrow\infty$ then the system is in the
(fully) ergodic phase~\cite{Nos}.

Note that similar to Eq.~(\ref{Anderson}), the averaging in Eq.~(\ref{Mott}) should
be done with the distribution truncated at $U_{{\rm max}}\sim O(1)$.
The reason for that is that rare large matrix elements $|H_{nm}|\gg O(1)$ split the
resonance pair of levels so much that they are pushed at the Lifshitz tail of
the spectrum and do not affect statistics of states in the body of
spectrum that we are studying~\cite{delta_typ-footnote}.

The physical meaning of Eq.~\eqref{Mott} is that the Breit-Wigner width $\Gamma$ that quantifies
the escape rate of a particle created at a given site $n$, is much
larger than the
spread of energy levels $W\sim O(1)$ due to disorder. In other words,
the fulfillment of the Mott's criterion implies that the width $\Gamma$ is
of the same order as the total spectral bandwidth and thus there are no
{\it mini-bands} (which width is $\Gamma$) in the local spectrum.
As the presence of such mini-bands is suggested~\cite{Pino-Ioffe-VEK, return, Nosov2019mixtures} as a
``smoking gun'' evidence of the {\it non-ergodic extended } (e.g. multifractal) phase,
the fulfillment of the Mott's criterion~\eqref{Mott} immediately implies that the system is in the {\it ergodic extended} phase.

The above {\it non-ergodic extended} (e.g., multifractal) phase realizes provided that in the limit $N\rightarrow\infty$:
\be
S_{1}\rightarrow \infty,\;\;\;S_{2}\to 0 \ .
\ee
The case of a finite $S_{2}$ in the limit $N\rightarrow\infty$ is more delicate and
may imply merely {\it weak ergodicity}~\cite{WE-footnote}.

\begin{figure}[t!]
\center{
\includegraphics[width=1.0 \linewidth,angle=0]{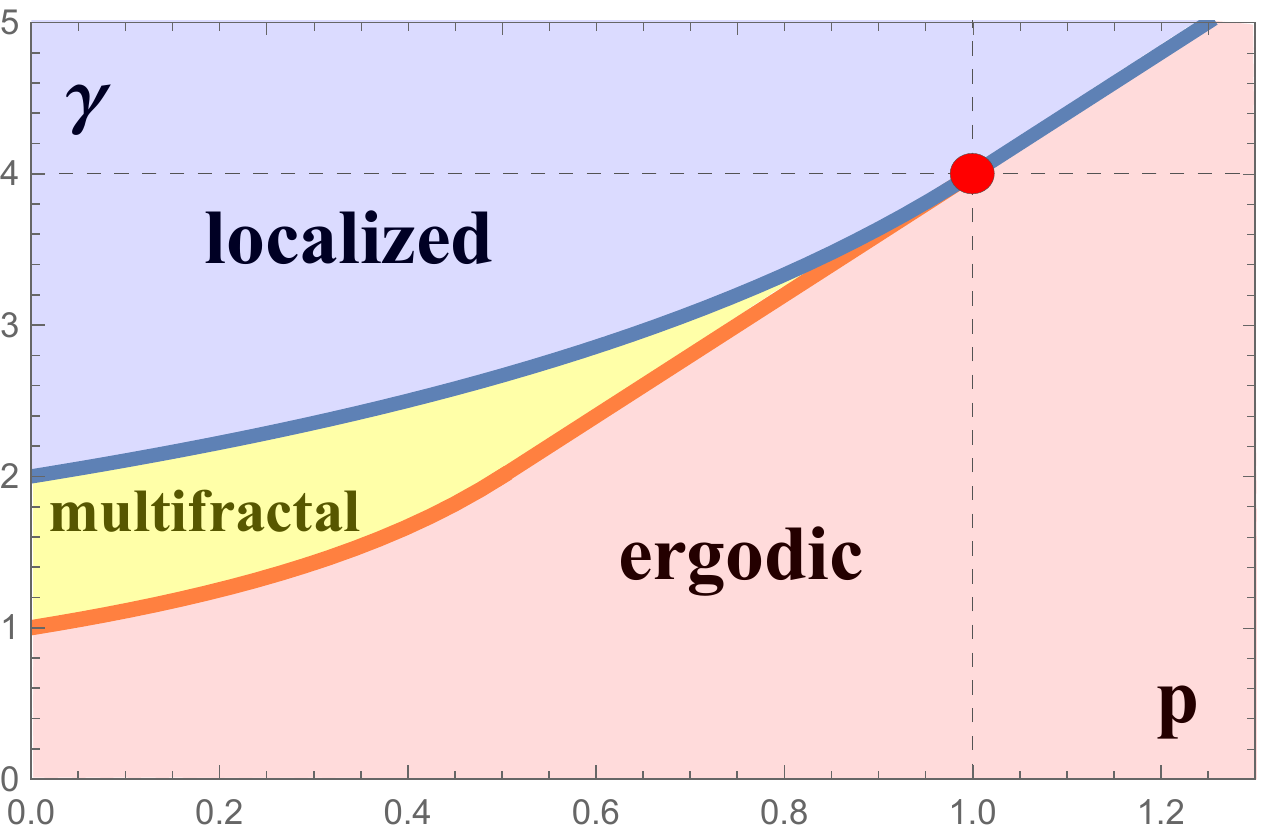}}
\caption{(Color online) \textbf{The phase diagram of the logarithmically-normal Rosenzweig-Porter
RMT.} The ergodic transition (orange) and the Anderson localization transition (blue) lines merge at the tricritical point
 $p=1$ (which is associated with RRG) and $\gamma=4$.
 This critical point corresponds
(see Eq.~(\ref{gamma-lambda}),~(\ref{typ-av})) to the Lyapunov exponent
$\lambda=2\ln K$, or  to
$\lambda_{{\it av}}\equiv\frac{1}{2}\gamma_{{\rm av}}\ln K=\frac{\gamma}{4}\ln K= \ln K$
which is the known criterion of the Anderson transition on a Cayley tree~\cite{AnnalRRG,Warzel2011PRL}.
For $p<1$ the transition from the localized to
the ergodic phase goes through the intermediate multifractal phase;
for $p\geq 1$ a direct transition happens from the localized to the ergodic phase.
 }
\label{Fig:phase_diagram}
\end{figure}

\begin{figure*}[t]
\includegraphics[width=0.24\linewidth]{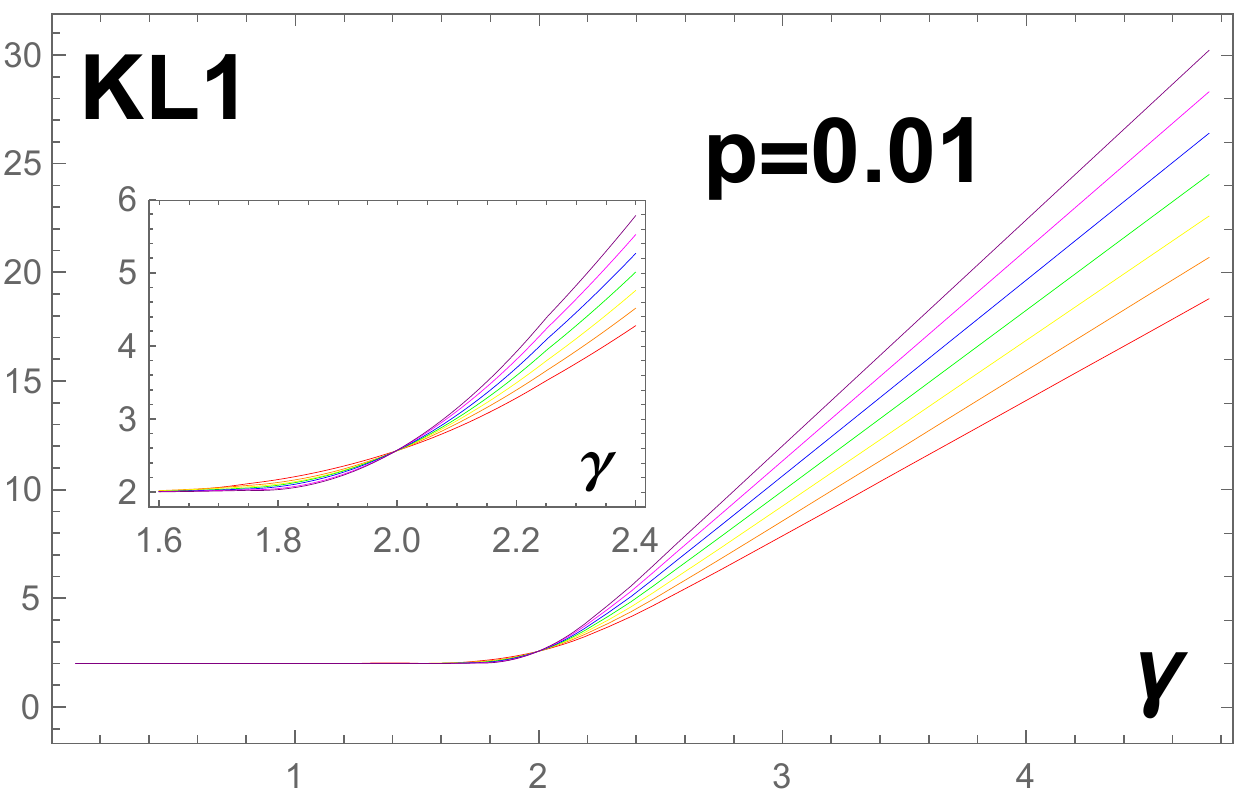}
\includegraphics[width=0.24\linewidth]{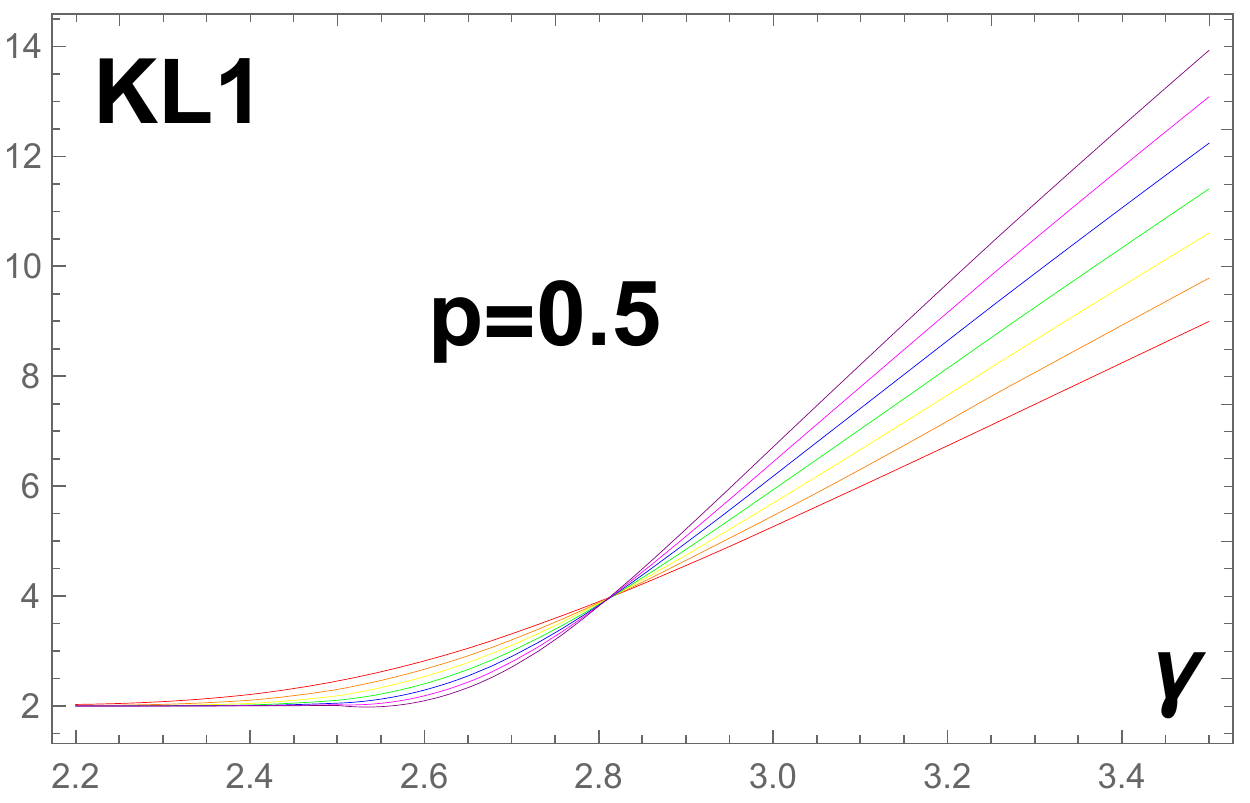}
\includegraphics[width=0.24\linewidth]{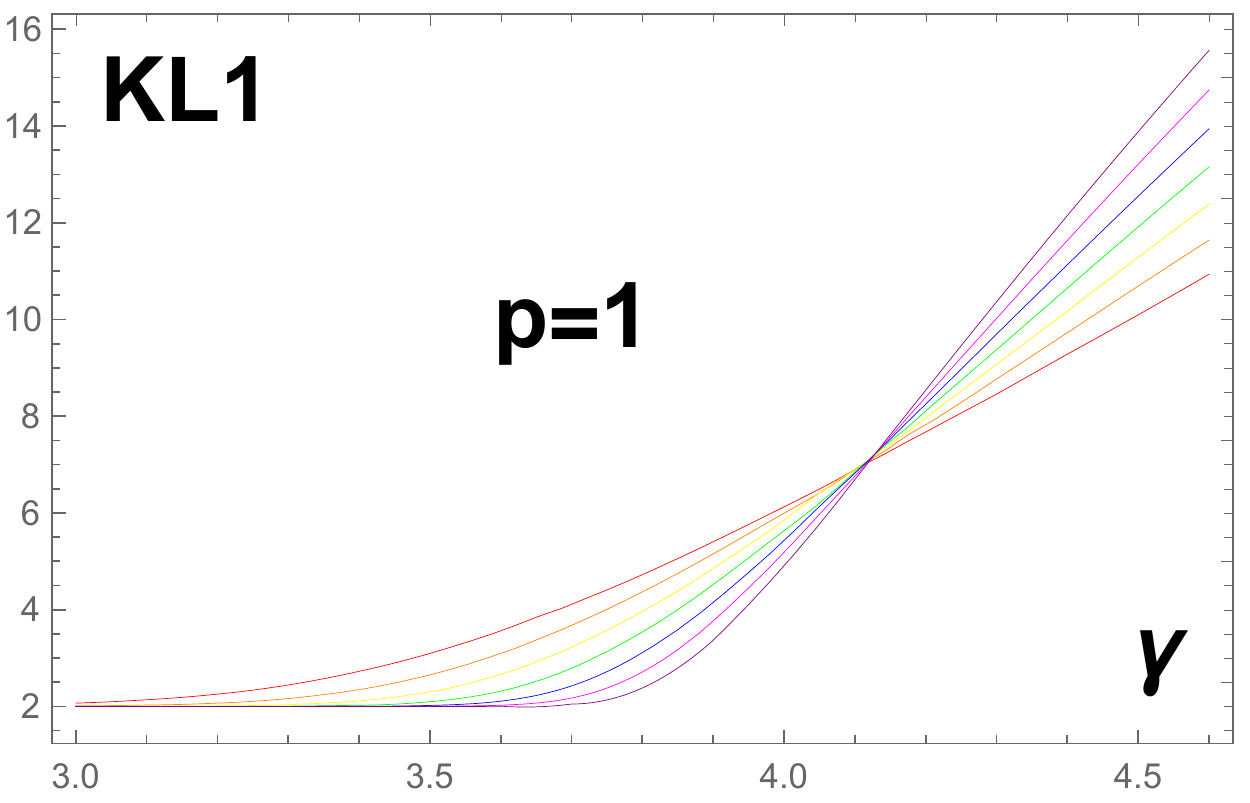}
\includegraphics[width=0.24\linewidth]{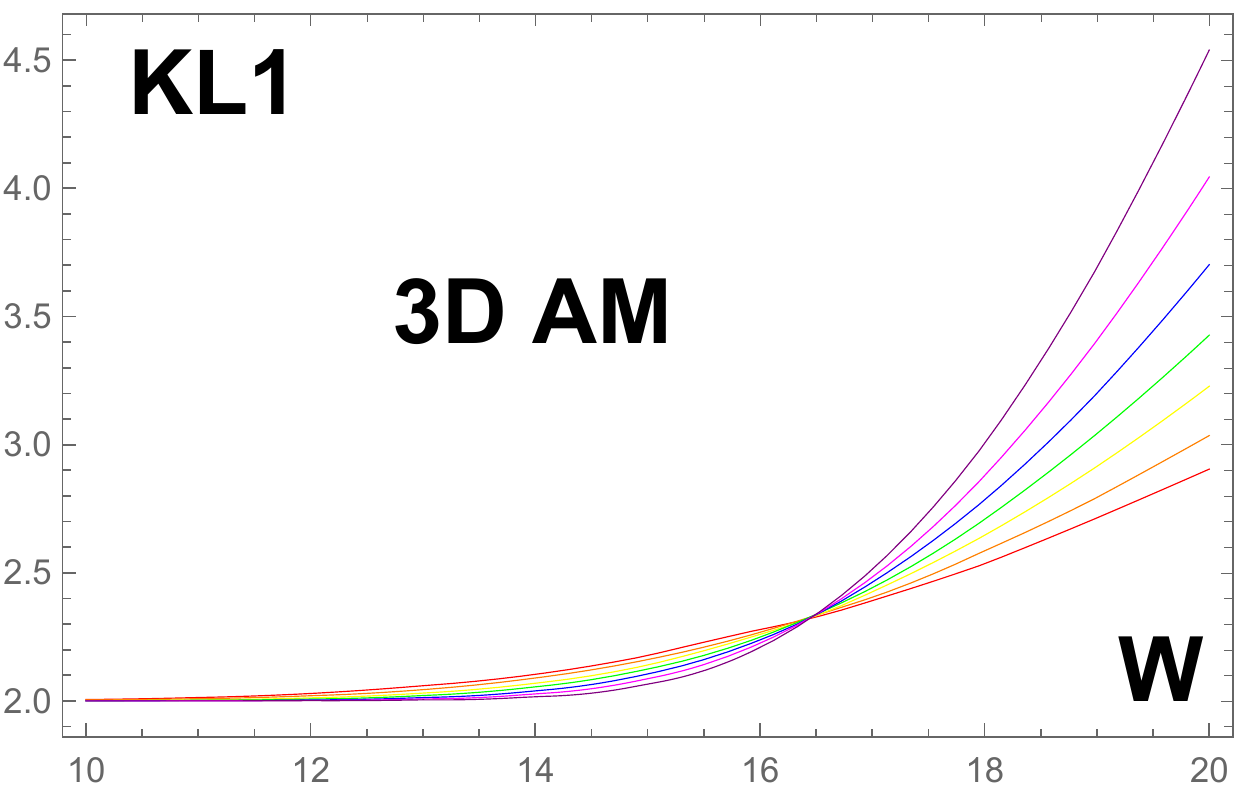}
\includegraphics[width=0.24\linewidth]{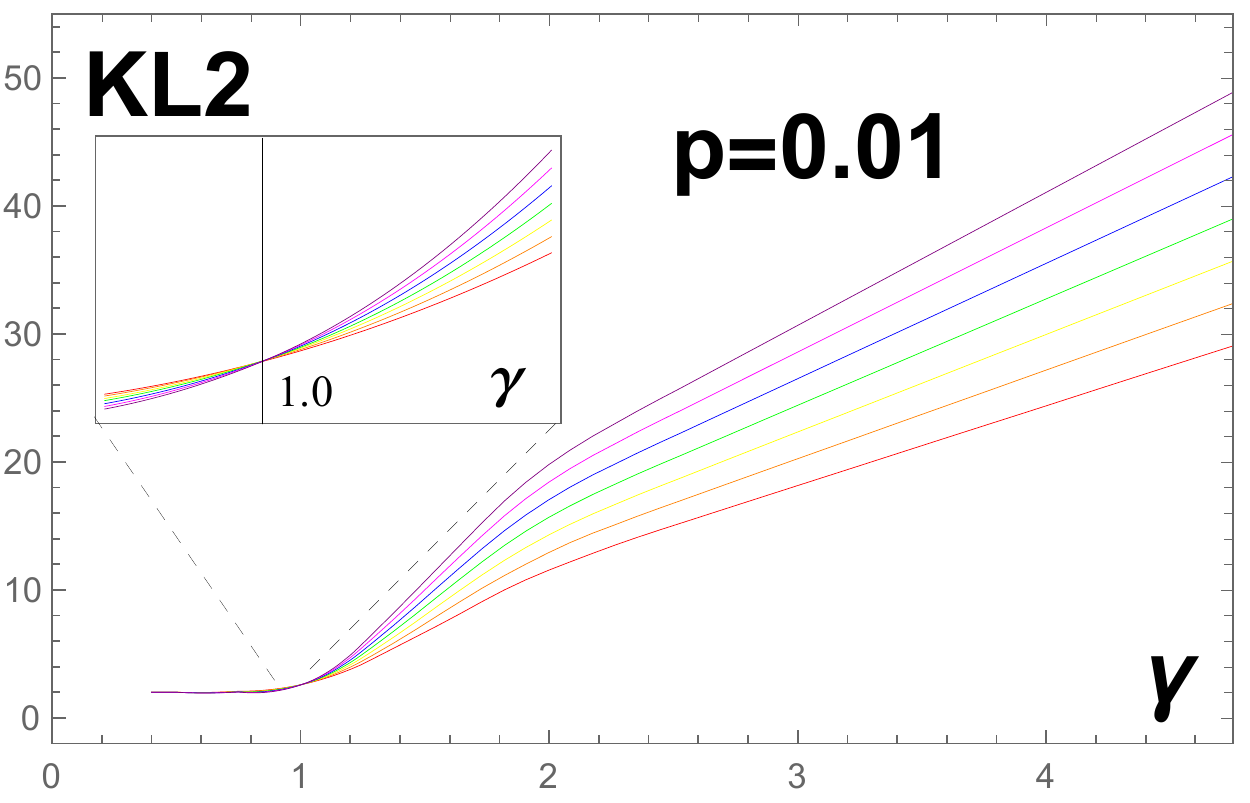}
\includegraphics[width=0.24\linewidth]{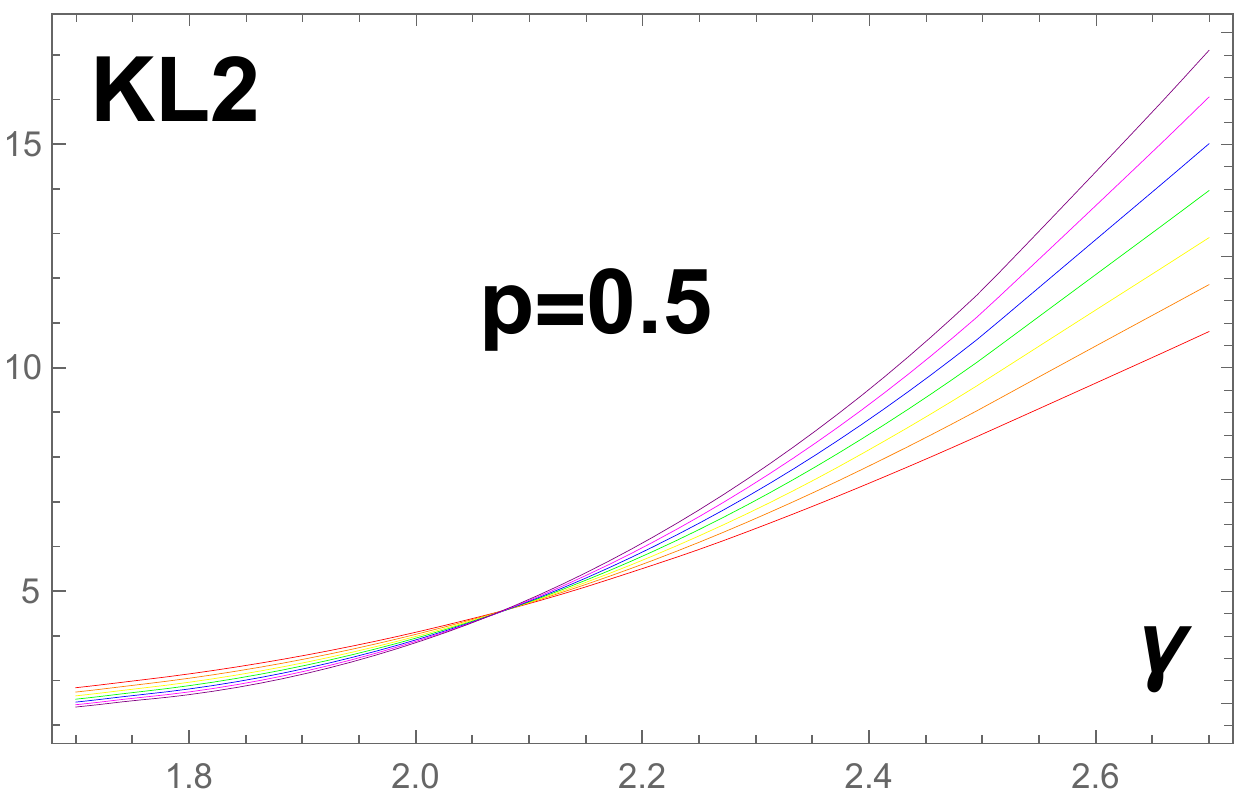}
\includegraphics[width=0.24\linewidth]{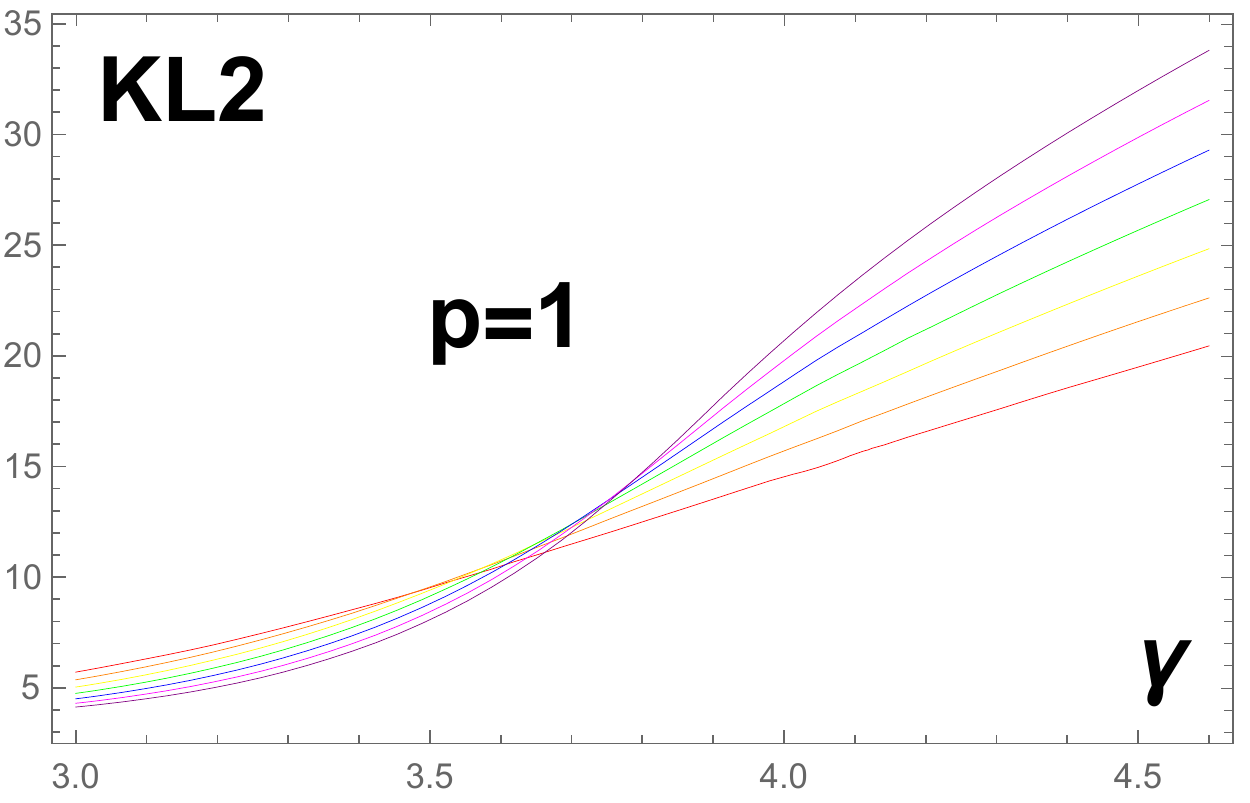}
\includegraphics[width=0.24\linewidth]{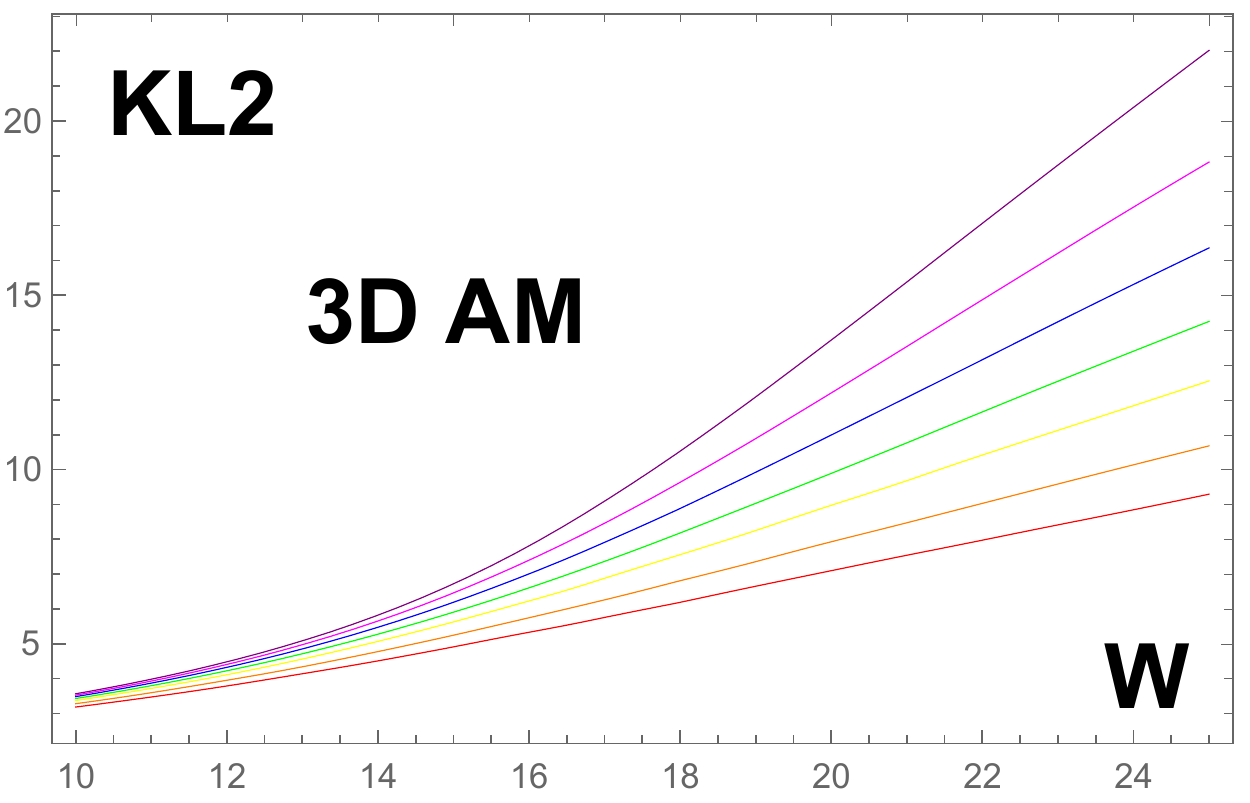}
\caption{(Color online)
\textbf{Plots of $KL1$ and $KL2$ vs. $\gamma$ for LN-RP model}
at 
$N=2^L$, with $L$ from $9$ to $15$ with the step $1$ (from red to violet)
and vs. $W$ for the $3$d 
Anderson model at 
$N=L^3$, with $L=8$, $10$, $13$, $16$, $20$, $25$, $32$.
The logarithmic in $N$ divergence of
$KL1$ for $\gamma>\gamma_{AT}\approx 2$ and of $KL2$ for $\gamma>\gamma_{ET}\approx 1$
is demonstrated in a wide interval of $\gamma$ for $p=0.01$, as well as
insensitivity of $KL1$ to the ergodic transition. Intersection for
$KL2(\gamma$) curves is sharp at the
isolated continuous ergodic
transition at $\gamma_{ET}\approx 1$ for $p=0.01$ and at $\gamma_{ET}\approx 2.1$
for $p=0.5$, it is smeared out for $p=1.0$ when the ergodic transition
merges with the localization transition, and it disappears completely for three-dimensional ($3$d) Anderson
model. Intersection of curves for $KL1$ at the Anderson localization transition
($\gamma_{AT}
\approx 2.0$ for $p=0.01$, $\gamma_{AT}\approx 2.8$ for $p=0.5$,
$\gamma_{AT}\approx 4.1$ for $p=1$, and $W\approx 16.6$ for 3D Anderson model)
is sharp in all the cases.
\label{Fig:intersect} }
\end{figure*}

The fact that it is the second moment of $|H_{nm}|=U$ which enters Eq.~(\ref{Mott}) is
related to the Fermi Golden Rule determining the Breit-Wigner width:
\be\label{FGR}
\Gamma=2\pi\,\sum_{m=1}^{N}\la \rho_{m}\,|H_{nm}|^{2}\ra_{W}\approx
2\pi\,\rho \sum_{m=1}^{N}\langle|H_{nm}|^{2}\rangle_{W},
\ee
where $\rho_{m}$ and $\rho \sim W^{-1}$ are the density of final states
and the
 density of on-site energies, respectively. The perturbative Eq.~(\ref{FGR}) is
valid as long as $\Gamma \lesssim W$ and one can neglect contribution of off-diagonal
matrix elements
$H_{nm}$ to the density of states. Then the total spectral bandwidth is limited by
$W$. In the opposite limit
$\Gamma\gg W $ the total spectral bandwidth $\rho^{-1} \simeq \Gamma$ is dominated by the off-diagonal matrix
elements and should be determined
self-consistently
\be
\rho^{2}\,\sum_{m=1}^{N}\langle |H_{nm}|^{2}\rangle_{W} \sim 1.
\ee
This means that in the fully ergodic phase the total spectral bandwidth is blowing
up with increasing $N$
\be
\rho^{-1} \sim [S_{2}(N)]^{\frac{1}{2}} \to \infty \ .
\ee

For the log-normal distribution~\eqref{LN-MF} one easily computes the moments~\eqref{Anderson} and~\eqref{Mott} truncated at $U_{{\rm max}}\sim O(1)$:
\be\label{mom_q}
\langle U^{q}\rangle_{W}=\left\{\begin{array}{ll}
N^{-\frac{\gamma q}{2}\,\left(1-\frac{pq}{2} \right)},& {\rm if}\;\; pq<1\cr
N^{-\frac{\gamma}{4p}}, & {\rm if}\;\; pq\geq 1\end{array}\right.
\ee
and using this equation finds:
\be\label{S1}
S_{1}=\left\{\begin{array}{ll}N^{1-\frac{\gamma}{2}\left(1-\frac{p}{2} \right)},&{\rm if}\;\;
p<1\cr N^{1-\frac{\gamma}{4p}},& {\rm if}\;\;p\geq 1\end{array}\right. \ ,
\ee
\be\label{S2}
S_{2}=\left\{\begin{array}{ll}N^{1- \gamma\,(1-p)},&{\rm if}\;\;
p<1/2\cr N^{1-\frac{\gamma}{4p}},& {\rm if}\;\;p\geq 1/2\end{array}\right. \ ,
\ee
leading to the critical points of the localization ($\gamma_{AT}$)
and ergodic ($\gamma_{ET}$) transitions from the conditions~\eqref{Anderson} and~\eqref{Mott} that $S_{1}$ or $S_{2}$,
respectively, are of order $O(1)$:
\be\label{AT}
\gamma_{AT}=\left\{\begin{array}{ll}\frac{4}{2-p}, & {\rm if}\;\; p<1\cr
4p,& {\rm if}\;\; p\geq 1 \end{array}\right.
\ee
\be\label{ET}
\gamma_{ET}=\left\{\begin{array}{ll} \frac{1}{1-p}, & {\rm if}\;\; p<1/2\cr
4p,& {\rm if}\;\;p\geq 1/2\end{array} \right.
\ee
The resulting phase diagram for the log-normal Rosenzweig-Porter ensemble, Eq.~\eqref{LN-MF},
which is the main result of this paper, is presented in Fig.~\ref{Fig:phase_diagram}.

It is remarkable that the point $p=1$ which corresponds to the Anderson model on RRG,
is the {\it tricritical} point on this phase diagram. At this point in the pure
log-normal
RP ensemble and in LN-RP ensemble with $P(U)$ truncated at
$U>U_{{\rm max}}\sim O(1)$, the multifractal phase vanishes.
However, as we demonstrate in Sec.~\ref{sec:Truncation}, the collapse of the
multifractal phase at $p\geq 1$
and thus the tricritical point,  is unstable with respect to any truncation of $P(U)$ with
$U_{{\rm max}}\ll O(1)$.

\section{Kullback-Leibler (KL) measure}\label{sec:KL}
The numerical verification of the phase diagram and determination of the
critical exponents
at the Anderson localization and ergodic transitions is done in this paper using the
Kullback-Leibler divergence (KL)~\cite{KLdiv, KLdiv_book, KLPino} of certain correlation
functions of random eigenstates (for more detailed multifractal analysis of this
model  see~\cite{KrKhay}).

The Kullback-Leibler correlation functions $KL1$ and $KL2$ are defined as follows
~\cite{KLPino}.
The first one is defined in terms of wave functions of two
{\it neighboring in energy} states:
\be\label{KL1}
KL1=\sum_{i}|\psi_{\alpha}(i)|^{2}\,\ln\left(\frac{|\psi_{\alpha}(i)|^{2}}
{|\psi_{\alpha+1}(i)|^{2}}\right).
\ee
The second one is similar but the states $\psi$ and $\tilde{\psi}$ correspond to
different (and totally uncorrelated) disorder realizations:
\be\label{KL2}
KL2=\sum_{i}|\psi (i)|^{2}\,\ln\left(\frac{|\psi (i)|^{2}}
{|\tilde{\psi}(i)|^{2}}\right).
\ee
The idea to define such two measures is the following. In the ergodic phase each of
the states has an amplitude $|\psi(i)|^{2}\sim N^{-1}$ of the same order of magnitude.
Then the logarithm of their ratio is of order $O(1)$, and for the normalized states
\be
KL1\sim KL2 \sim O(1).
\ee
For fully-ergodic states in the Wigner-Dyson limit the eigenfunction coefficients are
fully uncorrelated, even for the neighboring in energy states. Thus there is no
difference between $KL1$ and $KL2$. Using the Porter-Thomas distribution one finds:
\be\label{KL-EE}
KL1=KL2=2.
\ee

Deeply in the localized phase
$\ln |\psi_{\alpha}(i)|^{2}\sim -|i-i_{\alpha}|/\xi $, where $i_{\alpha}$
is the position of the
localization center.
Since the positions of localization centers $i_{\alpha}$ are not correlated
even for the states
neighboring in the energy, $KL1$ and $KL2$ are proportional to $L\sim \ln N$ and
divergent in the thermodynamic limit:
\be\label{KL-L}
KL1\sim KL2 \propto \ln N \rightarrow\infty.
\ee
The properties of KL, Eqs.~(\ref{KL-EE}),~(\ref{KL-L}) are fully confirmed by
numerics presented in Fig.~\ref{Fig:intersect}.

A qualitative difference between $KL1$ and $KL2$ is in the multifractal NEE phase.
In this phase the neighboring in energy states $|\psi_{\alpha}(i)|^{2}$ and
$|\psi_{\alpha+1}(i)|^{2}$ are
most probably belonging
to the same support set and hence they
are strongly overlapping: $|\psi_{\alpha}(i)|^{2}\sim |\psi_{\alpha+1}(i)|^{2}$.
Furthermore, eigenfunctions on the same fractal support set can be represented as:
$\psi_{\alpha}(i)=\Psi(i)\,\phi_{\alpha}(i)$, where $\Psi(i)$ is the multifractal
envelope on the support set and $\phi_{\alpha}(i)$ is the fast oscillating function
with the
Porter-Thomas statistics~\cite{DeLuca2014}. Thus the ratio
$|\psi_{\alpha}(i)|/|\psi_{\alpha+1}(i)|$ and hence $KL1$ has the same statistics
as in the ergodic phase. In other words, $KL1$ is {\it not sensitive}
to the {\it ergodic} transition but is {\it very sensitive} to the {\it localization}
one, Fig.~\ref{Fig:intersect}.

In contrast, the eigenfunctions $\psi(i)$ and
$\tilde{\psi}(i)$ corresponding to {\it different realizations} of a random Hamiltonian in $KL2$,
overlap very poorly. This is because the fractal support sets which
contain a vanishing fraction of all the sites,
do not typically overlap when taken at random. Therefore the ratio
$\ln(|\psi(i)|/|\tilde{\psi}(i)|)\sim \ln N$ in $KL2$ is divergent in
the thermodynamic limit in the {\it multifractal} phase, very much like in the localized one.
This makes $KL2$ {\it very sensitive} to the {\it ergodic} transition, Fig.~\ref{Fig:intersect}.

\begin{figure*}[t]
\includegraphics[width=0.32\linewidth]{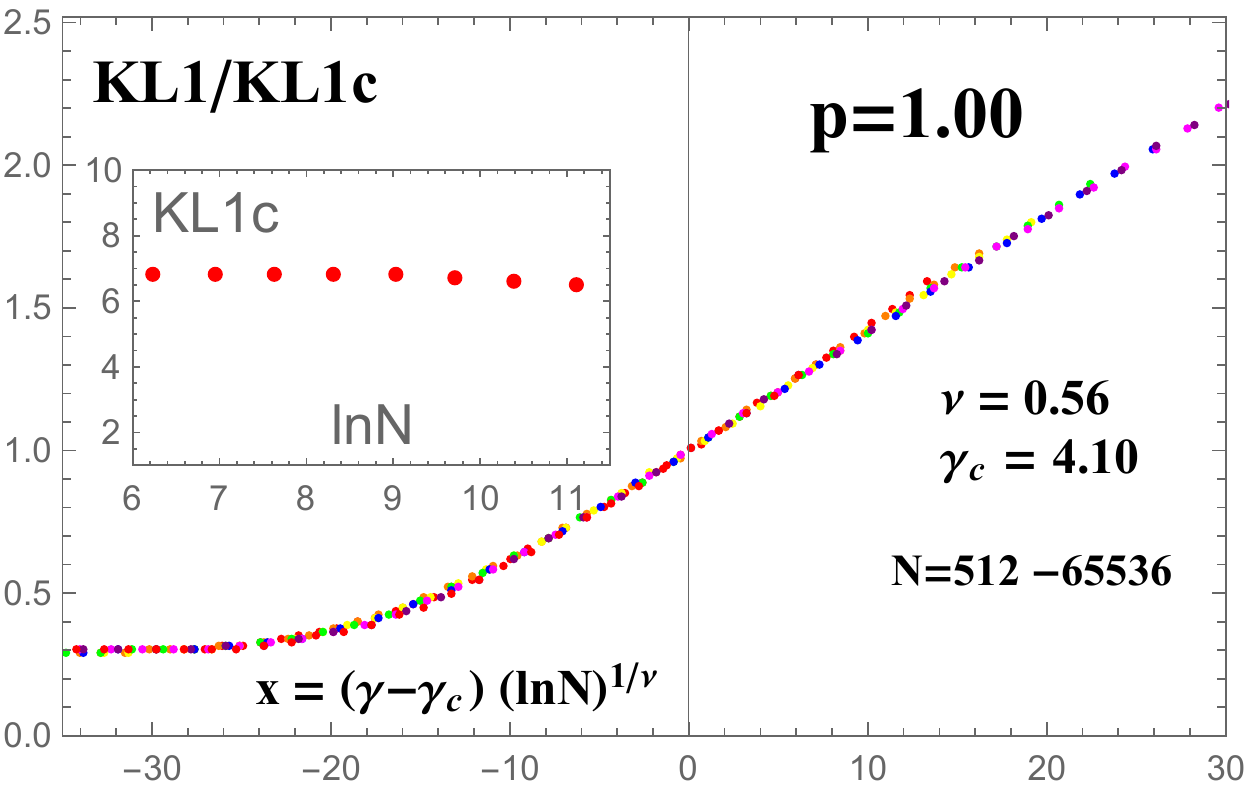}
\includegraphics[width=0.32\linewidth]{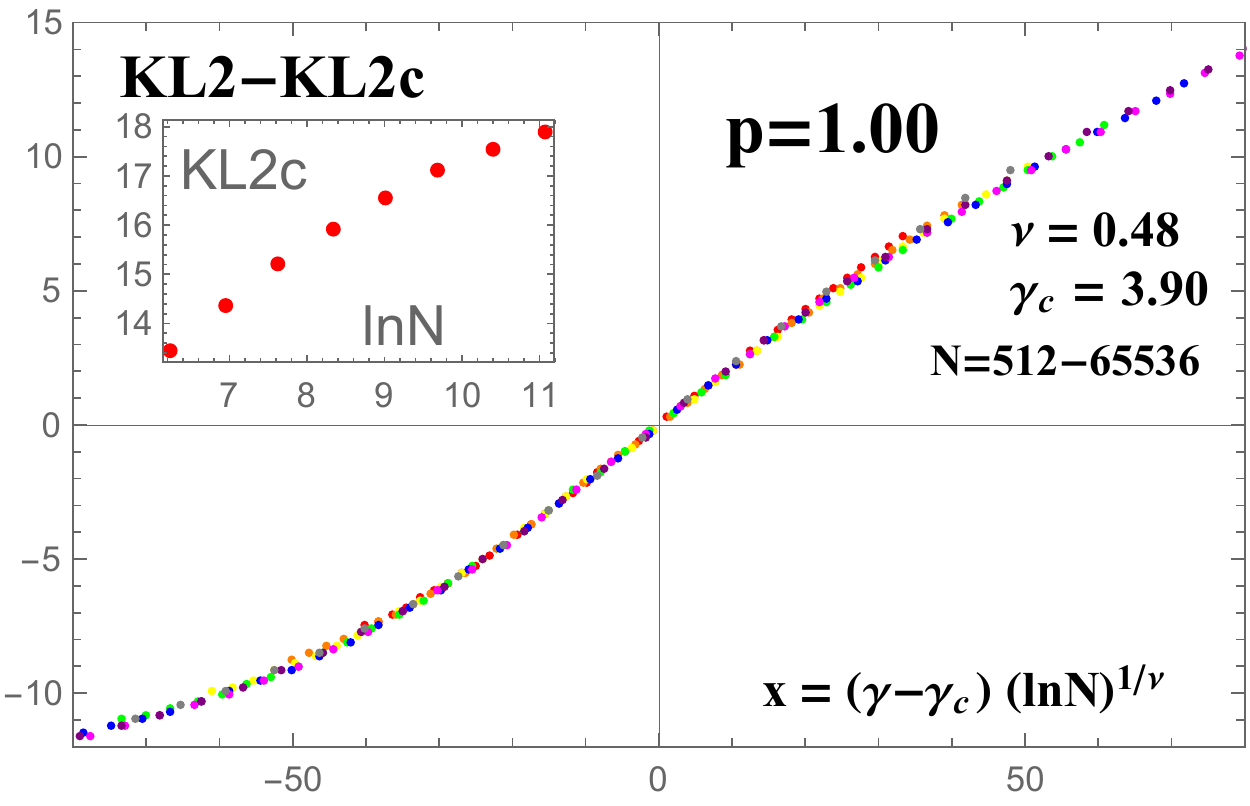}
\includegraphics[width=0.32\linewidth]{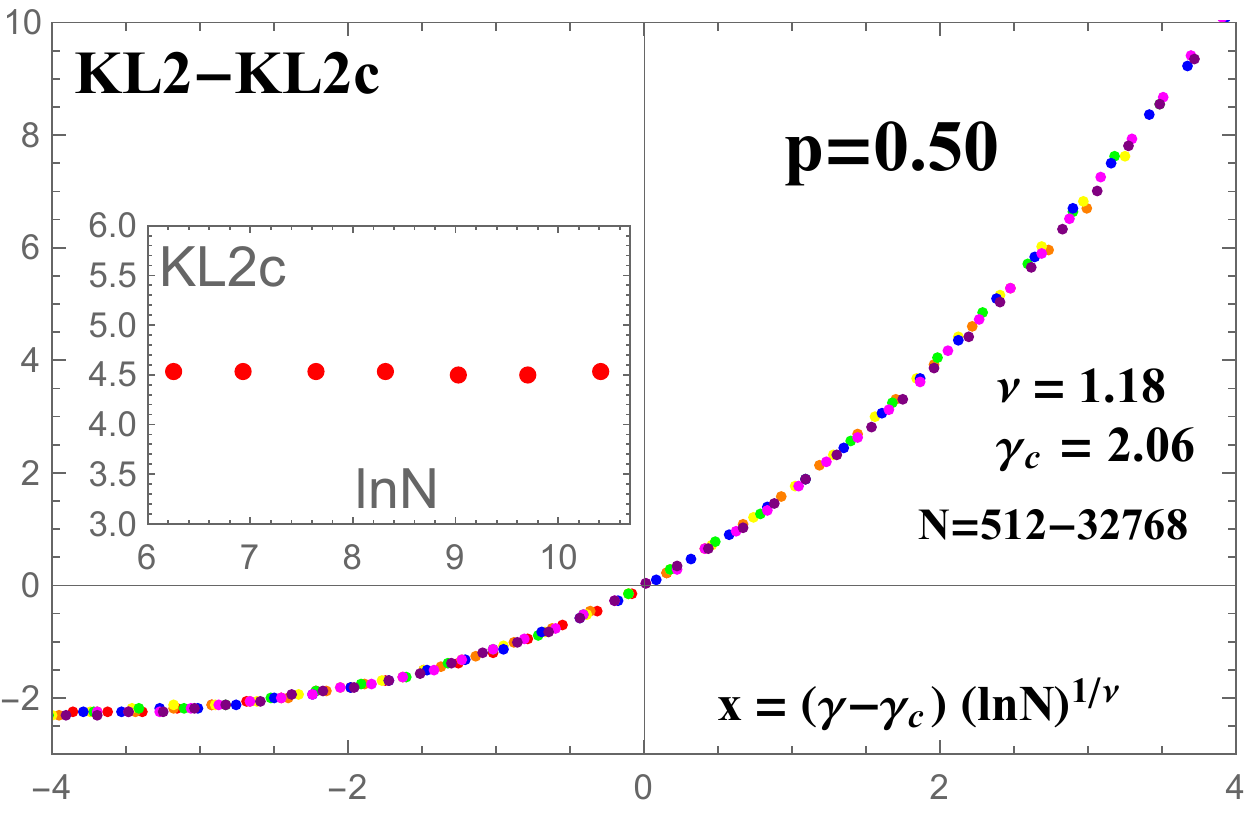}
\caption{(Color online) \textbf{The best collapse of the $KL1$ and $KL2$ data for LN-RP} with
$p=1$ and $p=0.5$. The collapse for $KL1$ and $KL2$ is done in the vicinity of
the localization (for $KL1$) and  ergodic (for $KL2$)
transitions by recursive procedure that finds $\gamma_{c}$ and $\nu$ by minimizing
the mean square deviation of data from a smooth scaling function which is updated at any
step of the procedure.
(insets) The critical value of $KL1$ and $KL2$ as a function of $\ln N$.
It stays almost a constant for $KL1$ and for $KL2$ at $p=0.5$ when the ergodic transition is continuous
and well separated from the localized one
but it grows logarithmically in $N$ at $p=1$ when the ergodic and localization
transitions merge together.
This growth is the reason of smearing of the intersection of
$KL2$ curves in Fig.~\ref{Fig:intersect}.
The exponent $\nu$ significantly depends on $p$ and is
consistent with $\nu_{1}\approx \nu_{2}=0.5$ at $p=1$ and $\nu_{2}=1$ at $p=0.5$.
\label{Fig:collapse_LN-RP} }
\end{figure*}

A more detailed theory of $KL1$ and $KL2$ in the multifractal phase
is given in Appendix~\ref{App_sec:theory_KL}. The main conclusion of the analysis done in
Appendix~\ref{App_sec:theory_KL} is that the curves for $KL1(\gamma,N$) for different $N$ have an
intersection point at
 the critical
point $\gamma=\gamma_{AT}$ of the Anderson localization transition.
 At the same time, the intersection point for curves for
$KL2(\gamma,N$) coincides with the ergodic transition~\cite{KLPino},
provided that it is continuous
and well separated from the Anderson localization transition. If the localization and ergodic
transition merge together and the multifractal state exists only at
the transition point (as in 3D Anderson model) then intersection of $KL2$ curves is
smeared out and may disappear whatsoever. However, the intersection of $KL1$ curves remains sharp
in this case too (see Fig.~\ref{Fig:intersect}).

\begin{figure}[b!]
\center{
\includegraphics[width=0.8 \linewidth,angle=0]{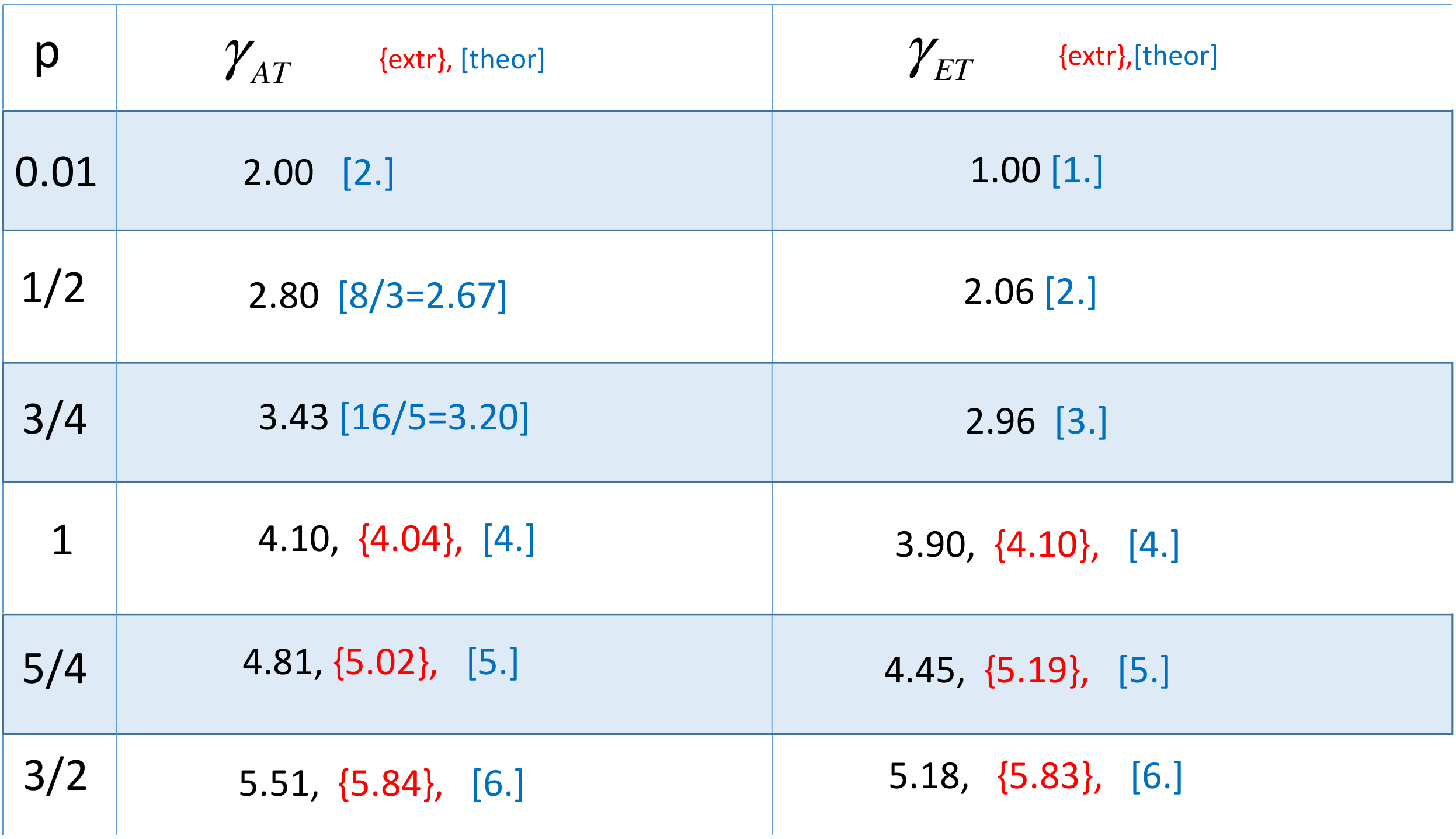}}
\caption{\textbf{Comparison of analytical predictions (blue), Eq.~(\ref{AT}),~(\ref{ET}), and numerical data for the transition points $\gamma_{AT}$ and
$\gamma_{ET}$ for LN-RP model.} Numerical data (black) is obtained by
exact diagonalization of LN-RP random matrices with $N=512-32768$
from the intersection points in $KL1$ and $KL2$, Fig.~\ref{Fig:intersect}
and from finite-size scaling, Fig.~\ref{Fig:collapse_LN-RP}.
For $p>1$ a linear in $1/\ln N$ extrapolation to $N\rightarrow \infty$ of
the position of the
intersection point for two consecutive $N$ is shown in red.
 }
\label{Fig:table1}
\end{figure}

The intersection of finite-size curves for $KL1$ and $KL2$ helps to locate numerically
the critical points $\gamma_{AT}$ and $\gamma_{ET}$.
First, we checked that for the well studied $3$d Anderson transition
the intersection point of $KL1$ curves exactly corresponds to the
known critical disorder $W\approx 16.56$, while $KL2$ curves show
no intersection whatsoever (see Fig.~\ref{Fig:intersect}). The results for $\gamma_{AT}$
and $\gamma_{ET}$ for LN-RP model are shown in the Table~\ref{Fig:table1}. They coincide (for $p>1$ after
the extrapolation) with the
theoretical prediction Eq.~(\ref{AT}), (\ref{ET}) with the deviation less than $6\%$.

\begin{figure}[b!]
\center{
\includegraphics[width=0.8 \linewidth,angle=0]{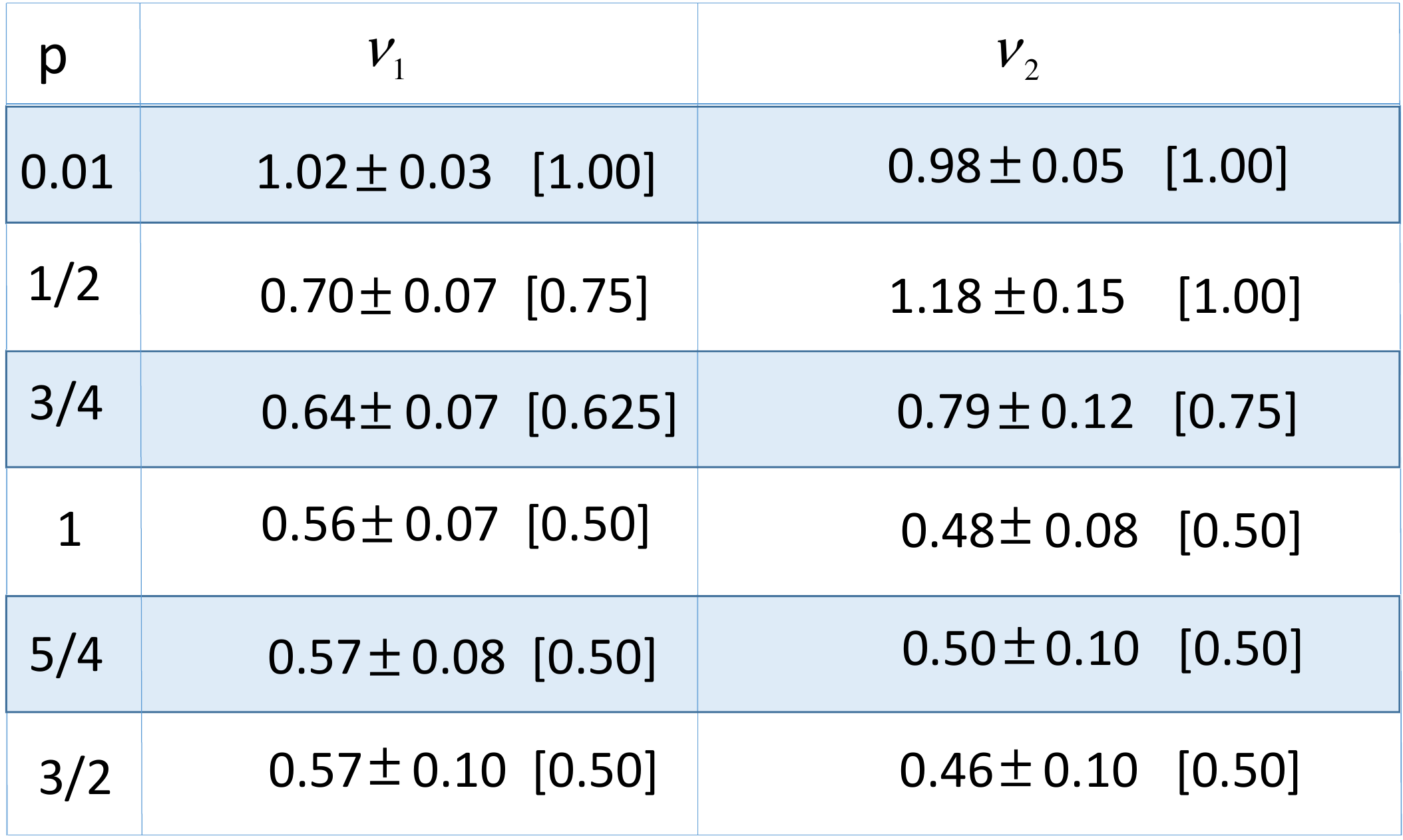}}
\caption{Critical exponents $\nu_{1}$ and $\nu_{2}$ in the
finite-size scaling, Eq.~(\ref{collapse-form}), obtained from the best collapse of $KL1$ and $KL2$ data,
respectively, see Fig.~\ref{Fig:collapse_LN-RP}.
In the $[...]$ are the conjectured values of $\nu_{1}$ and $\nu_{2}$.
 }
\label{Fig:tab_nu}
\end{figure}

The next step is to analyze the {\it finite-size scaling} (FSS) by a collapse of the
data for $KL1$ and $KL2$ at different $N$ in the
vicinity of the localization and ergodic transition, respectively.
To this end we use the form of FSS derived in Appendix~\ref{App_sec:theory_KL}:
\bes\label{collapse-form}
\begin{align}
{KL1} &=
\Phi_{1}(\ln N |\gamma-\gamma_{AT}|^{\nu_{1}}),\\
{KL2} - {KL2_c (N)} &=
\Phi_{2}(\ln N |\gamma-\gamma_{ET}|^{\nu_{2}}).
\end{align}
\ees
The input data for the collapse is $KL1$ and $KL2$ versus $\gamma$ and $W$ for $7$
values of $N$ is shown in Fig.~\ref{Fig:intersect}.
The fitting parameters extracted from the best collapse
are $\nu_{1}$ ($\nu_{2}$) and the critical points $\gamma_{AT}$ ($\gamma_{ET}$).
The critical value of ${KL2_c}(N)={KL2}(\gamma_{ET},N)$
is determined by the best fitting for $\gamma_{ET}$.
For the localization transition
where the critical point $\gamma_{AT}$ is well defined by
the intersection in $KL1$, one may look for the best collapse by fitting only $\nu_{1}$.

This procedure of the finite-size scaling has been tested for the 3D Anderson model with sizes
$L=8-32$. The results for the scaling collapse of data are presented in Appendix~\ref{App_sec:KL_collapse}. Note that in
this case there is no intersection in $KL2$ whatsoever (see Fig.~\ref{Fig:intersect}). Yet,
the best collapse corresponds to a well-defined $W_{c}\approx 17$ which is reasonably
close to the value $W_{c}=16.56$ found from the intersection in $KL1$ and known in the
literature. This encourages us to use the best collapse of $KL2$ data to determine $\gamma_{ET}$ and
$\nu_{2}$ for LN-RP model
where the intersection of $KL2$ curves does exists, albeit smeared out.

\begin{figure}[t!]
\center{
\includegraphics[width=0.8 \linewidth,angle=0]{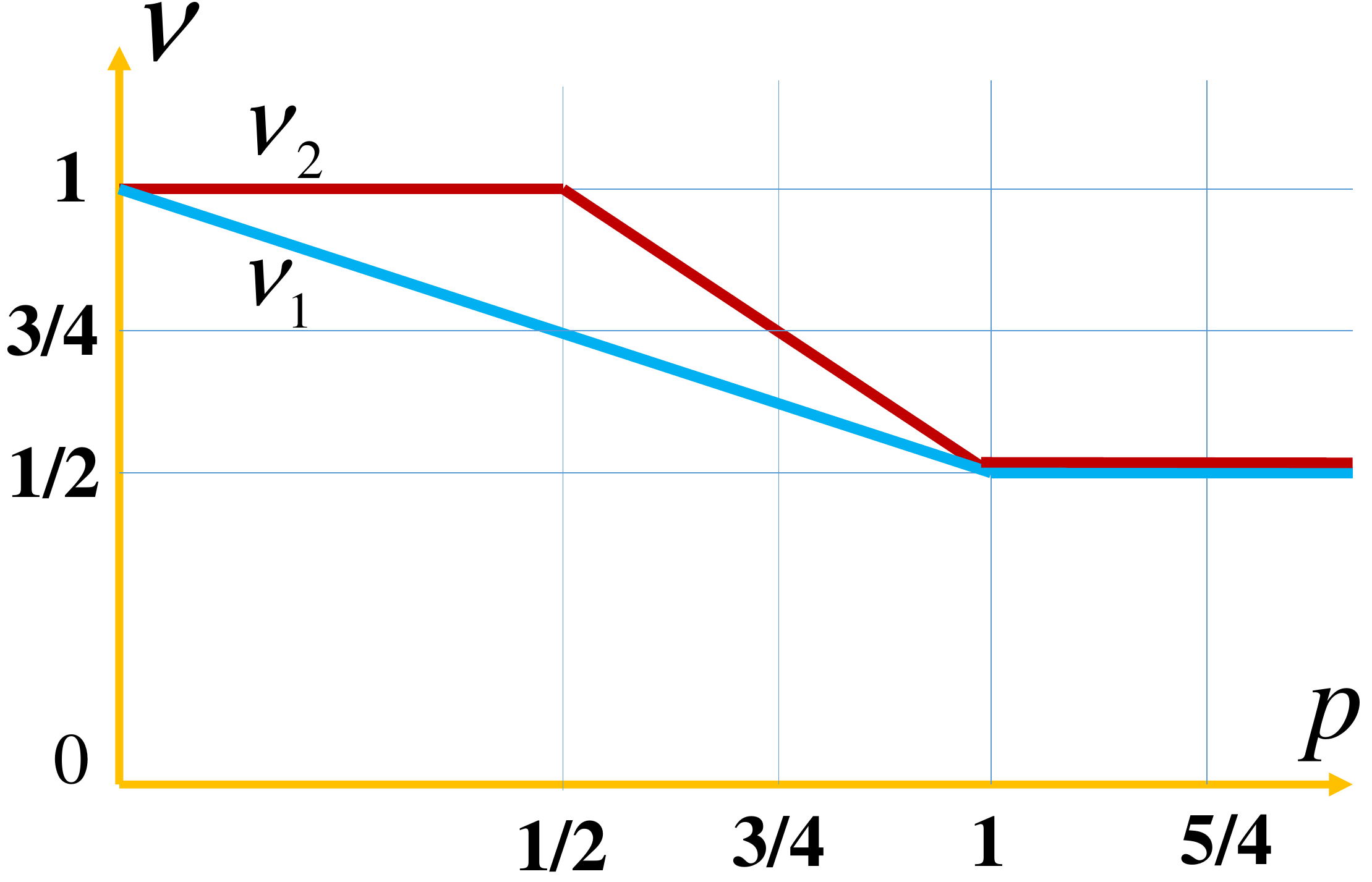}}
\caption{(Color online) \textbf{The conjectured dependence of the critical exponents $\nu_{1}$ and
$\nu_{2}$ on the symmetry parameter $p$ for 
LN-RP} at the localization and ergodic
transitions, respectively. In the limit $p\to 0$ the critical exponents approach
their values $\nu_{1}=\nu_{2}=1$ for the Gaussian RP model~\cite{KLPino}. For $p\geq 1$
we conjecture the mean-field values $\nu_{1}=\nu_{2}=1/2$. In the interval $0<p<1$
the critical
exponents of the Anderson ($\nu_{1}$) and ergodic ($\nu_{2}$)
transitions are different with $\nu_{2} >\nu_{1}$.
 }
\label{Fig:conj-nu}
\end{figure}

The results are shown in the Tables~\ref{Fig:table1},~\ref{Fig:tab_nu} while representative
samples of the data collapse are shown in Fig.~\ref{Fig:collapse_LN-RP}.
On the basis of these
numerical results we conjecture the dependence of the critical exponents
$\nu_{1}$ and $\nu_{2}$
on the symmetry parameter which is shown in Fig.~\ref{Fig:conj-nu}.

\section{Truncated LN-RP and fragility of ergodic phase.}\label{sec:Truncation}
\begin{figure*}
\includegraphics[width=0.45\linewidth]{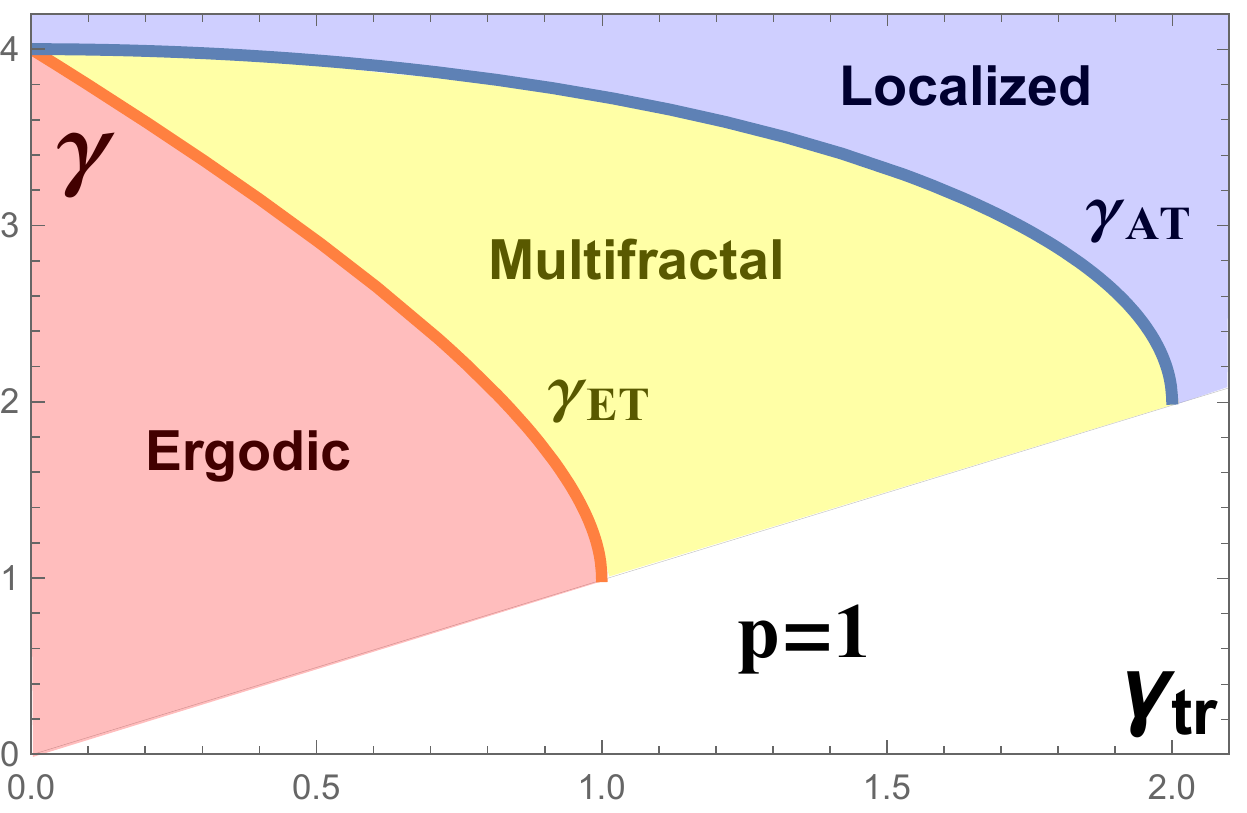}
\includegraphics[width=0.45\linewidth]{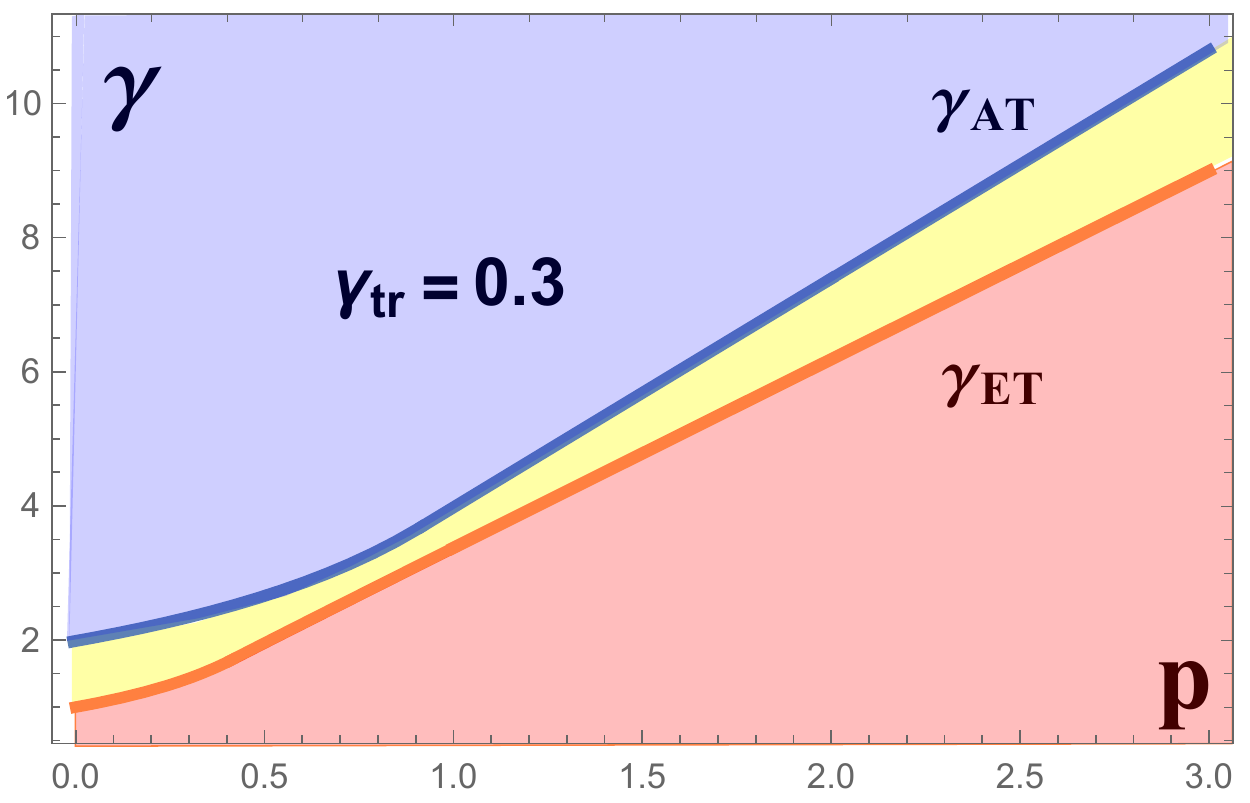}
\caption{(Color online) \textbf{Phase diagram of LN-RP model with
$U_{{\rm typ}}\sim N^{-\gamma}$ truncated at
$U_{{\rm max}}\sim N^{-\gamma_{tr}}$
($\gamma>\gamma_{tr}>0$).}
(Left~panel)~Phase diagram in the plane $\gamma-\gamma_{tr}$
at a fixed value $p=1$ of the symmetry parameter.
(Right~panel)~Phase diagram
in the plane $\gamma-p$ at fixed $\gamma_{tr}=0.3$. At any $\gamma_{tr}>0$ the
multifractal NEE phase  emerges at $p\geq 1$ and fills the gap between the ergodic and localized phases.
At a small $\gamma_{tr}$ the Anderson localization transition (blue line) is almost
unaffected by truncation,
while the ergodic transition (orange line) is pushed to smaller values of $\gamma$.
Thus the multifractal NEE phase substitutes
the ergodic one as the truncation parameter $\gamma_{tr}$ increases demonstrating
the fragility of the ergodic phase which existence is due to atypically large values of the
transition matrix elements $U$.
\label{Fig:phase_dia_trunc} }
\end{figure*}

The main result, Fig.~\ref{Fig:phase_diagram}, of Sec.~\ref{sec:Phase_diagram} confirmed numerically in
Sec.~\ref{sec:KL} is the collapse of the multifractal phase   at $p\geq 1$ and
existence of
the tricritical point in LN-RP which is associated, via the qualitative arguments
of Sec.~\ref{sec:RRG_motivation}, with the localization
transition on RRG.

In this section we show that the ergodic phase that emerges at the
localization transition in this tricritical point (and for all $p\geq 1$)
is unstable with respect to a deformation of LN-RP model such that $P(U)$ is cut from
above at:
\be
U_{\rm max}\sim N^{-\gamma_{tr}/2}\ll O(1), \;\;\;\;(\gamma_{tr}>0).
\ee
As the result of this truncation the multifractal phase re-appears by substituting
a part of the ergodic phase in a non-truncated LN-RP model
(see Fig.~\ref{Fig:phase_dia_trunc})~\cite{footnote:trunc}.
To this end we use the expression that generalizes Eq.~(\ref{mom_q}):
\begin{eqnarray}\label{us-id}
&&\int_{0}^{N^{-\gamma_{\rm tr}/2}}dU\,U^{q}\,P(U)\sim \nonumber \\ &&\sim \left\{
\begin{array}{ll} N^{-\frac{q\gamma}{2}\left(1-\frac{pq}{2} \right)}, &
\gamma_{\rm tr}<\gamma(1-pq)\cr N^{-\frac{1}{p\gamma} \left[\frac{
(\gamma-\gamma_{\rm tr})^{2}}{4}+\frac{1}{2}\,pq\,\gamma\gamma_{\rm tr}\right]},
& \gamma_{\rm tr}>\gamma(1-pq)
\end{array} \right.
\end{eqnarray}
and apply the same criteria Eq.~(\ref{Anderson}),~(\ref{Mott}) to find the critical points
of the localization and ergodic transitions. Then we obtain for the critical point
$\gamma_{AT}$ of
the Anderson localization transition:
\be\label{AT-trunc}
 \gamma_{AT} = 2p-(p-1)\gamma_{tr}+\sqrt{(2p-(p-1)\gamma_{tr})^{2}-\gamma_{tr}^{2}},
\ee
if $\gamma_{tr}>\gamma_{AT}\,(1-p)$. In the opposite case truncation does not affect
$\gamma_{AT}$.

For the critical point $\gamma_{ET}$ of the ergodic transition in the same way we find
for $\gamma_{tr}>\gamma_{ET}\,(1-2p)$:
\be\label{ET-trunc}
 \gamma_{ET}=2p-(2p-1)\gamma_{tr}+\sqrt{(2p-(2p-1)\gamma_{tr})^{2}-\gamma_{tr}^{2}}.
\ee
The results of Eq.~(\ref{AT-trunc}),~(\ref{ET-trunc}) are plotted in
Fig.~\ref{Fig:phase_dia_trunc}.

One can see that at {\it any} positive non-zero $\gamma_{tr}$
the multifractal NEE phase emerges at $p\geq 1$ in between of the localized and
ergodic ones. At small $\gamma_{tr}$ the line of localization transition is almost
insensitive to truncation, while the line of ergodic transition is pushed to
smaller values of $\gamma$ corresponding to larger typical
transition matrix elements $U$ (smaller effective disorder). This proves the fact that
the ergodic phase in LN-RP with $p\geq 1$ is very fragile and exists only due to atypically large
transition matrix elements. It is substituted by the multifractal NEE phase as soon
as such matrix elements are made improbable by truncation.

We believe that this scenario of the multifractal phase emergence at $p\geq 1$
is quite generic
and happens for the wide class of perturbations of the
LN-RP model~\cite{RRG_Gorsky-footnote}.
In the case of RRG  corresponding to the tricritical point, $p=1$,
of the non-truncated LN-RP model,
the effect of the local Cayley tree structure in the exact mapping of the Anderson model on
RRG onto LN-RP model
is unexplored in detail and might, in principle,
lead to an effective truncation of the above type.
In any scenario the tricritical nature of $p=1$
point in LN-RP makes this case (and the corresponding case of RRG) significantly
more complicated and different from the conventional Anderson localization transition
in finite dimensions. This is the reason, in our opinion, of the long-lasting
debates on the existence of NEE phase in the Anderson model on RRG (see the debates in~\cite{AnnalRRG, Tikh-Mir1, Tikh-Mir2, Scard-Par,
RRG_R(t),deTomasi2019subdiffusion} and references therein).

\section{Stability of non-ergodic states against hybridization}\label{sec:MF-hybridization}

\begin{figure}[t!]
\center{
\includegraphics[width=1.0 \linewidth,angle=0]{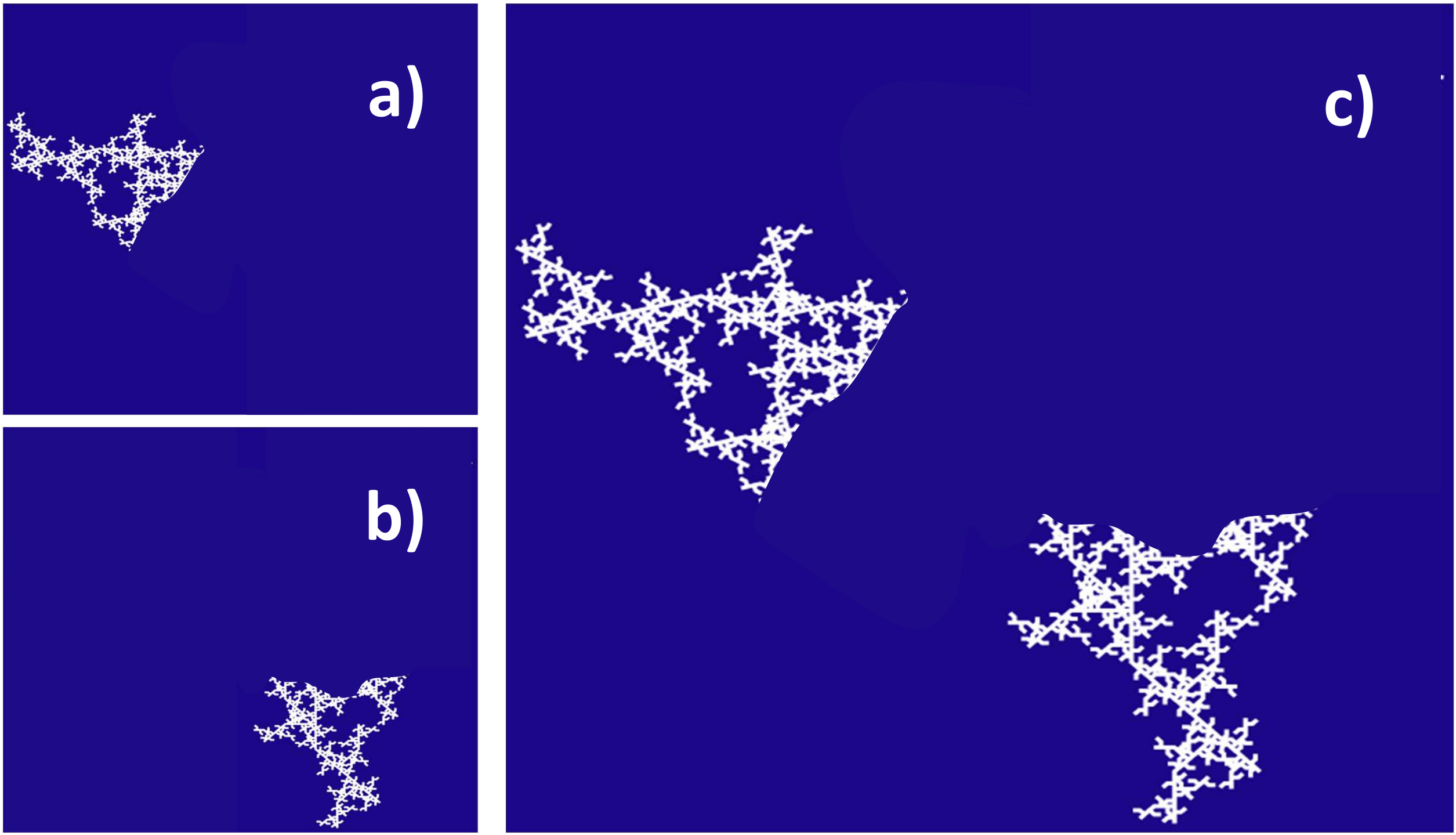}}
\caption{(Color online) \textbf{Hybridization of fractal support sets}
(a),~(b)~Two different fractal support sets,
(c)~The hybridized fractal support set.}
\label{Fig:hybrid}
\end{figure}
In this section we consider the stability of non-ergodic (multifractal and localized) states
 against hybridization. It allows us not only to derive expressions,
Eqs.~\eqref{Anderson} and~\eqref{Mott}, for the lines of the Anderson
localization and ergodic transitions in a different way
but also find in Sec.~\ref{sec:D_1}   the
fractal dimension $D_{1}(p,\gamma)$ of the
multifractal support set.  Furthermore, the new method
presented below is physically transparent and generic enough to be applied to analysis of the multifractal NEE
states in other systems.

Let us consider two states $\psi_{\mu}(i)$ and
$\psi_{\nu}(i)$ on different fractal support
sets as it is shown in Fig.~\ref{Fig:hybrid}(a) and~(b).
We assume that both states are multifractal with $m \sim N^{D_{1}}$ sites
on a fractal support set where $|\psi(i)|^{2}\sim N^{-D_{1}}$.

The key new element in the
theory we are introducing here is the transmission matrix element $V_{\mu,\nu}$
between the {\it states} and not between the {\it sites}
as we did in the previous sections
\be\label{U-mu-nu}
V_{\mu,\nu}=\sum_{i,j}G_{ij}\,\psi_{\mu}(i)\,\psi_{\nu}(j),
\ee
where $G_{ij}$ is the two-point Green's function.

Introducing $g_{ij}=-\ln G_{ij}/\ln N$ and suppressing the indices $i,j$ for brevity
we conveniently rewrite Eq.~(\ref{LN}) as follows:
\be\label{P-g}
{\cal P}(g)={\rm const} \, N^{-\frac{1}{p\gamma}\,\left( g-\frac{\gamma}{2}
\right)^{2}},\;\;\;(g\geq 0).
\ee
By the constraint $g\geq 0$ we implemented a cutoff at $G_{\rm max}\sim O(1)$
discussed in Section~\ref{sec:Phase_diagram} and Appendix~\ref{App_sec:Greens_funct_BL}.

The typical number of terms in the sum Eq.~(\ref{U-mu-nu}) with $g$ in the interval
$dg$
is $N^{D_1}N^{D_1}{\cal P}(g)\sim N^{\sigma(g,D_{1})}\,d g$ where
\be
\sigma(g,D_{1})=2D_{1}-\frac{1}{p\gamma}\,
\left( g-\frac{\gamma}{2} \right)^{2}.
\ee
If $\sigma(g,D_{1})<0$ (region I in Fig.~\ref{Fig:regions} of Appendix~\ref{App_sec:Stability}),
the sum, Eq.~(\ref{U-mu-nu}), is dominated by a single term with the largest $G_{ij}$.
For positive
$\sigma(g,D_{1})>0$ (region II in Fig.~\ref{Fig:regions} of Appendix~\ref{App_sec:Stability}), many terms
contribute to this sum and the distribution $P(V\equiv |V_{\mu,\nu}|)$
becomes Gaussian. In general, there are both contributions given by
\begin{multline}\label{distr-V}
 P(V)=\int_{g\in I}dg\,N^{\sigma(g,D_{1})}\,\delta(V-N^{-D_{1}-g})+
\\ \int_{g\in II} dg\,{\cal P}(g)\,\la \delta\left
(V-\left|\sum_{ij} G_{ij}\,\psi_{\mu}(i)\psi_{\nu}(j)\right| \right)\ra.
\end{multline}
The condition of stability of the multifractal phase against hybridization
is derived similar to the Anderson criteria of stability, Eq.~(\ref{Anderson}),
of the localized phase done in Sec.~\ref{sec:MF-hybridization}.
The difference is that now we have to replace the matrix element between the resonant
{\it sites} $U$ by the matrix element $V$ between the resonant {\it non-ergodic states}
and take into account that on each of $M=N^{1-D_{1}}$ different
support sets there are $m=N^{D_{1}}$ wave functions which belong to the same mini-band
and thus are {\it already in resonance} with each other.
Therefore the total number of {\it independent} states-candidates for hybridization
 with a given state should be smaller than the total number of states $M\,m=N$ and larger than the number of support sets $M$. This number is in fact equal to
their geometric mean $\sqrt{N M} = M\,\sqrt{m}=N^{1-\frac{D_{1}}{2}}$.

With this comment, the criterion of stability of the multifractal phase reads in the limit $N\rightarrow\infty$ as
\be\label{stab-MF}
N^{1-\frac{D_{1}}{2}}\int_{0}^{W} dV\, V\, P(V)< \infty \ .
\ee
The contribution of the Gaussian part of $P(V)$ to Eq.~(\ref{stab-MF}) is:
\be\label{stab-MF-Gauss}
N^{1-\frac{D_{1}}{2}}\,\sqrt{\langle V^{2} \rangle}= N^{1-\frac{D_{1}}{2}-
\frac{1}{2}\gamma_{{\rm eff}}(D_{1})}<\infty,
\ee
where
\be
\langle V^{2} \rangle \equiv N^{-\gamma_{{\rm eff}}}.
\ee
The contribution of the first (log-normal) term in Eq.~(\ref{distr-V}) to the stability
criterion is:
\be\label{stab-MF-LN}
N^{1-\frac{D_{1}}{2}} \int_{g\in I}dg\, N^{\sigma(g,D_{1})-g-D_{1}}
\equiv N^{1-\frac{D_{1}}{2}-\Delta(D_{1})}<\infty.
\ee
Thus the multifractal phase is stable against hybridization
if the following inequalities are both fulfilled
\bea\label{stab-MF-fin-Gauss}
 \frac{1}{2} D_{1}+\frac{1}{2}\gamma_{{\rm eff}}(D_{1}) &\geq& 1,\\
 \frac{1}{2} D_{1}+\Delta(D_{1}) &\geq & 1 \label{stab-MF-fin-LN}.
\eea
The function $\gamma_{{\rm eff}}(D_{1})$ and $\Delta(D_{1})$ are computed in
Appendix~\ref{App_sec:Stability}.
%

\begin{figure*}[t]
\includegraphics[width=0.3\linewidth]{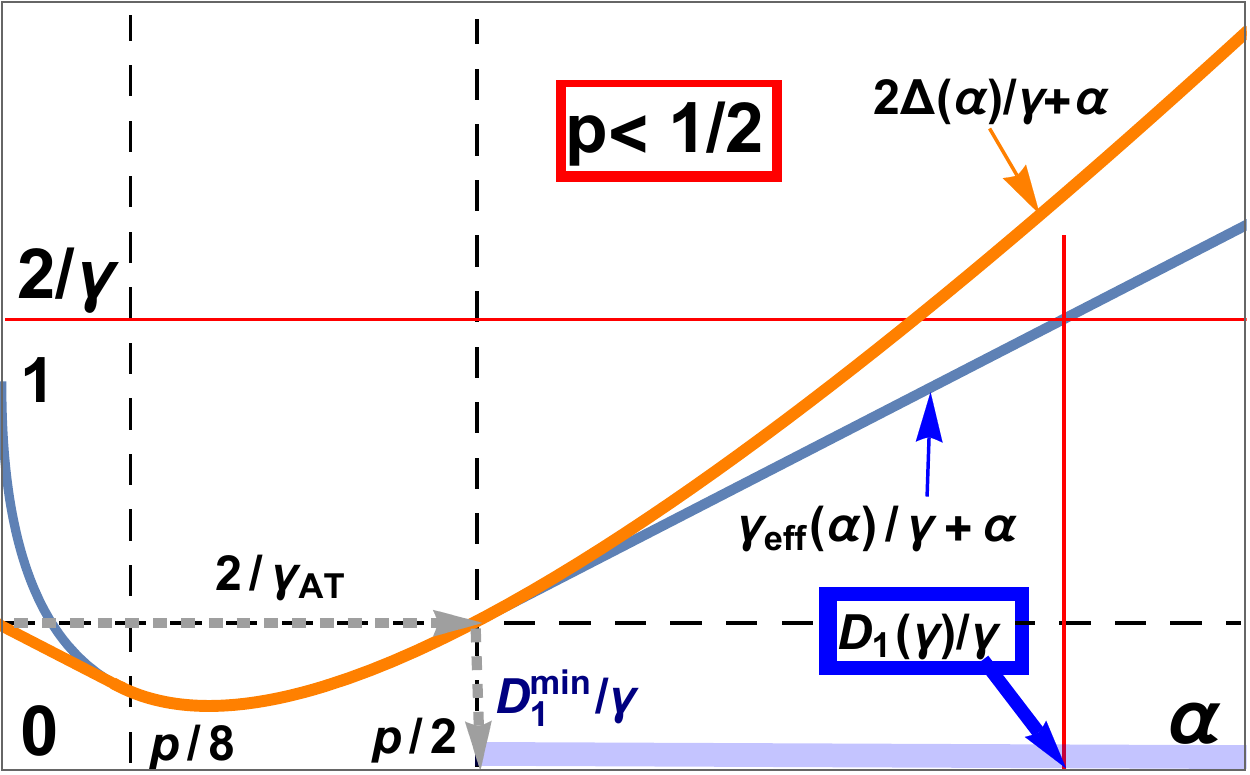}
\includegraphics[width=0.3\linewidth]{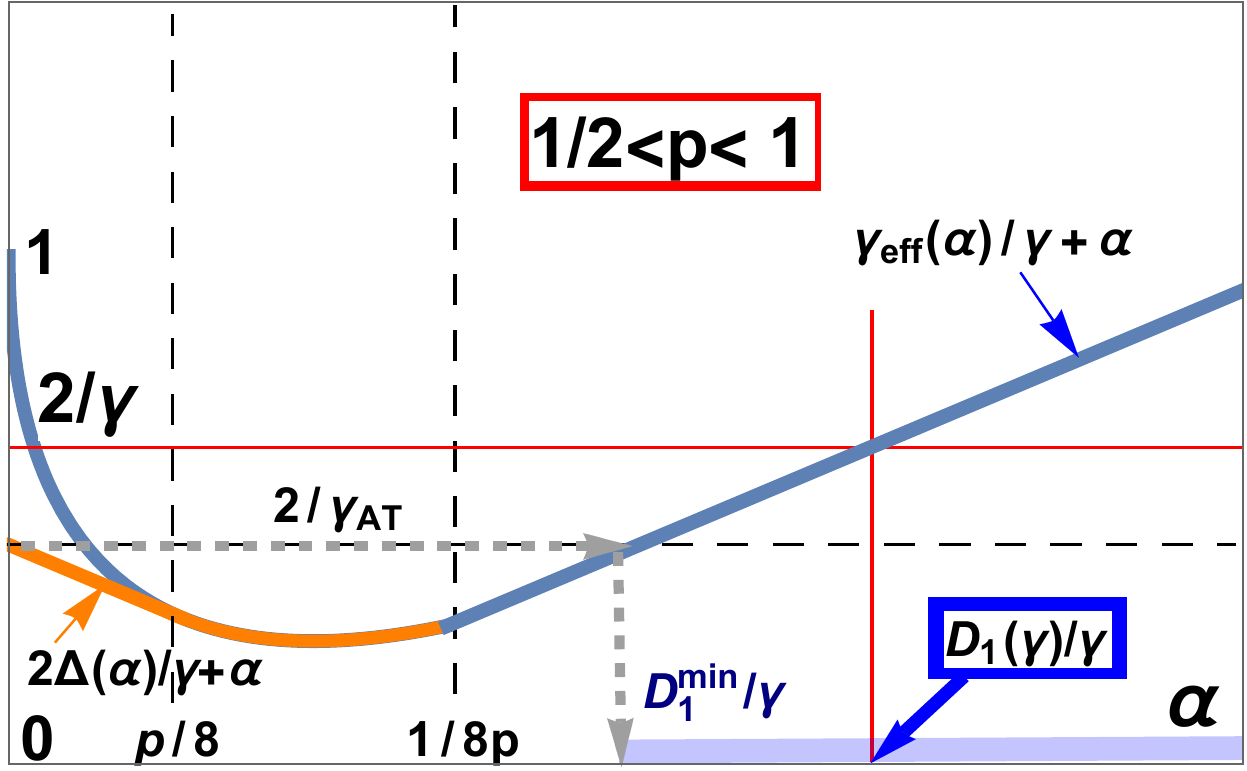}
\includegraphics[width=0.3\linewidth]{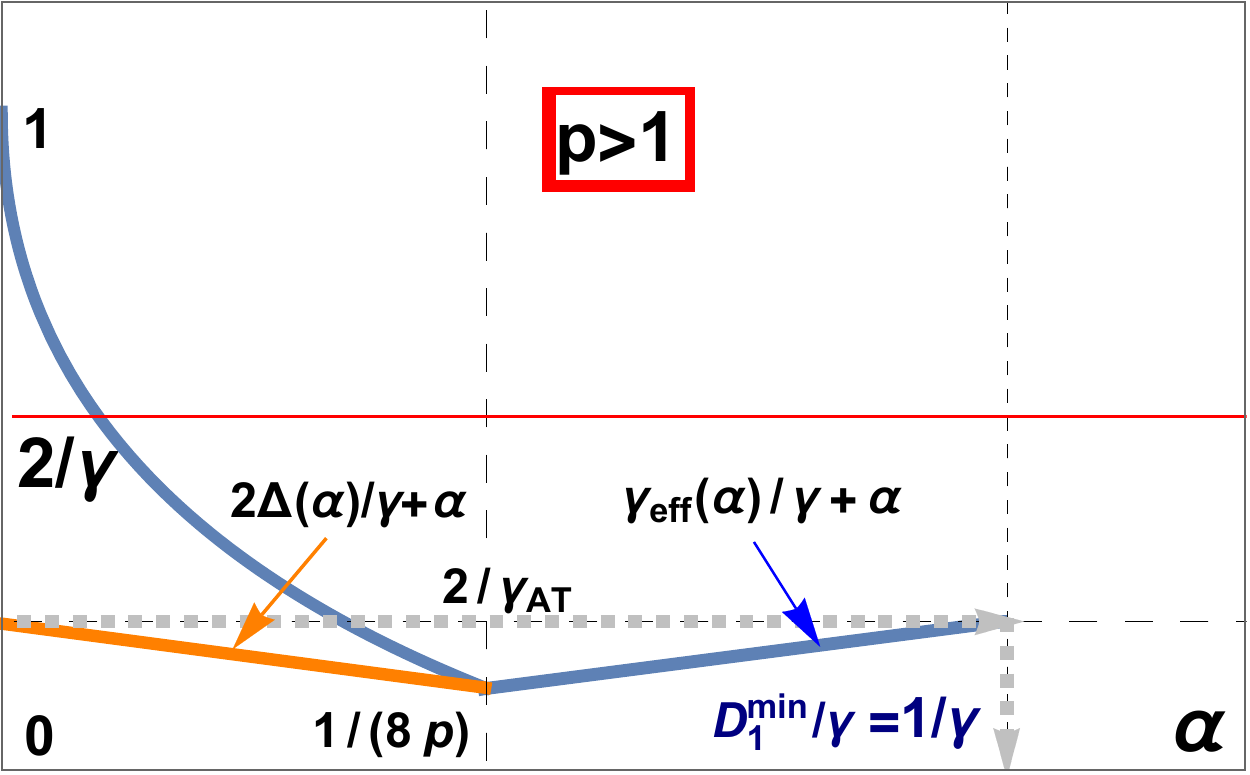}
\caption{(Color online) \textbf{The functions~\eqref{gamma_eff}
(blue curve) and~\eqref{Delta} 
(orange curve) entering inequalities
 Eqs.~(\ref{stab-MF-fin-Gauss}),~(\ref{stab-MF-fin-LN})
in different regions of $p$:}
(left) $p<1/2$; (middle) $1/2\leq p\leq 1$; (right) $p>1$.
Intervals of $\alpha = D_1/\gamma$ with different
functional dependence are shown by dashed vertical lines.
The Anderson localization transition corresponds to the lower of the blue and
orange curves equal to $2/\gamma$ at $\alpha=0$.
This transition is always determined by the orange curve representing the
log-normal part of the distribution $P(V)$.
On the contrary, the stable
fractal dimension $D_{1}(\gamma)=2-\gamma/\gamma_{ET}(p)$
for $\gamma\leq\gamma_{AT}$ is always determined by
the blue curve representing the Gaussian
part of the distribution $P(V)$.
The Anderson transition in all cases but $p=0$ is {\it discontinuous}, with the minimal
stable fractal dimension of the support set being
$D^{{\rm min}}_{1}=D_{1}(\gamma_{AT})=2-\gamma_{AT}/\gamma_{ET}(p)>0$
(shown by a gray dotted arrow).
The ergodic transition corresponds to $D_{1}(\gamma)=1$
and it is {\it continuous}.
For $p\geq 1$ there is no solution $D_{1}<1$ to the system of
inequalities Eqs.~(\ref{stab-MF-fin-Gauss}),~(\ref{stab-MF-fin-LN}) in the region
of parameters where the localized
phase is unstable. In this case the multifractal phase is absent.
\label{Fig:stability} }
\end{figure*}

A particular case $D_{1}=0$ of Eqs.~(\ref{stab-MF-fin-Gauss}),~(\ref{stab-MF-fin-LN}) describes the stability
criterion of the localized phase. If the localized phase is not stable, then
hybridization produces an avalanche of multifractal states living on fractal
support which
dimensionality grows until   inequalities
Eqs.~(\ref{stab-MF-fin-Gauss}),~(\ref{stab-MF-fin-LN})  are {\it   both first fulfilled}
for some $0<D_{1}<1$.
If this is possible in some parameter region then the multifractal state
is stable. If the only solution to the system of inequalities
Eq.~(\ref{stab-MF-fin-Gauss}),~(\ref{stab-MF-fin-LN}) corresponds to $D_{1}\geq 1$ then the only stable extended
phase is ergodic.

\section{Fractal dimension of the NEE support set}\label{sec:D_1}
In this section we re-consider the phase diagram of
non-truncated LN-RP, $\gamma_{tr}\leq 0$, from the viewpoint of
stability criteria given in the previous section by Eqs.~(\ref{stab-MF-fin-Gauss}),
(\ref{stab-MF-fin-LN})
and derive the expression for the fractal dimension $D_{1}(\gamma)$ of the support set of multifractal
wave functions.

To this end in Fig.~\ref{Fig:stability} we plot
\be\label{gamma_eff}
\frac{\gamma_{{\rm eff}}(\alpha)}{\gamma}+\alpha=
 \left\{\begin{array}{ll}1+3\alpha-2\sqrt{2\alpha p},&
 4\alpha< 2p, \frac{1}{2p} \\
 1/\gamma_{ET}(p)+\alpha,&
\text{otherwise}
\end{array}\right. \ ,
\ee
and
\be\label{Delta}
\frac{2\Delta(\alpha)}{\gamma}+\alpha=
 \left\{\begin{array}{ll}
 1+3\alpha-2\sqrt{2 \alpha p},&
p<8\alpha<\frac1p\\
 {2}/{\gamma_{AT}(p)}-\alpha,&8\alpha<p,\frac1p\\
 1+3\alpha+2\sqrt{2 \alpha p},&
8\alpha>\frac1p
\end{array}\right. \ ,
\ee
calculated in Appendix~\ref{App_sec:Stability} as functions of $\alpha=D_1/\gamma$.
Here $\gamma_{AT}(p)\geq 2$ and $\gamma_{ET}(p)\geq 1$ are given by Eqs.~\eqref{AT} and~\eqref{ET}, respectively.

According to the stability criteria Eqs.~(\ref{stab-MF-fin-Gauss}),~(\ref{stab-MF-fin-LN})
the functions Eqs.~(\ref{gamma_eff}),~(\ref{Delta})
should be compared to $2/\gamma$, see Fig.~\ref{Fig:stability}.
First, we note that the localized phase which formally corresponds to $D_{1}=0$,
is stable if the lowest of the blue and orange curves in Fig.~\ref{Fig:stability} is
higher than $2/\gamma$ at $\alpha=0$ and it is unstable otherwise.
One can see that at $\alpha=0$ for all values of $p$
the log-normal contribution (orange curve)
is lower than the Gaussian one (blue curve).
This means that the stability of the localized phase is always
determined   by the log-normal part of $P(V)$.
Moreover, since at $\alpha=0$
Eqs.~\eqref{gamma_eff},~\eqref{Delta}  reduce to
$\alpha+{\gamma_{{\rm eff}}(\alpha)}/{\gamma} = 1$ and
$\alpha+2\Delta(\alpha)/\gamma = 2/\gamma_{AT}$, respectively,
the stability of the localized phase  implies that
$\gamma>\gamma_{AT}(p)\geq 2$ in agreement with~\eqref{AT}.


If the localized phase is unstable then
different localized states hybridize and form a multifractal state with $D_{1}>0$.
Those states are, however, unstable until their support set reaches the fractal
dimension $D_{1}^{min}>0$ where Eqs.~(\ref{stab-MF-fin-Gauss}),
(\ref{stab-MF-fin-LN}) are first both fulfilled.

As the parameter $\gamma$ decreases below the critical value $\gamma_{AT}$,
the fractal dimension $D_{1}(\gamma)$ increases from $D^{{\rm min}}_{1}$
being always determined by the intersection of the horizontal line $y=2/\gamma>2/\gamma_{AT}(p)$ (red line in
Fig.~\ref{Fig:stability}) with the blue line.   Thus the stable fractal dimension
$D_{1}(\gamma)$ is always determined by the Gaussian part of $P(V)$  and  according to
the second line of Eq.~(\ref{gamma_eff}) and Fig.~\ref{Fig:stability} is equal to:
\be
D_{1}(\gamma)=2-\frac{\gamma}{\gamma_{ET}(p)} \ , \quad p\leq 1
 \ .
\ee
At $\gamma=\gamma_{ET}$
the fractal dimension $D_{1}(\gamma)$ reaches unity, and at this point a {\it continuous}
ergodic transition happens. Thus the critical point of ergodic transition
coincides with that determined by Eq.~(\ref{ET}).

Note that while $D_{1}(\gamma)$ is linear in $\gamma$, as
for the Gaussian RP model~\cite{gRP}, other fractal dimensions $D_{q}$ ($q>1$) are not
necessarily equal to $D_{1}(\gamma)$ as it was the case in Ref.~\cite{gRP}.
The calculation of $D_{q}$ with $q>1$ goes beyond the scope
of this paper and will be studied elsewhere~\cite{KrKhay}.

Note that, unlike the ergodic transition,
  the Anderson transition is
{\it discontinuous}: the stable fractal dimension $D_{1}(\gamma)$ is separated by a
finite gap $D_{1}^{min} = D_1(\gamma_{AT})$ from the localized state
$D_{1}=0$:
\be
D_{1}^{min}=
\left\{\begin{array}{ll}
2-\frac{\gamma_{AT}(p)}{\gamma_{ET}(p)},&0<p<1\cr
1, & p\geq 1
\end{array} \right.
\ee
This gap is shown by the gray dotted arrow in Fig.~\ref{Fig:stability}.
The right panel of Fig.~\ref{Fig:stability} demonstrates that for $p\geq 1$ the minimal
fractal dimension $D_{1}^{min}=1$, so that the multifractal phase is no
longer possible in LN-RP model with $\gamma_{tr}\leq 0$. However, as it is shown in
Sec.~\ref{sec:Truncation}, it appears if $\gamma_{tr}>0$.

\section{Conclusion and discussion}\label{sec:Conclusion}
In this paper we introduce
a log-normal Rosenzweig-Porter (LN-RP) random matrix ensemble  characterized by
a long-tailed distribution of off-diagonal matrix elements with the variance controlled by
the symmetry parameter $p$.
We calculate the phase diagram of LN-RP using the recently suggested
Anderson localization and Mott ergodicity criteria for random matrices.
An alternative  approach based on the analysis of stability with respect to hybridization of
multifractal wave functions developed
in this paper
gives results identical to those obtained from the above criteria and
consistent with
numerical calculations. It also helps to compute
the dimension $D_{1}$  of the eigenfunction fractal support set and show that
the Anderson localization
transition is discontinuous with $D_{1}^{{\rm min}}>0$ at all $p>0$.

This LN-RP model has many potential applications and  we use it
to develop an alternative approach
to the localization problem on random regular graph.
It is based on the partition of all sites on RRG into two groups:
(i)~the ``marked sites'' remote from each other
at the most abundant distance of the order of the graph diameter and
(ii)~the ``tree sites'' at much smaller distance from each other.
This partition is only meaningful for the graphs in which the marked sites take a
finite fraction
of all sites like in the graphs with a local tree structure.
Then we study an effective random matrix model involving only the marked sites
and show  that
it is a special case $p=1$ of
the log-normal Rosenzweig-Porter random matrix ensemble introduced in this paper.
An important result of this paper is that in this $p=1$ LN-RP model arising from the above
partition,
there is a direct transition from the localized to the
ergodic phase
similar to the one obtained in Refs.~\cite{Tikh-Mir1, Tikh-Mir2}.
However, the point $p=1$ appears to be very special: it
is a tricritical point of the LN-RP model
which is unstable to deformations of this model. In particular, it is unstable to
truncation of the far tail in the log-normal distribution considered in
Sec.~\ref{sec:Truncation} and sensitive to
modification of the loop statistics in RRG~\cite{Gorsky2019_RRG} leading to
the different (possibly non-convex) distribution
of off-diagonal matrix elements in the corresponding LN-RP model.

We would like to emphasize that our mapping of the localization problem on RRG to
LN-RP random matrix ensemble is an {\it approximation} which is justified
only qualitatively.
Note that the approach adopted in Ref.~\cite{Tikh-Mir1, Tikh-Mir2} is not
free of approximations too. While the local tree structure of RRG is treated exactly
in the framework of the supersymmetric sigma-model, the final solution rests on the
``self-consistency'' condition (Eqs.~(24) and (30) in Ref.\cite{Tikh-Mir2}):
\be\label{self-consist}
g_{0}(Q_{0})=\int {\cal D}Q_{0}'\,[g_{0}(Q_{0}')]^{K}\,
e^{-STr\left[-2g(Q_{0}-Q_{0}')^{2}+\eta\,\Lambda Q_{0}'\right]}.
\ee
This condition
is an {\it approximation}, as the corresponding equations do not take into account a
detailed topology of the graph (statistics of loop lengths, etc.) but only
(i)~the local tree structure (encoded in the non-linear term $[g_{0}(Q_{0}')]^{K}$), and
(ii)~the statistical homogeneity of the graph as a whole
(encoded in the fact that only
the {\it zero spatial mode}
component $Q_{0}$ of the supersymmetric $Q(i)$-field enters in Eq.~(\ref{self-consist})).

In a sense, this approximation is in many respects similar to our mapping onto
LN-RP model. Indeed,  the {\it zero spatial mode}
component $Q_{0}$  is known to
describe the Wigner-Dyson random matrix ensembles~\cite{Efetov_book}.
The Gaussian Rosenzweig-Porter model is a Wigner-Dyson random matrix ensemble with
parametrically enhanced fluctuations of diagonal matrix elements. The breakdown of
basis-rotation invariance by the special diagonal (like in the Rosenzweig-Porter
ensemble) is described in Eq.~(\ref{self-consist}) by the 'gradient'
term $-2g(Q_{0}-Q_{0}')^{2}$ which becomes non-zero (in contrast to the Wigner-Dyson case)
due to the presence of the ``external'' $Q_{0}$ supermatrix in the non-linear
integral equation~(\ref{self-consist}).

Therefore, one may conjecture that our mapping onto the $p=1$ LN-RP model
is {\it equivalent}
to the self-consistent approximation of Ref.~\cite{Tikh-Mir2} and earlier works
by Mirlin and Fyodorov (see Refs.~\cite{MF1991, MF1994} and references therein).
An additional support to this conjecture comes
from the fact that at $p=1$ the critical exponent $\nu_{1}=\nu_{2}\equiv \nu$ reaches its
mean-field value $\nu=1/2$ (see Table~\ref{Fig:tab_nu} and Fig.~\ref{Fig:conj-nu}).
Whether or not both approximations give correct predictions for the phases on RRG is,
in our opinion,  still an open issue.

%

\begin{acknowledgments}
V.E.K. and I.M.K are grateful for support and hospitality to GGI of INFN and University of Florence (Italy) where this work was initiated.
V.E.K and B.L.A. acknowledge the support and hospitality of Russian Quantum Center during the work on this paper and G. V. Shlyapnikov for illuminating discussions there.
V.E.K gratefully acknowledges support from the Simons Center for Geometry and Physics, Stony Brook University at which part of the research for this paper was performed.
This research was supported
by the DFG project KH~425/1-1 (I.~M.~K.),
by the Russian Foundation for Basic Research Grant No. 17-52-12044 (I. M. K.),
and by Google Quantum Research Award ``Ergodicity breaking in Quantum Many-Body Systems'' (V. E. K.).
\end{acknowledgments}

\appendix
\section{RRG-to-LN-RP correspondence.  }
\label{App_sec:RRG-LN-RP_correspondence}
As mentioned in Sec.~\ref{sec:RRG_motivation}, the local tree structure and the predominance of long loops on RRG
lead to ``condensation of large distances'' when most of the pairs of sites are located at a certain distance of the order of the graph diameter $d \simeq \ln N/\ln K$. 
This leads to the set of equally spaced sites on RRG with the most
abundant distance $r_*\approx d-4$.
Those marked sites interact with each other through the remaining
{\it tree sites} similar to the {\it indirect interaction} between Anderson impurities
in a metal.  This indirect interaction is {\it long-range} and the effective hopping
matrix elements $(H_{{\rm eff}})_{nm}$ of such a model can be found using the
Anderson impurity model.

Indeed, let us consider the marked sites $n$ and $m$ as Anderson impurities imbedded
into the Cayley tree which sites are connected by a hopping $V$. Those impurities
(which are not directly connected)
are supposed to be
connected with the neighboring tree sites by the same hopping $V$. Then the
impurity Green's function ${\cal G}=(E-H_{{\rm eff}})^{-1}$ can be expressed
through an exact
Green's function $G$ on a Cayley tree as follows:
\begin{equation}\label{Green}
 {\cal G}_{nm} =g_{n}\,V^{2}\,G_{m'n'}\,g_{m}+g_{n}\delta_{nm},
\end{equation}
where $n'$ and $m'$ are the sites on a tree neighboring to the marked sites
('impurities')
$m$ and $n$, respectively, and $\hat{g}_{nm}=g_{n}\delta_{nm}=(E-\varepsilon_{n})^{-1}\,
\delta_{nm}$ is the bare
impurity Green's function. Thus, inverting Eq.~(\ref{Green}) and assuming $V^{2}
||\hat{g}||\,||\hat{G}||\ll 1$ we obtain:
\be
{\cal G}^{-1}= \hat{g}^{-1}\,(1+V^{2}\,\hat{g}\,\hat{G})^{-1}
\approx\,\hat{g}^{-1}-V^{2}\,\hat{G}.
\ee
Substituting here ${\cal G}^{-1}=(E-\hat{H}_{{\rm eff}}) $ and $\hat{g}^{-1}=E-
\hat{\varepsilon}$ one obtains:
\be\label{hop-G}
\hat{H}_{{\rm eff}}(E)=\hat{\varepsilon} + V^{2}\,\hat{G}(E).
\ee
where $\hat{G}(E)$ is an exact Green's function on a Cayley tree and
$\hat{\varepsilon}={\rm diag}\{ \varepsilon_{n}\}$.

\section{Full and cavity Green's functions on a Cayley tree} \label{App_sec:Greens_funct_BL}
\begin{figure}[thb]
\includegraphics[width=1.0 \linewidth,angle=0]{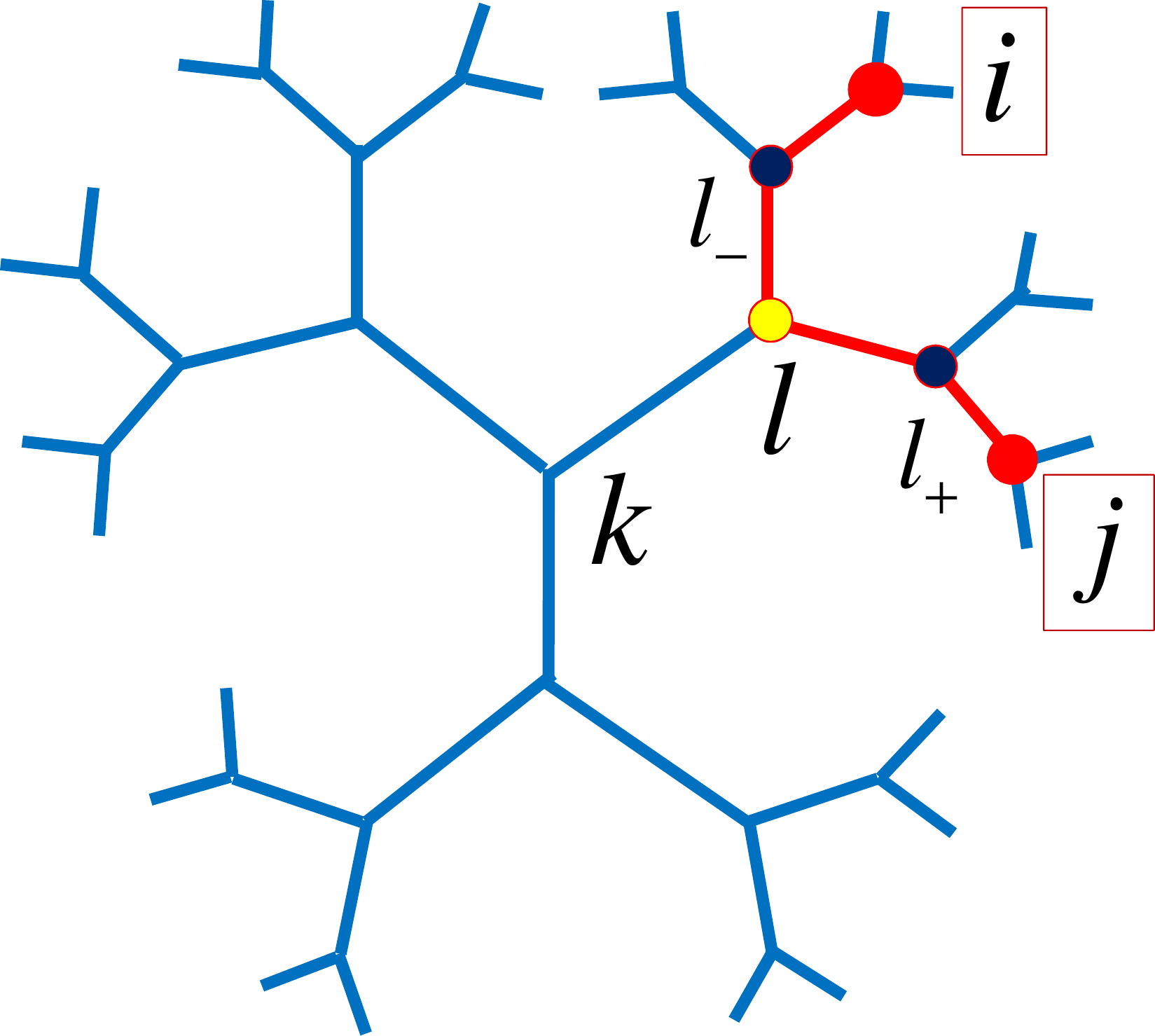}
\caption{(Color online) The full two-point Green's function  $G(i,j)$ on a Cayley
tree. The path from $i$ to $j$ is shown in red with the common
descendant $l$ of $i$ and $j$ and its nearest neighbors $l_{-}$ and $l_{+}$}.
\label{Fig:G_ij}
\end{figure}

Note that $\hat{G}(E)$ with components $G_{E}(i,j)\equiv G(i,j)$ in Eq.~(\ref{hop-G}) is the {\it full} two-point Green's
function on a Cayley
tree.  It is convenient to express it
  in terms of the product $\Pi(i,j)$ of the  {\it cavity}   Green's functions
$G_{p\rightarrow p_{-}}$, where $p_{-}$ is an immediate descendant of $p$ along the path from
$i$ to $j$ (see   Fig.~\ref{Fig:G_ij}):
\be\label{prod}
\Pi(i,j)=\prod_{i<p\leq j} G_{p\rightarrow p_{-}}.
\ee
Here $G_{i\rightarrow k}$ is the Green's function on a sub-tree such that the
link from the site $i$ to its immediate descendant $k$ is cut. The advantage of such an
object is that it can be found by a recursion relation:
\be\label{arrowed}
G^{-1}_{i\rightarrow k}=E-\varepsilon_{i}-\sum_{j(i)}^{K}G_{j\rightarrow i},
\ee
where $\varepsilon_{i}$ is the random on-site energy and $j(i)=1,...,K$ are immediate
predecessors  of $i$. Thus knowing the Green's functions on the outer sites of a tree
('leaves') one can find $G_{i\rightarrow k}$ at any other site by iteration.

The general expression for the two-point Green's function on a tree reads as follows
\cite{Aizenman2013}:
\be
\label{true-Green-two-different}
G(i,j)=\Pi(i,l)\,G(l,l)\,\Pi(j,l).
\ee
In terms of the cavity Green's functions Eq.(\ref{true-Green-two-different})
takes the form:
\bea\label{G-prod}
G(i,j)=G_{i\rightarrow i_{-}}...G_{l_{-}\rightarrow l}\,G(l,l)\,
G_{l_{+}\rightarrow l}...
G_{j\rightarrow j_{-}}
\eea
where $l_{-}$ and $l_{+}$ are the neighbors of $l$ along the paths from $i$ to $l$ and
from $l$ to $j$, respectively (see Fig.\ref{Fig:G_ij}), and
\be\label{G_ll}
G(l,l)=\frac{1}{E-\xi_{l}-\sum_{n(l)=1}^{K} G_{n(l)
\rightarrow l}-G_{k(l)\rightarrow l}}.
\ee
In Eq.(\ref{G-prod}) we denote the immediate predecessors of $l$ by $n(l)=1,...,K$,
while $k(l)$ is its descendant.  Note that two of $n(l)$ are necessarily $l_{-}$ and $l_{+}$
(see Fig.\ref{Fig:G_ij}). In a particular case when $i$ and $j>i$ lie on the
same path passing through
the root of the tree, one formally replaces in Eq.(\ref{G-prod}) $l$ by $i$ and sets $\Pi(i,i)=1$.

Eq.(\ref{G-prod}) is a product of the cavity Green's functions along the path
from $i$ to $j$, with the single exception that at the common descendant $l$ of $i$
and $j$ the cavity Green's function is replaced by the full one. This is the minimal
modification of the product which makes the Green's function $G(i,j)$
(as well as $G(l,l)$) singular at the
eigen-energy $E=E_{n}$.

However, this modification is crucial not only for the correct spectral properties.
It drastically restricts fluctuations of $|G(i,j)|$. Indeed, from the recursion relation
Eq.(\ref{arrowed}) it follows that if at some step $G_{j\rightarrow i}$
is anomalously large than at the next step $G_{i\rightarrow k}$ will be anomalously
small, so that the product $G_{j\rightarrow i}\,G_{i\rightarrow k}$ is of order one.
At the same time anomalously small $G_{j\rightarrow i}$ does not result
in an anomalously large $G_{i\rightarrow k}$. For this to happen {\it all} other
terms in r.h.s. of Eq.(\ref{arrowed}) should be anomalously small which is much less
probable.
This means that a product $\Pi(i,j)$ along a long path is typically small, as for
large number of terms in the product the probability to have a small $G_{j\rightarrow i}$
increases. However,
there are rare events when $\Pi(i,j)$  is anomalously large.
This happens {\it only}
when the last term in the product $G_{p\rightarrow p_{-}}$ with $p_{-}=i$
(which cannot be compensated)
is anomalously large. These rare events are responsible for the symmetric distribution
of $y=|G(i,j)|^{-1}$ discussed in Appendix \ref{App_sec:P(U)}.

Now let us consider Eq.(\ref{G-prod}). One can easily see
that   the anomalously large end-terms $G_{l_{-}\rightarrow l}$ and $G_{l_{+}\rightarrow l}$
in $\Pi(i,l)$ and $\Pi(j,l)$ cancel out by the corresponding terms in the
denominator of  Eq.(\ref{G_ll}),
if we neglect very improbable evens when they {\it both} are large.
With this restriction one may write:
\be\label{G_ij_appr}
r^{-1}\,\ln|G(i,j)|= \left\{\begin{array}{ll} r^{-1}\,\ln|\Pi(i,j)|, & {\rm if}\;\;
\ln|\Pi(i,j)|<0
\cr \approx 0, & {\rm otherwise}\end{array} \right.,
\ee
where $r=|i-j|\gg 1$.

Note that the above analysis applies also to one-dimensional Anderson model
which formally corresponds to $K=1$. Then Eq.(\ref{G_ij_appr}) is consistent
with the exact result \cite{Melnikov1981} that
$T=|G(i,j)|^{2}\leq 1$, where $T$ is the transmission
coefficient through a chain of the length $L$.

This {\it intrinsic} cutoff   at large values $|i-j|=d$ of $|G(i,j)|$ affects the
Anderson localization~\eqref{Anderson} and the Mott ergodicity~\eqref{Mott} principles
for the corresponding LN-RP random matrix model
 and leads to the phase diagram with the tricritical point at $p=1$.

\section{'Multifractal' distribution of $|\Pi(i,j)|$}\label{App_sec:P(U)}
Now   we  consider
  generic properties of the distribution of the product
	$\Pi_{r}\equiv|\Pi(i,j)|$ of cavity
	Green's functions
  $G_{p\rightarrow p_{-}}$ on a Cayley tree
at large distances $r\equiv |i-j|$.

As it is shown in Ref.~\cite{AnnalRRG}, in the limit of a long path $r\gg 1$
the distribution function ${\cal F}(y)$ of   $y = \Pi_{r}^{-1}$ has a
special symmetry:
\be\label{basic-sym}
{\cal F}(y)={\cal F}(1/y).
\ee
Correspondingly,  the distribution function
$P(\Pi_{r})$
 obeys the symmetry:
\be\label{sym}
 P( 1/\Pi_{r}) = \Pi_{r}^{4}\, P( \Pi_{r}).
\ee

In order to proceed further we make use of the expression for ${\cal F}(y)$ in terms of its
moments $I_{n}=\int {\cal F}(y) y^{-2m} dy$:
\be\label{Mellin}
{\cal F}(y)=\frac{2}{y}\int_{B}\frac{dm}{2\pi i}\,y^{2m}\,(I_{m})^{r},
\ee
where the integration is performed over the Bromwich $m=c+iz$ contour which goes
parallel to the imaginary axis ($z\in[-\infty,+\infty]$) on the positive side
of the real one ($c>0$).
Eq.~(\ref{Mellin}) is nothing but a Mellin transform which allows to restore the
distribution function, given that the (analytically continued) moments $I_{m}$
are known.

The moments $I_{m}$ at $m\in[0,1]$ obey
the following symmetry  which reflects the symmetry Eq.~(\ref{basic-sym})
\cite{AnnalRRG}:
\be\label{sym-I-m}
I_{m}=I_{1-m},
\ee
with a minimum at $m=1/2$ and:
\be\label{bc}
I_{0}=I_{1}=1,\;\;\;\;\partial_{m} I_{m}|_{m=1}=-\partial_{m} I_{m}|_{m=0}.
\ee
This symmetry is another representation of the basic $\beta$-symmetry on a Cayley tree
established in the seminal work \cite{AbouChacra}.

Computing the Mellin transform in the saddle-point approximation one finds
with the exponential accuracy:
\be\label{phase}
\ln(y {\cal F}(y))
=r\lp \ln I_{m}-m\partial_{m}\ln I_{m}\rp_{m=m_{*}},
\ee
where $m_{*}$ is found from the stationarity condition:
\be \label{stat}
\frac{1}{2}\lp\partial_{m}\ln I_{m}\rp_{m=m_{*}(y)}=-\,\frac{\ln y}{r}.
\ee
Eq.~(\ref{stat}) implies that $m_{*}$ is a function of the argument $\ln(y)/r$.
Then it follows from Eq.~(\ref{phase}) that:
\be\label{multifractal-form}
{\cal F}(y)\sim y^{-1}\,{{\rm exp}}\left[-r\,\mathfrak{G}\left(\frac{\ln y}{r}\right) \right],
\;\;\;(r\gg 1),
\ee
where $\mathfrak{G}(x)$ some function of $\ln(y)/r$.

The form Eq.~(\ref{multifractal-form}) is very special. A large parameter $r$ appears both
in front of $\mathfrak{G}(x)$ and in its argument in a reciprocal way. This form is know as the
{\it large deviation}, or {\it multifractal ansatz}. It appears in many different
problems of statistical mechanics (see e.g. Ref.~\cite{NatComm} and
references therein) and is a non-trivial generalization of Central Limit Theorem when
the logarithm of the fluctuating quantity is a sum of many terms with special correlations between them.

The simplest choice of the function $\mathfrak{G}(x)$ is a linear function which corresponds to a
power-law distribution.
A parabolic function $\mathfrak{G}(x)$ appears when $\ln y$
is the sum of nearly uncorrelated terms  which leads to the
logarithmically-normal distribution $P(\Pi_{r})$:
\be\label{LN}
P(\Pi_{r})=\frac{A(r)}{\Pi_{r}}\,{\rm exp}\left[-\frac{\ln^{2}(\Pi_{r}/\Pi_{{\rm typ}})}
{2p\,\ln(\Pi_{{\rm typ}}^{-1})}\right],
\;\; A(r)=\frac{1}{\sqrt{2\pi\,p\lambda\, r}}.
\ee
where   $\Pi_{{\rm typ}}\sim e^{-\lambda\,r}$, with the Lyapunov
exponent $\lambda$,
is the typical value of $\Pi_{r}$, and $p$ is the
symmetry parameter.
The symmetry Eq.~(\ref{sym}) corresponds to $p=1$. Any
$p\neq 1$ modifies the power of $\Pi_{r}$ in r.h.s. of the symmetry relation,
Eq.~(\ref{sym}).

Note also that the product  $\Pi_{r_{1}}\,\Pi_{r_{2}}$  has
the   log-normal distribution Eq.(\ref{LN}) with $r=r_{1}+r_{2}$, if both
$\Pi_{r_{1}}$ and $\Pi_{r_{2}}$ are distributed log-normally as in Eq.(\ref{LN})
with $r=r_{1}$
and $r_{2}$, respectively. This implies that the distribution of the product
$|\Pi(i,j)|$
depends only on the distance between the points $i$ and $j$, no matter whether $i$
and $j$ are on the same path passing through the root of a tree or they
are on two different branches (as in Fig.
\ref{Fig:G_ij}). The property $P(\Pi_{r_{1}}\Pi_{r_{2}})=P(\Pi_{r_{1}+r_{2}})$ is
required of {\it any} sensible distribution of the product of local quantities on a
Cayley  tree.

The distribution of the off-diagonal matrix elements $U=|G(i,j)|_{|i-j|=d}$
of the corresponding RP RMT
can be
found from $P(\Pi_{r})$ by setting  $r\approx d=\ln N/\ln K$, $U=\Pi_{r}$, and
employing Eq.(\ref{G_ij_appr}) which imposes a cutoff
at $U_{{\rm max}}\sim N^{0}$.

Here an important comment is at place. Eq.(\ref{LN}) with $p=1$ and $r=L$
applies also to a
one-dimensional Anderson model at weak disorder and localization length $\xi<<L$
much smaller than the system size $L$. Indeed, denoting
$\ln\Pi_{{\rm typ}}^{-1}=L/(2\xi)\equiv x/2$ one reduces Eq.(\ref{LN}) to
the distribution function of
$\sqrt{T}$ of Ref.\cite{Melnikov1981}, where $T\ll 1$ is the transmission coefficient.
The comparison with the exact result of  Ref.\cite{Melnikov1981} could give a more
precise limit of applicability of the log-normal distribution than
Eq.(\ref{G_ij_appr}). It shows that the
log-normal distribution of $ T=|G(i,j)|^{2}$ valid for $ \ln(1/G)\gtrsim x$,
is modified for
$1\lesssim\ln(1/G)\ll x$ (where $G=|G(i,j)|$ for brevity) by an additional
pre-exponential factor $x^{-1}\,\ln(1/G)$ which is not essential unless in the
vicinity of $G=1$. Thus in 1D case the essential cut-off of the log-normal
distribution, indeed, happens at $G=1$.

One can expect that on the Cayley tree the additional  factor which replaces
$x^{-1}\,\ln(1/G)$ is a function of
$\ell^{-1}=\ln(1/G)/\ln N$. Such a factor is of order $O(1)$ for any
$G\sim N^{-\gamma_{tr}}$, $\gamma_{tr}\sim 1$ and thus cannot lead to efficient truncation
of the log-normal distribution.


One can show that the distribution Eq.~(\ref{LN}) with $p=1$ is asymptotically
exact on an infinite Cayley tree in the limit of small disorder. In particular, for
a {\it granular}
Cayley tree described by the non-linear sigma model (NL$\sigma$M) the moments $I_{m}$ are given by
~\cite{Tikh-Mir1}:
\bea\label{I-m-NLSM}
I_{m}&=&\sqrt{\frac{2}{\pi\,\mathfrak{g}}}\,\left[K_{m+1/2}(\mathfrak{g})\,
\mathfrak{g}\sinh \mathfrak{g}\right. \\ \nonumber
&+& K_{m-1/2}(\mathfrak{g})\,
\left.(\mathfrak{g}\cosh \mathfrak{g} -m \sinh \mathfrak{g})\right],
\eea
where $\mathfrak{g}$ is the dimensionless conductance (the coefficient in front of
$({\nabla} Q)^{2}$ in the NL$\sigma$M).
In the limit of large
inter-grain conductance $\mathfrak{g}\gg 1$ one obtains:
\be\label{parabolic}
\ln(I_{m})\approx -(2\mathfrak{g})^{-1}\,m(1-m),
\ee
which according to Eq.~(\ref{Mellin})  implies the log-normal
distribution of $y$ and  $\Pi_{r}$.
The asymptotic expression Eq.~(\ref{parabolic}) reproduces
Eq.~(\ref{I-m-NLSM}) very accurately down to $\mathfrak{g}\sim 0.1$.

\begin{figure}[t!]
\center{
\includegraphics[width=0.9 \linewidth,angle=0]{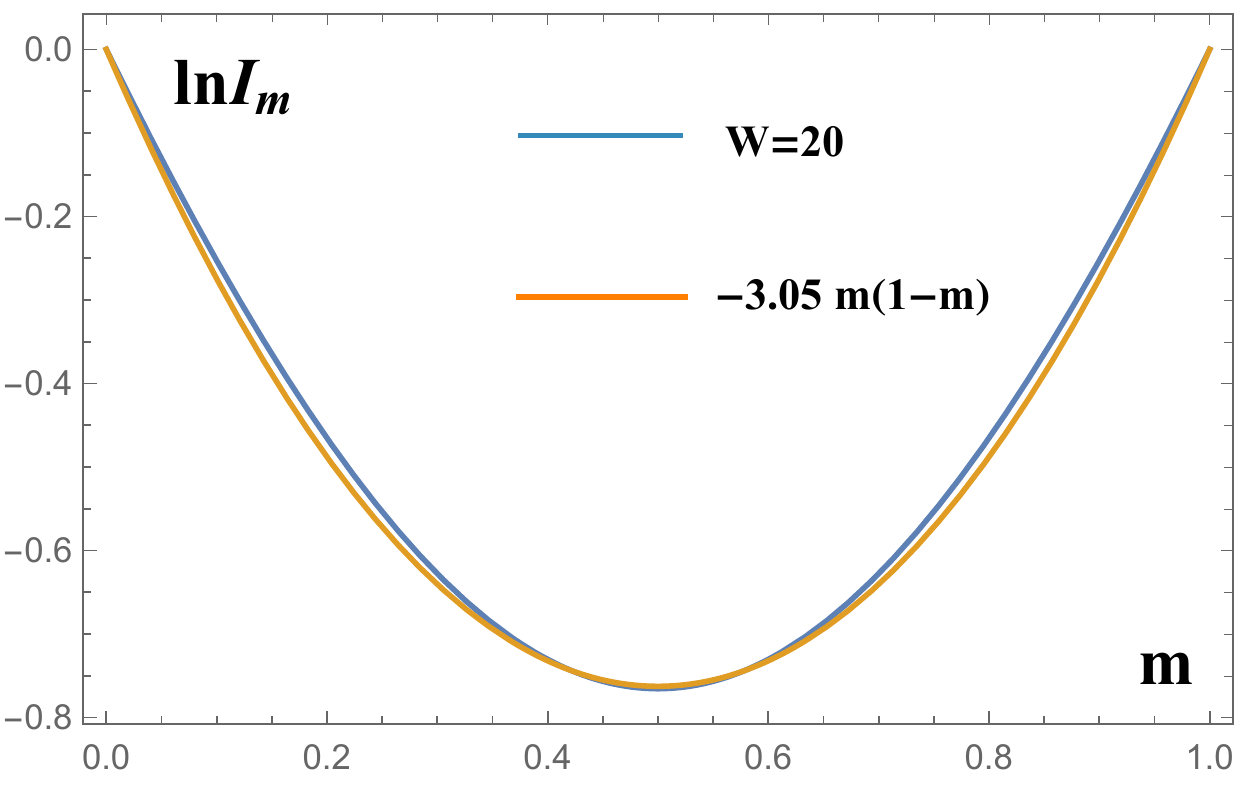}}
\caption{(Color online) Plots of $\ln I_{m}$ from Eq.~(\ref{two-briks}) for $W=20$
and the parabolic dependence $-3.05\,m(1-m)$. }
\label{Fig:parabola}
\end{figure}

The same is true for an ordinary
Cayley tree with a single orbital per site. In this case the 'two-brick'
approximation (Eq.~(90) in
Ref.~\cite{AnnalRRG}) gives for $I_{m}$:
\be\label{two-briks}
I_{m}=\frac{\sinh\left[(2m-1)\,\ln\left(\frac{W}{2} \right) \right]}{(2m-1)\,
\sinh\left[ \ln\left(\frac{W}{2} \right) \right]}.
\ee
One can show that $\ln I_{m}$ from Eq.~(\ref{two-briks}) is approaching Eq.~(\ref{parabolic})
with $(2\mathfrak{g})^{-1}\rightarrow
(W-2)^{2}/6$ for $W\rightarrow 2$ and remains an almost perfect parabola
in a broad interval $2<W\lesssim 30$ (see Fig.~\ref{Fig:parabola}). We conclude therefore that the log-normal distribution
of $G_{r}$ is {\it quantitatively} accurate in the whole range of disorder strengths up
to the Anderson transition point $W_{c}\sim K\ln K$ if the branching number $K\lesssim 12 \sim O(1)$.

\begin{figure}[t!]
\center{
\includegraphics[width=0.9 \linewidth,angle=0]{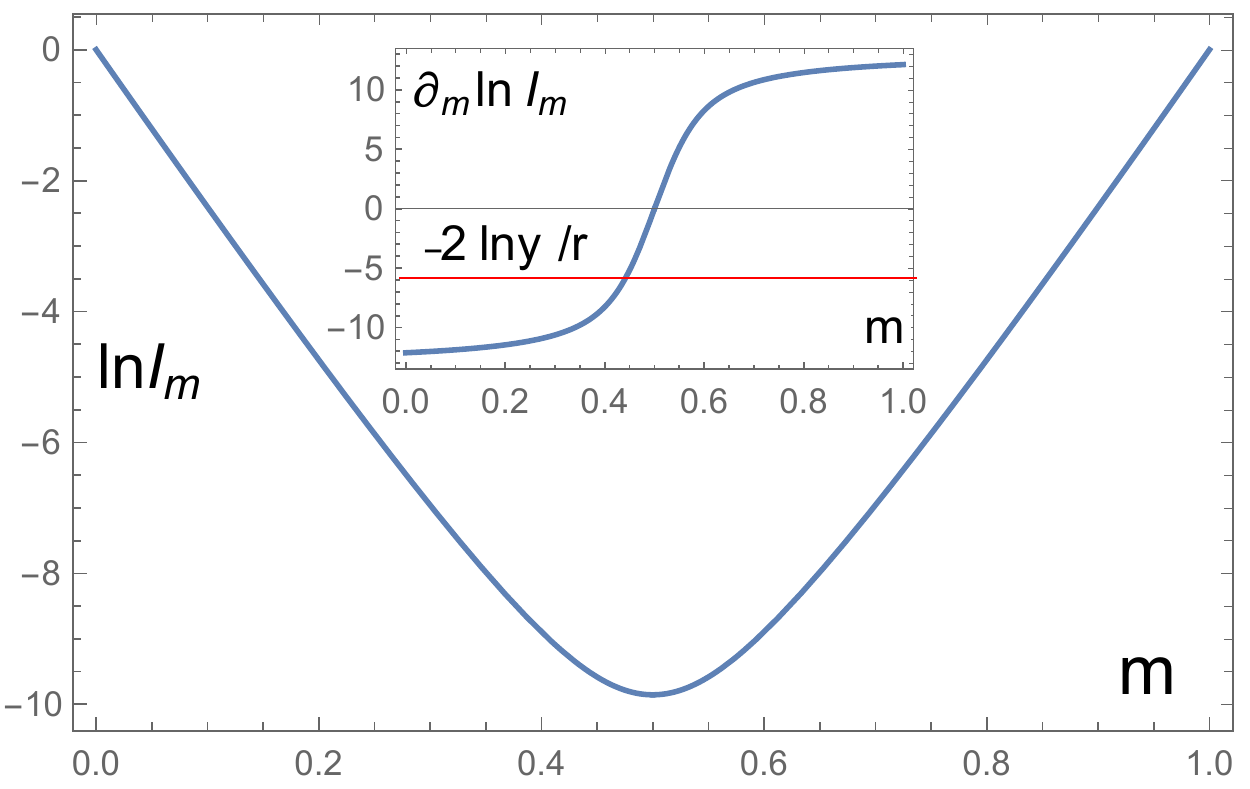}}
\caption{(Color online) Plots of $\ln I_{m}$ (main panel) and its derivative (inset)
from Eq.~(\ref{two-briks}) for $\ln(W/2)=13$. For $|\ln y|>r\,\ln(W/2)$ the saddle-point
Eq.~(\ref{stat}) does not have a solution.}
\label{Fig:large-W}
\end{figure}

However, for large $\ln(W/2)\gg 1$
\be
\ln I_{m}\approx (|2m-1|-1)\,\ln(W/2)
\ee
is linear in
$m$ everywhere except for a small interval of the width $\sim 1/\ln(W/2)$
in the vicinity of the minimum at $m=1/2$ in which $\ln I_{m}$ can be approximated
by a parabola (see Fig.~\ref{Fig:large-W}).
In this case the saddle point Eq.~(\ref{stat}) does not have a solution for
$|\ln y|>r\,\ln(W/2)$ (see Fig.~\ref{Fig:large-W}), and the distribution ${\cal F}(y)$ is truncated. For
$|\ln y|<r\,\ln(W/2)$ we obtain:
\be
{\cal F}(y) \sim C\,{\rm exp}\left[-\frac{\ln^{2}y}{\Sigma}\right],\;\;\;
(|\ln y|<r\,\ln(W/2)),
\ee
where $C=e^{-r\,\ln(W/2)}$ and $\Sigma=(2/3)\, r\, \ln^{2}(W/2)$.

Then from the results of Appendix~\ref{App_sec:Greens_funct_BL} it follows that
$P(U)$ is truncated from below at
$U_{{\rm min}}\sim N^{-\frac{\ln(W/2)}{\ln K}}$
and from above at $U_{\rm max}\sim 1$.
Between these limits it can be approximated by:
\be
P(U)\sim \frac{A}{U}\,{\rm exp}\left
[- \frac{\ln^{2}\lp U\rp}{(2/3)\,d(N)\,\ln^{2}(W/2)}\right]\,\left( \frac{U_{{\rm typ}}}{U}
\right),
\ee
where $U_{{\rm typ}}\sim N^{-\frac{\ln(W/2)}{\ln K}}$ and
$d(N)=\ln N/\ln K$.

One can see that the probability to find $U$ larger than the typical one
is considerably smaller than the one resulting from the forward scattering approximation (FSA):
\be
P_{{\rm FSA}}\sim \frac{A'}{U}\,{\rm exp}\left[-\frac{\ln^{2}(U/U_{{\rm typ}})}
{d(N)\,\ln^{2}(W/2)} \right].
\ee
Furthermore, because of the resonances (neglected in the FSA but captured by
Eq.~(\ref{two-briks}))
the transition matrix
elements $U$ are never described by the FSA, no matter how large is $\ln(W/2)$.

\section{Kullback-Leibler measures in the multifractal phase}\label{App_sec:theory_KL}
In this section we give a more detailed quantitative description of $KL2$.
In order to do this we employ the ansatz for the wavefunction moments:
\be\label{moment-ansatz}
M_{q}=\la \sum_{i}|\psi(i)|^{2q}\ra= N^{-D_{q}(q-1)}\,f_{q}
 (L/\xi_{q}),
\ee
where $D_{q}$ is the fractal dimension and $f_{q}(x)$ is the crossover scaling function:
\bea\label{asympt}
&&f_{q}(L/\xi_{q}\rightarrow \infty)\rightarrow
\nonumber\\ && \left\{\begin{array}{ll}{\rm const}. &\;\;{\rm multifractal\;\; phase}\cr
{\rm const }.\, N^{(q-1)(D_{q}-1)},& \;\; {\rm ergodic\;\; phase}\cr
{\rm const }.\, N^{(q-1)D_{q}} &\;\;{\rm localized\;\; phase}
\end{array}\right.
\eea
Note that for the graphs with the local tree structure the length scale $L\propto \ln N$,
so that
the scaling function is in general a function of {\it two arguments} $\ln N/\xi_{q}$
and $N/e^{\xi_{q}}$ representing the {\it length}- and {\it volume} scaling~\cite{Lemarie2017,Lemarie2020_2loc_lengths}. On
the finite-dimensional lattices $N\propto L^{d}$, and the volume scaling can be represented
as the length scaling in the modified scaling function. In this case a single argument
$L/\xi_{q}$ is sufficient.

When $L\propto \ln N$ the volume scaling is the
leading one for $L\gg \xi_{q}$, and
it is this scaling that provides
the asymptotic behavior Eq.~(\ref{asympt}).
The length scaling is
important in the crossover region $L\lesssim \xi_{q}$. Below for brevity we will use the short-hand notation
$L/\xi_{q}$ in all the cases.

There are two trivial cases:
$M_{0}=N$ and $M_{1}=1$ (which follows from the normalization of wave function).
As a consequence we have $D_{0}=1$ and
\be
f_{0}(x)=f_{1}(x)\equiv 1.
\ee
\begin{figure*}[t!]
\includegraphics[width=0.44\linewidth]{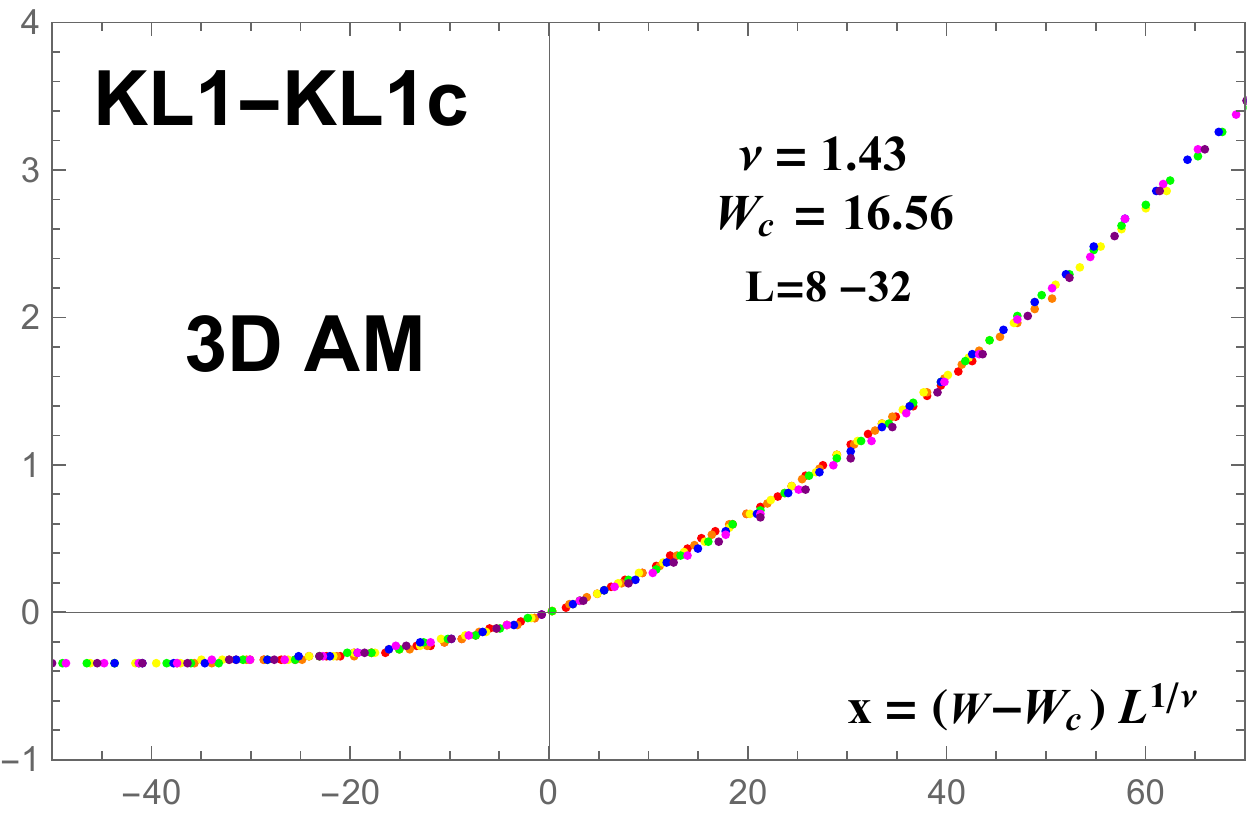}
\includegraphics[width=0.45\linewidth]{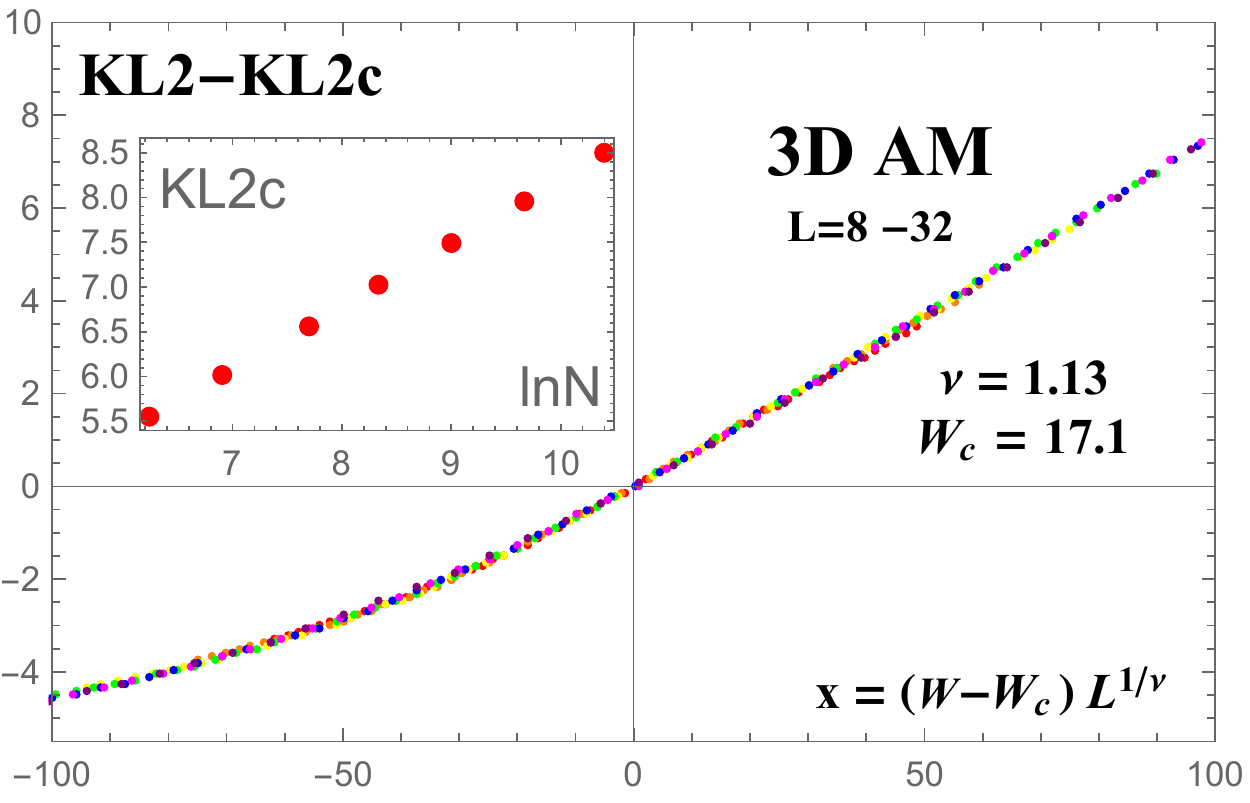}
\caption{(Color online) Collapse of $KL1$ and $KL2$ data in the vicinity of the
localization transition in $3$d 
Anderson model. The transition point
in the $KL2$ collapse and the corresponding critical exponent $\nu$ were found by the
best collapse with the minimal $\chi^{2}$ deviation from the scaling
function which was updated at any step of iterative collapse process.
Such process converges
and gives the optimal values of $\gamma_{c}$ and $\nu$, as well as the scaling function
(parameterized by a 6-order polynomial), despite there is no intersection in the $KL2$ curves
for different $N$ (see Fig.~\ref{Fig:intersect}).
\label{Fig:collapse_3DAM} }
\end{figure*}
Next using the statistical independence of $\psi$ and $\tilde{\psi}$ in Eq.~\eqref{KL2} and normalization
of wave functions we represent
\be\label{expr-KL2}
KL2 =\la \sum_{i}|\psi(i)|^{2}\,\ln|\psi(i)|^{2} \ra
-N^{-1}\la \sum_{i} \ln|\psi(i)|^{2}\ra.
\ee
Now we express both terms in Eq.~(\ref{expr-KL2}) in terms of $M_{q}$ using the identity:
\be\label{log-expr}
\ln |\psi_{\alpha}(i)|^{2} = \lim_{\epsilon\rightarrow 0}
\epsilon^{-1}\,(|\psi_{\alpha}(i)|^{\epsilon}-1)
\ee
The first term
is equal to:
\be
\la\sum_{i}\lim_{\epsilon\rightarrow\infty}\frac{|\psi(i)|^{2(1+\epsilon)}-
|\psi(i)|^{2}}
{\epsilon}\ra=\lim_{\epsilon\rightarrow\infty}\left[\frac{1}{\epsilon}\left(
M_{1+\epsilon}-1\right)\right].
\ee
The second term can be expressed as:
\be
-\frac{1}{N}\la\sum_{i}\lim_{\epsilon\rightarrow\infty}\frac{
|\psi(i)|^{2\epsilon}-1}{\epsilon}\ra=-\lim_{\epsilon\rightarrow 0}\left[\frac{1}
{\epsilon}\left( N^{-1}M_{\epsilon}-1\right)\right].
\ee
Now expanding $M_{1+\epsilon}$ and $M_{\epsilon}$ in the vicinity of $q=0,1$ and defining
\bea
f_{1+\epsilon}(x)&=&1+\epsilon\,\varphi_{1}(x)+O(\epsilon^2);\\
 f_{\epsilon}(x)&=&1-\epsilon\,\varphi_{0}(x)+O(\epsilon^2),
\eea
we obtain:
\be
KL2=KL2_{c}(N) +\varphi_{0}(L/\xi_{0})+\varphi_{1}(L/\xi_{1}),
\ee
where $KL2_{c}$ is logarithmically divergent:
\bea\label{KL2_crit}
KL2_{c}&=&\ln N \,(1-\partial_{\epsilon} D_{\epsilon}|_{\epsilon=0} -
D_{1})+{\rm const}.\\ \nonumber &=&\ln N\,(\alpha_{0}-D_{1})+{\rm const}.
\eea
Here we used the identity for $\alpha_{0}$ describing the typical value of the
wave function amplitude $|\psi|_{typ}^{2}=N^{-\alpha_{0}}$:
\bea
\alpha_{0}=\frac{d\tau_{\epsilon}}{d\epsilon}\left|_{\epsilon=0}\right.
=\partial_{\epsilon}[D_{\epsilon}(\epsilon-1)]\left|_{\epsilon=0}\right..
\eea

 Note that, generally speaking, the characteristic lengths
$\xi_{0}\sim |\gamma-\gamma_{c}|^{-\nu^{(0)}}$ and $\xi_{1}
\sim|\gamma-\gamma_{c}|^{-\nu^{(1)}}$ may have different critical exponents
$\nu^{(0)}$ and $\nu^{(1)}$. If this is the case, the smallest one will dominate
the finite-size corrections near the critical point:
\be\label{KL2_fin}
KL2 - KL2_c (N) =
\Phi_{2}(L |\gamma-\gamma_c|^{\nu_{2}}),\;\;\;
 \nu_{2}={\rm min}\{\nu^{(0)},\nu^{(1)} \}.
\ee

Eq.~(\ref{KL2_fin}) is used in this paper for the numerical characterization of
the phases. Deeply in the multifractal phase and at $L\gg 1$ the scaling
function $\Phi_{2}(x)$ according to
Eq.~(\ref{asympt}) is a constant. Then $KL2_c(N)$ and $KL2$ are both
logarithmically divergent, as
$\alpha_{0}>1$ and $D_{1}<1$ in
Eq.~(\ref{KL2_crit}) in the multifractal phase.

The scaling function $\Phi_{2}(x)$ is also a constant deeply in the ergodic phase
but in
this case $\alpha_{0}=D_{1}=1$ and the logarithmic divergence of $KL2_c$ is gone. As the
result $KL2=2$ is independent of $N$ deeply in the ergodic phase.

At the {\it continuous}
ergodic transition $\alpha_{0}=D_{1}=1$, and
the critical value $KL2_c(N)$ of $KL2$ is independent of $N$. This results in
{\it crossing} at $\gamma=\gamma_{ET}$ of all the curves for $KL2$ at different values of
$N$ which helps to identify the {\it ergodic} transition~\cite{KLPino}.

However, if the ergodic transition coincides with the Anderson localization transition
and is {\it discontinuous}, (i.e. $\alpha_{0}$ and $D_{1}$
are not equal to 1 at the transition), the critical value $KL2_c (N)$
is no longer $N$-independent. In this case the crossing is smeared out and can
disappear whatsoever. Nonetheless, by subtracting $KL2_c$ from $KL2$ one can still
identify
the transition from the condition of the best collapse by choosing an optimal $\gamma_{c}$
in Eq.~(\ref{KL2_fin}). However, it is safer to use $KL1$ in this case.

The derivation of finite size scaling (FSS) for $KL1$ proceeds in the same way by plugging the identity
Eq.~(\ref{log-expr})
into:
\begin{multline}\label{expr-KL1}
KL1 =\la \sum_{i}|\psi_{\alpha}(i)|^{2}\,\ln|\psi_{\alpha}(i)|^{2}\ra\\
-\la \sum_{i} |\psi_{\alpha}|^{2}\,\ln|\psi_{\alpha+1}(i)|^{2}\ra.
\end{multline}
and employing the ansatz:
\begin{multline}
\la\sum_{i}|\psi_{E}(i)|^{2q_{1}}\,|\psi_{E+\omega}(i)|^{2q_{2}}
\ra \sim N^{1+\beta}\,N_{\omega}^{\alpha}\\ \times \,F_{q_{1},q_{2}}
(L/\xi_{q_{1}},L/\xi_{q_{2}}),
\end{multline}
where $N_{\omega}=1/(\rho\omega)$ and $\rho$ is the mean DoS.

Applying for large $\omega\sim \rho^{-1}$ ($N_\omega\simeq 1$) the ``decoupling rule'':
\begin{multline}
\la\sum_{i}|\psi_{E}(i)|^{2q_{1}}\,|\psi_{E+\omega}(i)|^{2q_{2}}
\ra \sim \\ \sum_{i}\la |\psi_{E}(i)|^{2q_{1}}\ra\,
\la|\psi_{E+\omega}(i)|^{2q_{2}}
\ra,
\end{multline}
and for small $\omega\sim \delta$ ($N_\omega \simeq N$) the ``fusion rule'':
\begin{multline}
\la\sum_{i}|\psi_{E}(i)|^{2q_{1}}\,|\psi_{E+\omega}(i)|^{2q_{2}}
\ra \sim \\\la\sum_{i} |\psi_{E}(i)|^{2q_{1}+2q_{2}}\ra,
\end{multline}
one easily finds:
\bea
\beta &=& -2 +D_{q_{1}}(1-q_{1})+D_{q_{2}}(1-q_{2}),\\ \nonumber
\alpha+\beta &=& -1+D_{q_{1}+q_{2}} (1-q_{1}-q_{2}).
\eea
Due to the ``fusion rule'' for $\psi_{\alpha}$ and $\psi_{\alpha+1}$ we obtain from
Eq.~(\ref{moment-ansatz}):
\bea\label{fuss}
 &&\la\sum_{i}|\psi_{\alpha}(i)|^{2q_{1}}\,|\psi_{\alpha+1}(i)|^{2q_{2}}
\ra \sim F_{q_{1},q_{2}}
(L/\xi_{q_{1}+q_{2}} )\nonumber \\ &&\times
 N^{-D_{q_{1}+q_{2}}(q_{1}+q_{2}-1)}.
\eea

Substituting Eq.~(\ref{fuss}) in Eqs.~(\ref{log-expr}),~(\ref{expr-KL1})
we observe cancelation of the leading logarithmic in $N$ terms in $KL1$ deeply in
the multifractal phase:
\be
KL1_c ={\rm const}.
\ee
We obtain:
\be\label{KL1_fin}
KL1 = \Phi_{1}(L|\gamma-\gamma_c|^{\nu_{1}}).
\ee
where $\nu_{1} = \nu^{(1)} \geq \nu_{2}$ and the crossover scaling function $\Phi_{1}(x)$ is:
\be
\Phi_{1}(x)=\partial_{\epsilon}f_{1+\epsilon}(x)-\partial_{\epsilon}f_{1,\epsilon}(x)|
_{\epsilon=0}.
\ee
As it is seen from Eq.~(\ref{KL1_fin}), $KL1$ is independent of $N$ at the Anderson
transition point
$\gamma=\gamma_{AT}$. Thus all curves for $KL1$ at different values of N intersect at
$\gamma=\gamma_{AT}$. This gives us a powerful instrument to identify the Anderson
localization transition point.
\section{Finite-size scaling collapse for $KL1$ and $KL2$ for 3D Anderson model}\label{App_sec:KL_collapse}

In Fig.~\ref{Fig:collapse_3DAM} we present the result for the data collapse for $KL1$ and
$KL2$
in the vicinity of the localization transition in the $3$d 
Anderson model.
This result demonstrates that our iteration procedure is convergent and gives a good
approximation for the critical point from the collapse of $KL2$ data which do not show
any intersection of $KL2$ vs. $W$ curves at the critical point. From this collapse we
found the critical exponents:
\be\label{nu_3D}
\nu_{1}=1.43\pm 0.15,\;\;\;\nu_{2}=1.13\pm 0.2.
\ee
Note that from the results of Appendix~\ref{App_sec:theory_KL} it follows that quite
generally at the {\it same} critical point:
\be\label{ineq_nu}
\nu_{2}\leq \nu_{1},
\ee
since $\nu_{2}$ is given by the minimal of the two values $\nu^{(0)}$ and $\nu^{(1)}$
(see Eq.~(\ref{KL2_fin})) corresponding
to the wave function moments Eq.~(\ref{moment-ansatz}) with $q=0$ and $q=1$,
respectively. At the same time, $\nu_{1}=\nu^{(1)}$. Our result Eq.~(\ref{nu_3D})
satisfies the inequality Eq.~(\ref{ineq_nu}). On the theory side it follows from nowhere that there is only
one single critical exponent $\nu$ of {\it any} FSS in a situation where there is a continuous
multitude of multifractal dimensions. In our opinion, it is more natural to assume that
the exponent $\nu$ is specific to the quantity which FSS is studied, as it is shown in the
Appendix~\ref{App_sec:theory_KL} for $KL1$ and $KL2$.

However, our samples are too small and our FSS analysis
is too simplistic (e.g. it does not take into account irrelevant scaling exponents) to
claim that $\nu_{1}$ and $\nu_{2}$ are really different.

Note that for different ergodic and Anderson localization transitions
the inequality~\eqref{ineq_nu} is not valid in general and thus, cannot be
applied to the LN-RP at $p<1$, while for $p\geq 1$ it is saturated, see Table~\ref{Fig:tab_nu} and Fig.~\ref{Fig:conj-nu}.

\section{Analysis of stability}\label{App_sec:Stability}
\begin{figure}[b!]
\center{
\includegraphics[width=0.7\linewidth,angle=0]{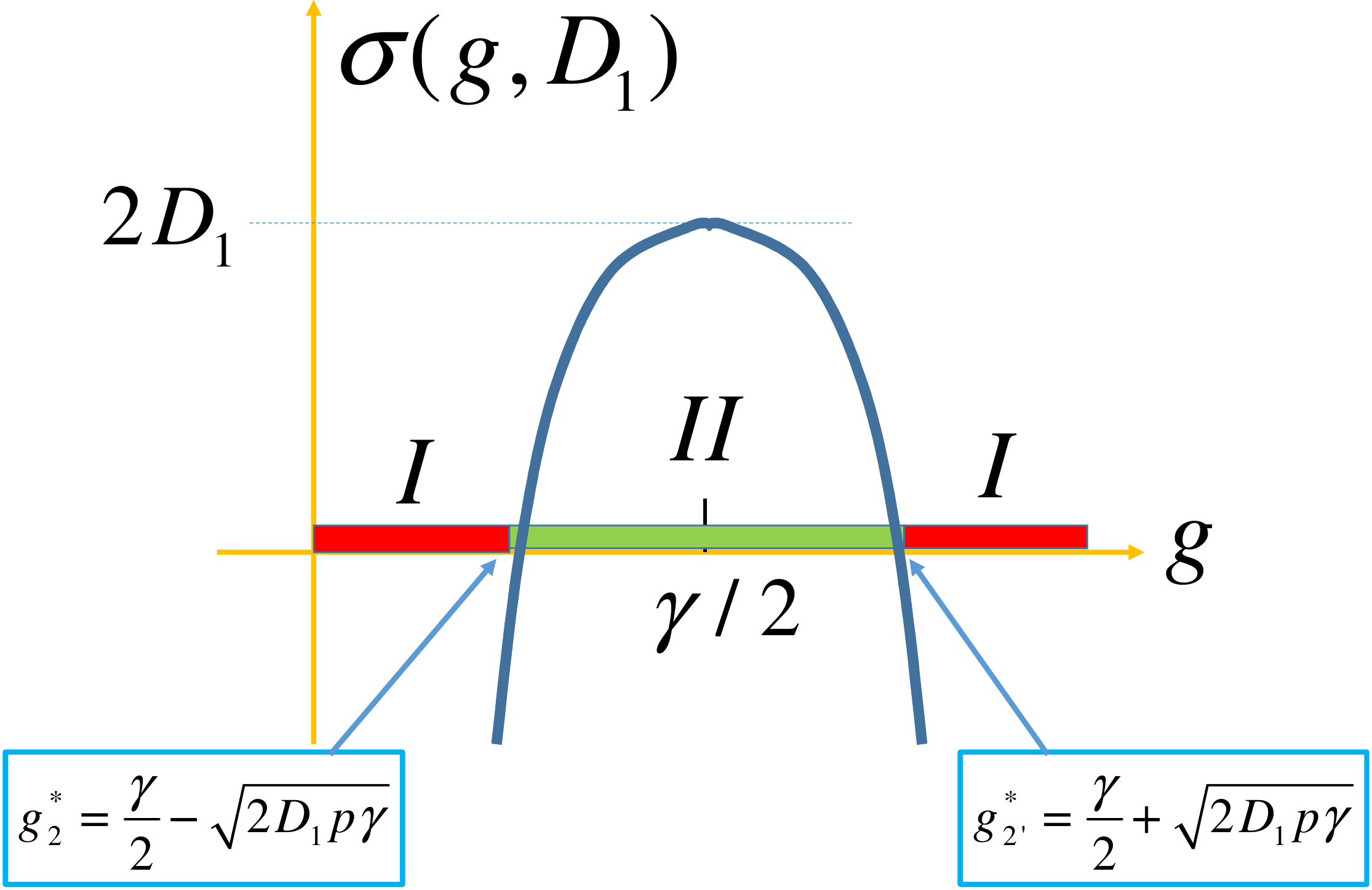}}
\caption{(Color online) Regions of $g$ contributing to the log-normal (I) and
Gaussian (II) parts of the distribution function $P(U_{\mu,\nu})$. }
\label{Fig:regions}
\end{figure}

In this section we calculate the contributions to $P(V)$ from the log-normal and Gaussian parts to Eq.~\eqref{distr-V}.

\begin{figure*}[t]
\center{
\includegraphics[width=0.49 \linewidth,angle=0]{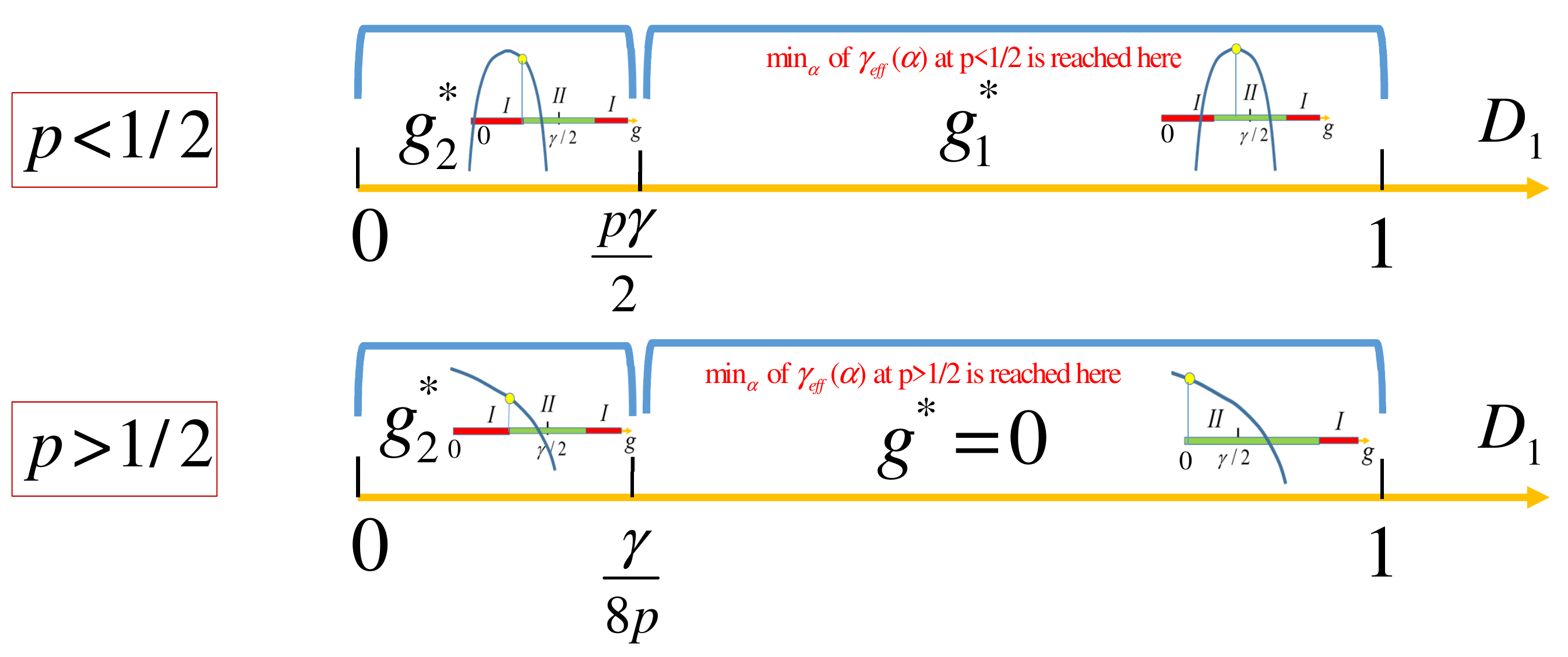}
\includegraphics[width=0.49 \linewidth,angle=0]{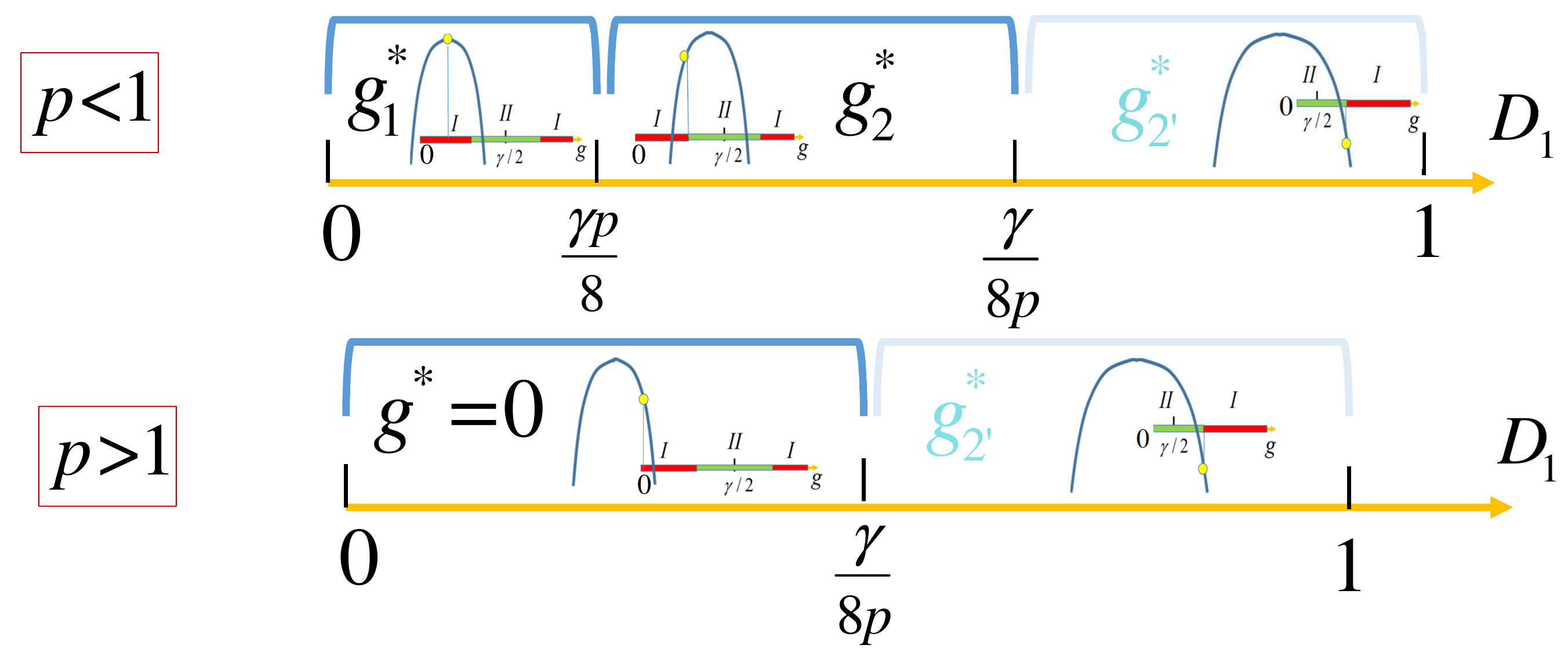}}
\caption{(Color online) (Left panel) Different possible positions $g^{*}_{1}$, $g^{*}_{2}$
or $g^{*}=0$ that maximize
Eq.~(\ref{max}) in region II depending on $p$, $\gamma$ and $D_{1}$. The configuration
of maximum realized in each sector of parameters is shown by an ikon
in the corresponding sector.
(Right panel) Different possible positions $g^{*}_{1}$, $g^{*}_{2}$
or $g^{*}=0$ that maximize
Eq.~(\ref{max-LN}) in region I. The configuration
of maximum realized in each sector of parameters is shown by an ikon
in the corresponding sector. For $D_{1}>\gamma/8p$ the maximum in
Eq.~(\ref{max-LN}) is reached at the
edge of the {\it right} segment of region I, $g=g_{2'}^{*}$ (not to be
confused with the edge of the {\it left} segment $g=g_{2}^{*}$,
see Fig.~\ref{Fig:regions} ). It leads to a higher branch of
the orange curve $\Delta(\alpha)+\alpha/2$ in Fig.~\ref{Fig:stability}
(not shown in Fig.~\ref{Fig:stability})
which is separated by a gap from the blue curve in Fig.~\ref{Fig:stability}
and thus is irrelevant for our analysis.
\label{Fig:maximize_Gauss_LN}}
\end{figure*}

One can easily compute the variance of the Gaussian part of $P(V)$
leaving in it only the
bi-diagonal terms with $i=i'$ and $j=j'$:
\bea\label{max}
&&\langle V^{2} \rangle =\int_{g\in II} dg\, N^{-\frac{1}{p\gamma}
\left(g-\frac{\gamma}{2} \right)^{2} -2g}\\ \nonumber
&\sim & {\rm max}_{g\in II}\left\{ N^{-\frac{1}{p\gamma}
\left(g-\frac{\gamma}{2} \right)^{2} -2g}\right\}\equiv
N^{-\gamma_{{\rm eff}}}.
\eea
The maximum in Eq.~(\ref{max}) at $g$ belonging to region II in Fig.~\ref{Fig:regions}
can be reached
(i) inside the region II at $g=g^{*}_{1}$, (ii) at the border of this region at
$g=g^{*}_{2}$,
and (iii) at the
cut-off of $P(g)$ at $g^{*}=0$ (see Fig.~\ref{Fig:regions} and
Fig.~\ref{Fig:maximize_Gauss_LN}(left)).

The expression for $\gamma_{{\rm eff}}(D_{1)}$ takes the form:
\be\label{gamma_eff_App}
 \gamma_{{\rm eff}}(D_{1})=\left\{\begin{array}{ll}
 \gamma(1-p),& \frac{p\gamma}{2} <D_{1}<1,\; p<\frac{1}{2}\cr
 2D_{1}+\gamma-2\sqrt{2D_{1}\gamma p},& D_{1}<\min\lp\frac{p\gamma}{2},\frac{\gamma}{8p}\rp\cr
 \frac{\gamma}{4p},&\frac{\gamma}{8p}<D_{1}<1,\;
p\geq\frac{1}{2}\end{array}\right..
\ee

Next we compute the function
\be\label{max-LN}
\Delta(D_{1})=-{\rm max}_{g\in I}\left\{ \sigma(g,D_{1})-g-D_{1}\right\}.
\ee
in Eq.~(\ref{stab-MF-LN}).

The details of the calculation which is similar to calculation of
$\gamma_{{\rm eff}}(D_{1})$ in Eq.~(\ref{max}) are illustrated in
Fig.~\ref{Fig:maximize_Gauss_LN}(right). The resulting expression for $\Delta(D_{i})$
is:
\be\label{Delta_App}
 \Delta(D_{1})=\left\{\begin{array}{ll}
 \frac{\gamma}{2}\left(1-\frac{p}{2}\right)-D_{1},&0<D_{1}<\frac{\gamma p}{8},\;p<1\cr
 D_{1}+\frac{\gamma}{2}-\sqrt{2D_{1}\gamma p},&\frac{\gamma p}{8}<D_{1}<\frac{\gamma}{8p},\; p<1\cr
 \frac{\gamma}{4p}-D_{1},& 0<D_{1}<\frac{\gamma}{8p},\;
p\geq 1\end{array}\right..
\ee

\onecolumngrid
\twocolumngrid

\bibliography{LN-RP}

\begin{thebibliography}{48}%
\makeatletter
\providecommand \@ifxundefined [1]{%
 \@ifx{#1\undefined}
}%
\providecommand \@ifnum [1]{%
 \ifnum #1\expandafter \@firstoftwo
 \else \expandafter \@secondoftwo
 \fi
}%
\providecommand \@ifx [1]{%
 \ifx #1\expandafter \@firstoftwo
 \else \expandafter \@secondoftwo
 \fi
}%
\providecommand \natexlab [1]{#1}%
\providecommand \enquote  [1]{``#1''}%
\providecommand \bibnamefont  [1]{#1}%
\providecommand \bibfnamefont [1]{#1}%
\providecommand \citenamefont [1]{#1}%
\providecommand \href@noop [0]{\@secondoftwo}%
\providecommand \href [0]{\begingroup \@sanitize@url \@href}%
\providecommand \@href[1]{\@@startlink{#1}\@@href}%
\providecommand \@@href[1]{\endgroup#1\@@endlink}%
\providecommand \@sanitize@url [0]{\catcode `\\12\catcode `\$12\catcode
  `\&12\catcode `\#12\catcode `\^12\catcode `\_12\catcode `\%12\relax}%
\providecommand \@@startlink[1]{}%
\providecommand \@@endlink[0]{}%
\providecommand \url  [0]{\begingroup\@sanitize@url \@url }%
\providecommand \@url [1]{\endgroup\@href {#1}{\urlprefix }}%
\providecommand \urlprefix  [0]{URL }%
\providecommand \Eprint [0]{\href }%
\providecommand \doibase [0]{http://dx.doi.org/}%
\providecommand \selectlanguage [0]{\@gobble}%
\providecommand \bibinfo  [0]{\@secondoftwo}%
\providecommand \bibfield  [0]{\@secondoftwo}%
\providecommand \translation [1]{[#1]}%
\providecommand \BibitemOpen [0]{}%
\providecommand \bibitemStop [0]{}%
\providecommand \bibitemNoStop [0]{.\EOS\space}%
\providecommand \EOS [0]{\spacefactor3000\relax}%
\providecommand \BibitemShut  [1]{\csname bibitem#1\endcsname}%
\let\auto@bib@innerbib\@empty
\bibitem [{\citenamefont {Basko}\ \emph {et~al.}(2006)\citenamefont {Basko},
  \citenamefont {Aleiner},\ and\ \citenamefont {Altshuler}}]{BAA}%
  \BibitemOpen
  \bibfield  {author} {\bibinfo {author} {\bibfnamefont {D.M.}\ \bibnamefont
  {Basko}}, \bibinfo {author} {\bibfnamefont {I.~L.}\ \bibnamefont {Aleiner}},
  \ and\ \bibinfo {author} {\bibfnamefont {Boris~L.}\ \bibnamefont
  {Altshuler}},\ }\bibfield  {title} {\enquote {\bibinfo {title}
  {{Metal-insulator transition in a weakly interacting many-electron system
  with localized single-particle states}},}\ }\href {\doibase
  10.1016/j.aop.2005.11.014} {\bibfield  {journal} {\bibinfo  {journal} {Ann.
  Phys. (N. Y).}\ }\textbf {\bibinfo {volume} {321}},\ \bibinfo {pages}
  {1126--1205} (\bibinfo {year} {2006})}\BibitemShut {NoStop}%
\bibitem [{\citenamefont {Altshuler}\ \emph {et~al.}(1997)\citenamefont
  {Altshuler}, \citenamefont {Gefen}, \citenamefont {Kamenev},\ and\
  \citenamefont {Levitov}}]{AGKL}%
  \BibitemOpen
  \bibfield  {author} {\bibinfo {author} {\bibfnamefont {Boris~L.}\
  \bibnamefont {Altshuler}}, \bibinfo {author} {\bibfnamefont {Yuval}\
  \bibnamefont {Gefen}}, \bibinfo {author} {\bibfnamefont {Alex}\ \bibnamefont
  {Kamenev}}, \ and\ \bibinfo {author} {\bibfnamefont {Leonid~S.}\ \bibnamefont
  {Levitov}},\ }\bibfield  {title} {\enquote {\bibinfo {title} {{Quasiparticle
  Lifetime in a Finite System: A Nonperturbative Approach}},}\ }\href {\doibase
  10.1103/PhysRevLett.78.2803} {\bibfield  {journal} {\bibinfo  {journal}
  {Phys. Rev. Lett.}\ }\textbf {\bibinfo {volume} {78}},\ \bibinfo {pages}
  {2803--2806} (\bibinfo {year} {1997})}\BibitemShut {NoStop}%
\bibitem [{\citenamefont {De~Luca}\ \emph {et~al.}(2014)\citenamefont
  {De~Luca}, \citenamefont {Altshuler}, \citenamefont {Kravtsov},\ and\
  \citenamefont {Scardicchio}}]{DeLuca2014}%
  \BibitemOpen
  \bibfield  {author} {\bibinfo {author} {\bibfnamefont {Andrea}\ \bibnamefont
  {De~Luca}}, \bibinfo {author} {\bibfnamefont {BL}~\bibnamefont {Altshuler}},
  \bibinfo {author} {\bibfnamefont {VE}~\bibnamefont {Kravtsov}}, \ and\
  \bibinfo {author} {\bibfnamefont {A}~\bibnamefont {Scardicchio}},\ }\bibfield
   {title} {\enquote {\bibinfo {title} {{Anderson localization on the Bethe
  lattice: Nonergodicity of extended states}},}\ }\href {\doibase
  10.1103/PhysRevLett.113.046806} {\bibfield  {journal} {\bibinfo  {journal}
  {Phys. Rev. Lett.}\ }\textbf {\bibinfo {volume} {113}},\ \bibinfo {pages}
  {046806} (\bibinfo {year} {2014})}\BibitemShut {NoStop}%
\bibitem [{\citenamefont {V.E.Kravtsov}\ \emph {et~al.}(2018)\citenamefont
  {V.E.Kravtsov}, \citenamefont {B.L.Altshuler},\ and\ \citenamefont
  {L.B.Ioffe}}]{AnnalRRG}%
  \BibitemOpen
  \bibfield  {author} {\bibinfo {author} {\bibnamefont {V.E.Kravtsov}},
  \bibinfo {author} {\bibnamefont {B.L.Altshuler}}, \ and\ \bibinfo {author}
  {\bibnamefont {L.B.Ioffe}},\ }\bibfield  {title} {\enquote {\bibinfo {title}
  {{Non-ergodic delocalized phase in Anderson model on Bethe lattice and
  regular graph}},}\ }\href {\doibase
  https://doi.org/10.1016/j.aop.2017.12.009} {\bibfield  {journal} {\bibinfo
  {journal} {Annals of Physics}\ }\textbf {\bibinfo {volume} {389}},\ \bibinfo
  {pages} {148--191} (\bibinfo {year} {2018})}\BibitemShut {NoStop}%
\bibitem [{\citenamefont {Smelyanskiy}\ \emph {et~al.}(2020)\citenamefont
  {Smelyanskiy}, \citenamefont {Kechedzhi}, \citenamefont {Boixo},
  \citenamefont {Isakov}, \citenamefont {Neven},\ and\ \citenamefont
  {Altshuler}}]{ASm1}%
  \BibitemOpen
  \bibfield  {author} {\bibinfo {author} {\bibfnamefont {Vadim~N.}\
  \bibnamefont {Smelyanskiy}}, \bibinfo {author} {\bibfnamefont {Kostyantyn}\
  \bibnamefont {Kechedzhi}}, \bibinfo {author} {\bibfnamefont {Sergio}\
  \bibnamefont {Boixo}}, \bibinfo {author} {\bibfnamefont {Sergei~V.}\
  \bibnamefont {Isakov}}, \bibinfo {author} {\bibfnamefont {Hartmut}\
  \bibnamefont {Neven}}, \ and\ \bibinfo {author} {\bibfnamefont {Boris}\
  \bibnamefont {Altshuler}},\ }\bibfield  {title} {\enquote {\bibinfo {title}
  {Nonergodic delocalized states for efficient population transfer within a
  narrow band of the energy landscape},}\ }\href {\doibase
  10.1103/PhysRevX.10.011017} {\bibfield  {journal} {\bibinfo  {journal} {Phys.
  Rev. X}\ }\textbf {\bibinfo {volume} {10}},\ \bibinfo {pages} {011017}
  (\bibinfo {year} {2020})}\BibitemShut {NoStop}%
\bibitem [{\citenamefont {Kechedzhi}\ \emph {et~al.}(2018)\citenamefont
  {Kechedzhi}, \citenamefont {Smelyanskiy}, \citenamefont {McClean},
  \citenamefont {Denchev}, \citenamefont {Mohseni}, \citenamefont {Isakov},
  \citenamefont {Boixo}, \citenamefont {Altshuler},\ and\ \citenamefont
  {Neven}}]{ASm2}%
  \BibitemOpen
  \bibfield  {author} {\bibinfo {author} {\bibfnamefont {K.}~\bibnamefont
  {Kechedzhi}}, \bibinfo {author} {\bibfnamefont {V.~N.}\ \bibnamefont
  {Smelyanskiy}}, \bibinfo {author} {\bibfnamefont {J.~R}\ \bibnamefont
  {McClean}}, \bibinfo {author} {\bibfnamefont {V.~S}\ \bibnamefont {Denchev}},
  \bibinfo {author} {\bibfnamefont {M.}~\bibnamefont {Mohseni}}, \bibinfo
  {author} {\bibfnamefont {S.~V.}\ \bibnamefont {Isakov}}, \bibinfo {author}
  {\bibfnamefont {S.}~\bibnamefont {Boixo}}, \bibinfo {author} {\bibfnamefont
  {B.~L.}\ \bibnamefont {Altshuler}}, \ and\ \bibinfo {author} {\bibfnamefont
  {H.}~\bibnamefont {Neven}},\ }\href@noop {} {\enquote {\bibinfo {title}
  {Efficient population transfer via non-ergodic extended states in quantum
  spin glass},}\ } (\bibinfo {year} {2018}),\ \Eprint
  {http://arxiv.org/abs/1807.04792} {arXiv:1807.04792} \BibitemShut {NoStop}%
\bibitem [{\citenamefont {Tikhonov}\ and\ \citenamefont
  {Mirlin}(2016)}]{Tikh-Mir1}%
  \BibitemOpen
  \bibfield  {author} {\bibinfo {author} {\bibfnamefont {K.~S.}\ \bibnamefont
  {Tikhonov}}\ and\ \bibinfo {author} {\bibfnamefont {A.~D.}\ \bibnamefont
  {Mirlin}},\ }\bibfield  {title} {\enquote {\bibinfo {title} {{Fractality of
  wave functions on a Cayley tree: Difference between tree and locally treelike
  graph without boundary}},}\ }\href {\doibase 10.1103/PhysRevB.94.184203}
  {\bibfield  {journal} {\bibinfo  {journal} {Phys. Rev. B}\ }\textbf {\bibinfo
  {volume} {94}},\ \bibinfo {pages} {184203} (\bibinfo {year}
  {2016})}\BibitemShut {NoStop}%
\bibitem [{\citenamefont {Tikhonov}\ and\ \citenamefont
  {Mirlin}(2019)}]{Tikh-Mir2}%
  \BibitemOpen
  \bibfield  {author} {\bibinfo {author} {\bibfnamefont {K.~S.}\ \bibnamefont
  {Tikhonov}}\ and\ \bibinfo {author} {\bibfnamefont {A.~D.}\ \bibnamefont
  {Mirlin}},\ }\bibfield  {title} {\enquote {\bibinfo {title} {Statistics of
  eigenstates near the localization transition on random regular graphs},}\
  }\href {\doibase 10.1103/PhysRevB.99.024202} {\bibfield  {journal} {\bibinfo
  {journal} {Phys. Rev. B}\ }\textbf {\bibinfo {volume} {99}},\ \bibinfo
  {pages} {024202} (\bibinfo {year} {2019})}\BibitemShut {NoStop}%
\bibitem [{\citenamefont {Parisi}\ \emph {et~al.}(2019)\citenamefont {Parisi},
  \citenamefont {Pascazio}, \citenamefont {Pietracaprina}, \citenamefont
  {Ros},\ and\ \citenamefont {Scardicchio}}]{Scard-Par}%
  \BibitemOpen
  \bibfield  {author} {\bibinfo {author} {\bibfnamefont {Giorgio}\ \bibnamefont
  {Parisi}}, \bibinfo {author} {\bibfnamefont {Saverio}\ \bibnamefont
  {Pascazio}}, \bibinfo {author} {\bibfnamefont {Francesca}\ \bibnamefont
  {Pietracaprina}}, \bibinfo {author} {\bibfnamefont {Valentina}\ \bibnamefont
  {Ros}}, \ and\ \bibinfo {author} {\bibfnamefont {Antonello}\ \bibnamefont
  {Scardicchio}},\ }\bibfield  {title} {\enquote {\bibinfo {title} {{Anderson
  transition on the Bethe lattice: an approach with real energies}},}\ }\href
  {\doibase 10.1088/1751-8121/ab56e8} {\bibfield  {journal} {\bibinfo
  {journal} {Journal of Physics A: Mathematical and Theoretical}\ }\textbf
  {\bibinfo {volume} {53}},\ \bibinfo {pages} {014003} (\bibinfo {year}
  {2019})}\BibitemShut {NoStop}%
\bibitem [{\citenamefont {Bera}\ \emph {et~al.}(2018)\citenamefont {Bera},
  \citenamefont {De~Tomasi}, \citenamefont {Khaymovich},\ and\ \citenamefont
  {Scardicchio}}]{RRG_R(t)}%
  \BibitemOpen
  \bibfield  {author} {\bibinfo {author} {\bibfnamefont {S.}~\bibnamefont
  {Bera}}, \bibinfo {author} {\bibfnamefont {G.}~\bibnamefont {De~Tomasi}},
  \bibinfo {author} {\bibfnamefont {I.~M.}\ \bibnamefont {Khaymovich}}, \ and\
  \bibinfo {author} {\bibfnamefont {A.}~\bibnamefont {Scardicchio}},\
  }\bibfield  {title} {\enquote {\bibinfo {title} {{Return probability for the
  Anderson model on the random regular graph}},}\ }\href {\doibase
  10.1103/PhysRevB.98.134205} {\bibfield  {journal} {\bibinfo  {journal} {Phys.
  Rev. B}\ }\textbf {\bibinfo {volume} {98}},\ \bibinfo {pages} {134205}
  (\bibinfo {year} {2018})}\BibitemShut {NoStop}%
\bibitem [{\citenamefont {{De Tomasi}}\ \emph {et~al.}(2019)\citenamefont {{De
  Tomasi}}, \citenamefont {{Bera}}, \citenamefont {{Scardicchio}},\ and\
  \citenamefont {{Khaymovich}}}]{deTomasi2019subdiffusion}%
  \BibitemOpen
  \bibfield  {author} {\bibinfo {author} {\bibfnamefont {Giuseppe}\
  \bibnamefont {{De Tomasi}}}, \bibinfo {author} {\bibfnamefont {Soumya}\
  \bibnamefont {{Bera}}}, \bibinfo {author} {\bibfnamefont {Antonello}\
  \bibnamefont {{Scardicchio}}}, \ and\ \bibinfo {author} {\bibfnamefont
  {Ivan~M.}\ \bibnamefont {{Khaymovich}}},\ }\href
  {https://journals.aps.org/prb/accepted/e2074YdbW1a1166624a35e7563f71e4dc037b0b5b}
  {\enquote {\bibinfo {title} {{Sub-diffusion in the Anderson model on random
  regular graph}},}\ } (\bibinfo {year} {2019}),\ \bibinfo {note} {accepted for
  publication in PRB(R)},\ \Eprint {http://arxiv.org/abs/1908.11388}
  {arXiv:1908.11388} \BibitemShut {NoStop}%
\bibitem [{\citenamefont {Garc\'{\i}a-Mata}\ \emph {et~al.}(2017)\citenamefont
  {Garc\'{\i}a-Mata}, \citenamefont {Giraud}, \citenamefont {Georgeot},
  \citenamefont {Martin}, \citenamefont {Dubertrand},\ and\ \citenamefont
  {Lemari\'e}}]{Lemarie2017}%
  \BibitemOpen
  \bibfield  {author} {\bibinfo {author} {\bibfnamefont {I.}~\bibnamefont
  {Garc\'{\i}a-Mata}}, \bibinfo {author} {\bibfnamefont {O.}~\bibnamefont
  {Giraud}}, \bibinfo {author} {\bibfnamefont {B.}~\bibnamefont {Georgeot}},
  \bibinfo {author} {\bibfnamefont {J.}~\bibnamefont {Martin}}, \bibinfo
  {author} {\bibfnamefont {R.}~\bibnamefont {Dubertrand}}, \ and\ \bibinfo
  {author} {\bibfnamefont {G.}~\bibnamefont {Lemari\'e}},\ }\bibfield  {title}
  {\enquote {\bibinfo {title} {Scaling theory of the anderson transition in
  random graphs: Ergodicity and universality},}\ }\href {\doibase
  10.1103/PhysRevLett.118.166801} {\bibfield  {journal} {\bibinfo  {journal}
  {Phys. Rev. Lett.}\ }\textbf {\bibinfo {volume} {118}},\ \bibinfo {pages}
  {166801} (\bibinfo {year} {2017})}\BibitemShut {NoStop}%
\bibitem [{\citenamefont {Garc\'{\i}a-Mata}\ \emph {et~al.}(2020)\citenamefont
  {Garc\'{\i}a-Mata}, \citenamefont {Martin}, \citenamefont {Dubertrand},
  \citenamefont {Giraud}, \citenamefont {Georgeot},\ and\ \citenamefont
  {Lemari\'e}}]{Lemarie2020_2loc_lengths}%
  \BibitemOpen
  \bibfield  {author} {\bibinfo {author} {\bibfnamefont {I.}~\bibnamefont
  {Garc\'{\i}a-Mata}}, \bibinfo {author} {\bibfnamefont {J.}~\bibnamefont
  {Martin}}, \bibinfo {author} {\bibfnamefont {R.}~\bibnamefont {Dubertrand}},
  \bibinfo {author} {\bibfnamefont {O.}~\bibnamefont {Giraud}}, \bibinfo
  {author} {\bibfnamefont {B.}~\bibnamefont {Georgeot}}, \ and\ \bibinfo
  {author} {\bibfnamefont {G.}~\bibnamefont {Lemari\'e}},\ }\bibfield  {title}
  {\enquote {\bibinfo {title} {Two critical localization lengths in the
  anderson transition on random graphs},}\ }\href {\doibase
  10.1103/PhysRevResearch.2.012020} {\bibfield  {journal} {\bibinfo  {journal}
  {Phys. Rev. Research}\ }\textbf {\bibinfo {volume} {2}},\ \bibinfo {pages}
  {012020} (\bibinfo {year} {2020})}\BibitemShut {NoStop}%
\bibitem [{\citenamefont {Rosenzweig}\ and\ \citenamefont {Porter}(1960)}]{RP}%
  \BibitemOpen
  \bibfield  {author} {\bibinfo {author} {\bibfnamefont {N.}~\bibnamefont
  {Rosenzweig}}\ and\ \bibinfo {author} {\bibfnamefont {C.~E.}\ \bibnamefont
  {Porter}},\ }\bibfield  {title} {\enquote {\bibinfo {title} {{"Repulsion of
  Energy Levels" in Complex Atomic Spectra}},}\ }\href {\doibase
  10.1103/PhysRev.120.1698} {\bibfield  {journal} {\bibinfo  {journal} {Phys.
  Rev. B}\ }\textbf {\bibinfo {volume} {120}},\ \bibinfo {pages} {1698}
  (\bibinfo {year} {1960})}\BibitemShut {NoStop}%
\bibitem [{\citenamefont {Kravtsov}\ \emph {et~al.}(2015)\citenamefont
  {Kravtsov}, \citenamefont {Khaymovich}, \citenamefont {Cuevas},\ and\
  \citenamefont {Amini}}]{gRP}%
  \BibitemOpen
  \bibfield  {author} {\bibinfo {author} {\bibfnamefont {V.~E.}\ \bibnamefont
  {Kravtsov}}, \bibinfo {author} {\bibfnamefont {I.~M.}\ \bibnamefont
  {Khaymovich}}, \bibinfo {author} {\bibfnamefont {E.}~\bibnamefont {Cuevas}},
  \ and\ \bibinfo {author} {\bibfnamefont {M.}~\bibnamefont {Amini}},\
  }\bibfield  {title} {\enquote {\bibinfo {title} {A random matrix model with
  localization and ergodic transitions},}\ }\href {\doibase
  10.1088/1367-2630/17/12/122002} {\bibfield  {journal} {\bibinfo  {journal}
  {New J. Phys.}\ }\textbf {\bibinfo {volume} {17}},\ \bibinfo {pages} {122002}
  (\bibinfo {year} {2015})}\BibitemShut {NoStop}%
\bibitem [{\citenamefont {von Soosten}\ and\ \citenamefont
  {Warzel}(2018)}]{Warzel}%
  \BibitemOpen
  \bibfield  {author} {\bibinfo {author} {\bibfnamefont {Per}\ \bibnamefont
  {von Soosten}}\ and\ \bibinfo {author} {\bibfnamefont {Simone}\ \bibnamefont
  {Warzel}},\ }\bibfield  {title} {\enquote {\bibinfo {title} {{Non-ergodic
  delocalization in the Rosenzweig--Porter model}},}\ }\href {\doibase
  10.1007/s11005-018-1131-7} {\bibfield  {journal} {\bibinfo  {journal}
  {Letters in Mathematical Physics}\ ,\ \bibinfo {pages} {1--18}} (\bibinfo
  {year} {2018})}\BibitemShut {NoStop}%
\bibitem [{\citenamefont {Facoetti}\ \emph {et~al.}(2016)\citenamefont
  {Facoetti}, \citenamefont {Vivo},\ and\ \citenamefont {Biroli}}]{Biroli_RP}%
  \BibitemOpen
  \bibfield  {author} {\bibinfo {author} {\bibfnamefont {D.}~\bibnamefont
  {Facoetti}}, \bibinfo {author} {\bibfnamefont {P.}~\bibnamefont {Vivo}}, \
  and\ \bibinfo {author} {\bibfnamefont {G.}~\bibnamefont {Biroli}},\
  }\bibfield  {title} {\enquote {\bibinfo {title} {From non-ergodic
  eigenvectors to local resolvent statistics and back: A random matrix
  perspective},}\ }\href {\doibase 10.1209/0295-5075/115/47003} {\bibfield
  {journal} {\bibinfo  {journal} {Europhys. Lett.}\ }\textbf {\bibinfo {volume}
  {115}},\ \bibinfo {pages} {47003} (\bibinfo {year} {2016})}\BibitemShut
  {NoStop}%
\bibitem [{\citenamefont {Truong}\ and\ \citenamefont
  {Ossipov}(2016)}]{Ossipov_EPL2016_H+V}%
  \BibitemOpen
  \bibfield  {author} {\bibinfo {author} {\bibfnamefont {K.}~\bibnamefont
  {Truong}}\ and\ \bibinfo {author} {\bibfnamefont {A.}~\bibnamefont
  {Ossipov}},\ }\bibfield  {title} {\enquote {\bibinfo {title} {Eigenvectors
  under a generic perturbation: Non-perturbative results from the random matrix
  approach},}\ }\href {\doibase 10.1209/0295-5075/116/37002} {\bibfield
  {journal} {\bibinfo  {journal} {Europhys. Lett.}\ }\textbf {\bibinfo {volume}
  {116}},\ \bibinfo {pages} {37002} (\bibinfo {year} {2016})}\BibitemShut
  {NoStop}%
\bibitem [{\citenamefont {Amini}(2017)}]{Amini2017}%
  \BibitemOpen
  \bibfield  {author} {\bibinfo {author} {\bibfnamefont {M.}~\bibnamefont
  {Amini}},\ }\bibfield  {title} {\enquote {\bibinfo {title} {Spread of wave
  packets in disordered hierarchical lattices},}\ }\href {\doibase
  10.1209/0295-5075/117/30003} {\bibfield  {journal} {\bibinfo  {journal}
  {Europhys. Lett.}\ }\textbf {\bibinfo {volume} {117}},\ \bibinfo {pages}
  {30003} (\bibinfo {year} {2017})}\BibitemShut {NoStop}%
\bibitem [{\citenamefont {Monthus}(2017)}]{Monthus}%
  \BibitemOpen
  \bibfield  {author} {\bibinfo {author} {\bibfnamefont {C.}~\bibnamefont
  {Monthus}},\ }\bibfield  {title} {\enquote {\bibinfo {title} {Statistical
  properties of the green function in finite size for anderson localization
  models with multifractal eigenvectors},}\ }\href {\doibase
  10.1088/1751-8121/aa5ad2} {\bibfield  {journal} {\bibinfo  {journal} {J.
  Phys. A: Math. Theor.}\ }\textbf {\bibinfo {volume} {50}},\ \bibinfo {pages}
  {295101} (\bibinfo {year} {2017})}\BibitemShut {NoStop}%
\bibitem [{\citenamefont {Cizeau}\ and\ \citenamefont
  {Bouchaud}(1994)}]{Bouchard_Levy_Mat}%
  \BibitemOpen
  \bibfield  {author} {\bibinfo {author} {\bibfnamefont {P.}~\bibnamefont
  {Cizeau}}\ and\ \bibinfo {author} {\bibfnamefont {J.~P.}\ \bibnamefont
  {Bouchaud}},\ }\bibfield  {title} {\enquote {\bibinfo {title} {Theory of
  l\'evy matrices},}\ }\href {\doibase 10.1103/PhysRevE.50.1810} {\bibfield
  {journal} {\bibinfo  {journal} {Phys. Rev. E}\ }\textbf {\bibinfo {volume}
  {50}},\ \bibinfo {pages} {1810--1822} (\bibinfo {year} {1994})}\BibitemShut
  {NoStop}%
\bibitem [{\citenamefont {Tarquini}\ \emph {et~al.}(2016)\citenamefont
  {Tarquini}, \citenamefont {Biroli},\ and\ \citenamefont
  {Tarzia}}]{Biroli_Levy_Mat}%
  \BibitemOpen
  \bibfield  {author} {\bibinfo {author} {\bibfnamefont {E.}~\bibnamefont
  {Tarquini}}, \bibinfo {author} {\bibfnamefont {G.}~\bibnamefont {Biroli}}, \
  and\ \bibinfo {author} {\bibfnamefont {M.}~\bibnamefont {Tarzia}},\
  }\bibfield  {title} {\enquote {\bibinfo {title} {Level statistics and
  localization transitions of l\'evy matrices},}\ }\href {\doibase
  10.1103/PhysRevLett.116.010601} {\bibfield  {journal} {\bibinfo  {journal}
  {Phys. Rev. Lett.}\ }\textbf {\bibinfo {volume} {116}},\ \bibinfo {pages}
  {010601} (\bibinfo {year} {2016})}\BibitemShut {NoStop}%
\bibitem [{\citenamefont {Abou-Chacra}\ \emph {et~al.}(1973)\citenamefont
  {Abou-Chacra}, \citenamefont {Anderson},\ and\ \citenamefont
  {Thouless}}]{AbouChacra}%
  \BibitemOpen
  \bibfield  {author} {\bibinfo {author} {\bibfnamefont {R}~\bibnamefont
  {Abou-Chacra}}, \bibinfo {author} {\bibfnamefont {P.W.}\ \bibnamefont
  {Anderson}}, \ and\ \bibinfo {author} {\bibfnamefont {D.J.}\ \bibnamefont
  {Thouless}},\ }\bibfield  {title} {\enquote {\bibinfo {title} {{ A
  selfconsistent theory of localization}},}\ }\href {\doibase ?} {\bibfield
  {journal} {\bibinfo  {journal} {J. Phys. C.}\ }\textbf {\bibinfo {volume}
  {6}},\ \bibinfo {pages} {1734} (\bibinfo {year} {1973})}\BibitemShut
  {NoStop}%
\bibitem [{\citenamefont {Kullback}\ and\ \citenamefont
  {Leibler}(1951)}]{KLdiv}%
  \BibitemOpen
  \bibfield  {author} {\bibinfo {author} {\bibfnamefont {Solomon}\ \bibnamefont
  {Kullback}}\ and\ \bibinfo {author} {\bibfnamefont {Richard~A}\ \bibnamefont
  {Leibler}},\ }\bibfield  {title} {\enquote {\bibinfo {title} {On information
  and sufficiency},}\ }\href {\doibase 10.1214/aoms/1177729694} {\bibfield
  {journal} {\bibinfo  {journal} {The annals of mathematical statistics}\
  }\textbf {\bibinfo {volume} {22}},\ \bibinfo {pages} {79--86} (\bibinfo
  {year} {1951})}\BibitemShut {NoStop}%
\bibitem [{\citenamefont {Kullback}(1959)}]{KLdiv_book}%
  \BibitemOpen
  \bibfield  {author} {\bibinfo {author} {\bibfnamefont {Solomon}\ \bibnamefont
  {Kullback}},\ }\href@noop {} {\emph {\bibinfo {title} {{Information Theory
  and Statistics}}}}\ (\bibinfo  {publisher} {John Riley and Sons},\ \bibinfo
  {year} {1959})\BibitemShut {NoStop}%
\bibitem [{\citenamefont {Pino}\ \emph {et~al.}(2019)\citenamefont {Pino},
  \citenamefont {Tabanera},\ and\ \citenamefont {Serna}}]{KLPino}%
  \BibitemOpen
  \bibfield  {author} {\bibinfo {author} {\bibfnamefont {M}~\bibnamefont
  {Pino}}, \bibinfo {author} {\bibfnamefont {J}~\bibnamefont {Tabanera}}, \
  and\ \bibinfo {author} {\bibfnamefont {P}~\bibnamefont {Serna}},\ }\bibfield
  {title} {\enquote {\bibinfo {title} {{From ergodic to non-ergodic chaos in
  Rosenzweig--Porter model}},}\ }\href {\doibase 10.1088/1751-8121/ab4b76}
  {\bibfield  {journal} {\bibinfo  {journal} {Journal of Physics A:
  Mathematical and Theoretical}\ }\textbf {\bibinfo {volume} {52}},\ \bibinfo
  {pages} {475101} (\bibinfo {year} {2019})}\BibitemShut {NoStop}%
\bibitem [{\citenamefont {Ruderman}\ and\ \citenamefont {Kittel}(1954)}]{RKKY}%
  \BibitemOpen
  \bibfield  {author} {\bibinfo {author} {\bibfnamefont {M.~A.}\ \bibnamefont
  {Ruderman}}\ and\ \bibinfo {author} {\bibfnamefont {C.}~\bibnamefont
  {Kittel}},\ }\bibfield  {title} {\enquote {\bibinfo {title} {{Indirect
  Exchange Coupling of Nuclear Magnetic Moments by Conduction Electrons}},}\
  }\href {\doibase doi:10.1103/PhysRev.96.99} {\bibfield  {journal} {\bibinfo
  {journal} {Phys. Rev.}\ }\textbf {\bibinfo {volume} {96}},\ \bibinfo {pages}
  {99} (\bibinfo {year} {1954})}\BibitemShut {NoStop}%
\bibitem [{\citenamefont {Khaymovich}\ \emph {et~al.}(2015)\citenamefont
  {Khaymovich}, \citenamefont {Koski}, \citenamefont {Saira}, \citenamefont
  {Kravtsov},\ and\ \citenamefont {Pekola}}]{NatComm}%
  \BibitemOpen
  \bibfield  {author} {\bibinfo {author} {\bibfnamefont {I.~M.}\ \bibnamefont
  {Khaymovich}}, \bibinfo {author} {\bibfnamefont {J.~V.}\ \bibnamefont
  {Koski}}, \bibinfo {author} {\bibfnamefont {O.-P.}\ \bibnamefont {Saira}},
  \bibinfo {author} {\bibfnamefont {V.~E.}\ \bibnamefont {Kravtsov}}, \ and\
  \bibinfo {author} {\bibfnamefont {J.~P.}\ \bibnamefont {Pekola}},\ }\bibfield
   {title} {\enquote {\bibinfo {title} {{Multifractality of random
  eigenfunctions and generalization of Jarzynski equality}},}\ }\href {\doibase
  10.1038/ncomms8010} {\bibfield  {journal} {\bibinfo  {journal} {Nature
  Comms.}\ }\textbf {\bibinfo {volume} {6}},\ \bibinfo {pages} {7010} (\bibinfo
  {year} {2015})}\BibitemShut {NoStop}%
\bibitem [{\citenamefont {Bogomolny}\ and\ \citenamefont
  {Sieber}(2018)}]{BogomolnyPLRBM2018}%
  \BibitemOpen
  \bibfield  {author} {\bibinfo {author} {\bibfnamefont {E.}~\bibnamefont
  {Bogomolny}}\ and\ \bibinfo {author} {\bibfnamefont {M.}~\bibnamefont
  {Sieber}},\ }\bibfield  {title} {\enquote {\bibinfo {title} {Power-law random
  banded matrices and ultrametric matrices: Eigenvector distribution in the
  intermediate regime},}\ }\href {\doibase 10.1103/PhysRevE.98.042116}
  {\bibfield  {journal} {\bibinfo  {journal} {Phys. Rev. E}\ }\textbf {\bibinfo
  {volume} {98}},\ \bibinfo {pages} {042116} (\bibinfo {year}
  {2018})}\BibitemShut {NoStop}%
\bibitem [{\citenamefont {Nosov}\ \emph {et~al.}(2019)\citenamefont {Nosov},
  \citenamefont {Khaymovich},\ and\ \citenamefont {Kravtsov}}]{Nos}%
  \BibitemOpen
  \bibfield  {author} {\bibinfo {author} {\bibfnamefont {P.~A.}\ \bibnamefont
  {Nosov}}, \bibinfo {author} {\bibfnamefont {I.~M.}\ \bibnamefont
  {Khaymovich}}, \ and\ \bibinfo {author} {\bibfnamefont {V.~E.}\ \bibnamefont
  {Kravtsov}},\ }\bibfield  {title} {\enquote {\bibinfo {title}
  {Correlation-induced localization},}\ }\href
  {https://doi.org/10.1103/PhysRevB.99.104203} {\bibfield  {journal} {\bibinfo
  {journal} {Physical Review B}\ }\textbf {\bibinfo {volume} {99}},\ \bibinfo
  {pages} {104203} (\bibinfo {year} {2019})}\BibitemShut {NoStop}%
\bibitem [{del()}]{delta_typ-footnote}%
  \BibitemOpen
  \href@noop {} {}\bibinfo {note} {Note that in~\cite{Nos} this criterion has
  been modified in order to exclude measure zero of modes with atypically large
  hopping energies.}\BibitemShut {Stop}%
\bibitem [{\citenamefont {Pino}\ \emph {et~al.}(2017)\citenamefont {Pino},
  \citenamefont {Kravtsov}, \citenamefont {Altshuler},\ and\ \citenamefont
  {Ioffe}}]{Pino-Ioffe-VEK}%
  \BibitemOpen
  \bibfield  {author} {\bibinfo {author} {\bibfnamefont {M.}~\bibnamefont
  {Pino}}, \bibinfo {author} {\bibfnamefont {V.~E.}\ \bibnamefont {Kravtsov}},
  \bibinfo {author} {\bibfnamefont {B.~L.}\ \bibnamefont {Altshuler}}, \ and\
  \bibinfo {author} {\bibfnamefont {L.~B.}\ \bibnamefont {Ioffe}},\ }\bibfield
  {title} {\enquote {\bibinfo {title} {{Multifractal metal in a disordered
  Josephson junctions array}},}\ }\href {\doibase 10.1103/PhysRevB.96.214205}
  {\bibfield  {journal} {\bibinfo  {journal} {Phys. Rev. B}\ }\textbf {\bibinfo
  {volume} {96}},\ \bibinfo {pages} {214205} (\bibinfo {year}
  {2017})}\BibitemShut {NoStop}%
\bibitem [{\citenamefont {de~Tomasi}\ \emph {et~al.}(2019)\citenamefont
  {de~Tomasi}, \citenamefont {Amini}, \citenamefont {Bera}, \citenamefont
  {Khaymovich},\ and\ \citenamefont {Kravtsov}}]{return}%
  \BibitemOpen
  \bibfield  {author} {\bibinfo {author} {\bibfnamefont {Giuseppe}\
  \bibnamefont {de~Tomasi}}, \bibinfo {author} {\bibfnamefont {Moshen}\
  \bibnamefont {Amini}}, \bibinfo {author} {\bibfnamefont {Soumya}\
  \bibnamefont {Bera}}, \bibinfo {author} {\bibfnamefont {Ivan~M.}\
  \bibnamefont {Khaymovich}}, \ and\ \bibinfo {author} {\bibfnamefont
  {Vladimir~E.}\ \bibnamefont {Kravtsov}},\ }\bibfield  {title} {\enquote
  {\bibinfo {title} {{Survival probability in Generalized Rosenzweig-Porter
  random matrix ensemble}},}\ }\href {\doibase 10.21468/SciPostPhys.6.1.014}
  {\bibfield  {journal} {\bibinfo  {journal} {SciPost Phys.}\ }\textbf
  {\bibinfo {volume} {6}},\ \bibinfo {pages} {014} (\bibinfo {year}
  {2019})}\BibitemShut {NoStop}%
\bibitem [{\citenamefont {Nosov}\ and\ \citenamefont
  {Khaymovich}(2019)}]{Nosov2019mixtures}%
  \BibitemOpen
  \bibfield  {author} {\bibinfo {author} {\bibfnamefont {P.~A.}\ \bibnamefont
  {Nosov}}\ and\ \bibinfo {author} {\bibfnamefont {I.~M.}\ \bibnamefont
  {Khaymovich}},\ }\bibfield  {title} {\enquote {\bibinfo {title} {Robustness
  of delocalization to the inclusion of soft constraints in long-range random
  models},}\ }\href {\doibase 10.1103/PhysRevB.99.224208} {\bibfield  {journal}
  {\bibinfo  {journal} {Phys. Rev. B}\ }\textbf {\bibinfo {volume} {99}},\
  \bibinfo {pages} {224208} (\bibinfo {year} {2019})}\BibitemShut {NoStop}%
\bibitem [{WE-()}]{WE-footnote}%
  \BibitemOpen
  \href@noop {} {}\bibinfo {note} {Here we call wave-function ``weakly
  ergodic'' if it occupies a finite fraction of the total Hilbert space. Such
  states play an important role in several recent
  works~\cite{BogomolnyPLRBM2018,RRG_R(t),deTomasi2019subdiffusion,Behemoths2019,Nos,Baecker2019,luitz2019multifractality,Nosov2019mixtures}.}\BibitemShut
  {Stop}%
\bibitem [{\citenamefont {Aizenman}\ and\ \citenamefont
  {Warzel}(2011)}]{Warzel2011PRL}%
  \BibitemOpen
  \bibfield  {author} {\bibinfo {author} {\bibfnamefont {Michael}\ \bibnamefont
  {Aizenman}}\ and\ \bibinfo {author} {\bibfnamefont {Simone}\ \bibnamefont
  {Warzel}},\ }\bibfield  {title} {\enquote {\bibinfo {title} {{Extended States
  in a Lifshitz Tail Regime for Random Schr\"odinger Operators on Trees}},}\
  }\href {\doibase 10.1103/PhysRevLett.106.136804} {\bibfield  {journal}
  {\bibinfo  {journal} {Phys. Rev. Lett.}\ }\textbf {\bibinfo {volume} {106}},\
  \bibinfo {pages} {136804} (\bibinfo {year} {2011})}\BibitemShut {NoStop}%
\bibitem [{\citenamefont {Khaymovich}\ and\ \citenamefont
  {Kravtsov}(2020)}]{KrKhay}%
  \BibitemOpen
  \bibfield  {author} {\bibinfo {author} {\bibfnamefont {I.~M.}\ \bibnamefont
  {Khaymovich}}\ and\ \bibinfo {author} {\bibfnamefont {V.~E.}\ \bibnamefont
  {Kravtsov}},\ }\href@noop {} {} (\bibinfo {year} {2020}),\ \bibinfo {note}
  {(unpublished)}\BibitemShut {NoStop}%
\bibitem [{foo()}]{footnote:trunc}%
  \BibitemOpen
  \href@noop {} {}\bibinfo {note} {Note that the truncation at $U_{{\rm
  max}}\gtrsim O(1)$, $\gamma_{tr}\leq 0$, does not alter the phase diagram in
  Fig.~\ref{Fig:phase_diagram}.}\BibitemShut {Stop}%
\bibitem [{RRG()}]{RRG_Gorsky-footnote}%
  \BibitemOpen
  \href@noop {} {}\bibinfo {note} {One possible perturbation of the Anderson
  model on RRG with respect to its structure considered
  in~\cite{Gorsky2019_RRG} explicitly shows the above mentioned emergence of
  the multifractal phase.}\BibitemShut {Stop}%
\bibitem [{\citenamefont {Avetisov}\ \emph {et~al.}(2019)\citenamefont
  {Avetisov}, \citenamefont {Gorsky}, \citenamefont {Nechaev},\ and\
  \citenamefont {Valba}}]{Gorsky2019_RRG}%
  \BibitemOpen
  \bibfield  {author} {\bibinfo {author} {\bibfnamefont {V}~\bibnamefont
  {Avetisov}}, \bibinfo {author} {\bibfnamefont {A}~\bibnamefont {Gorsky}},
  \bibinfo {author} {\bibfnamefont {S}~\bibnamefont {Nechaev}}, \ and\ \bibinfo
  {author} {\bibfnamefont {O}~\bibnamefont {Valba}},\ }\bibfield  {title}
  {\enquote {\bibinfo {title} {{Localization and non-ergodicity in clustered
  random networks}},}\ }\href {\doibase 10.1093/comnet/cnz026} {\bibfield
  {journal} {\bibinfo  {journal} {Journal of Complex Networks}\ } (\bibinfo
  {year} {2019}),\ 10.1093/comnet/cnz026},\ \bibinfo {note}
  {cnz026}\BibitemShut {NoStop}%
\bibitem [{\citenamefont {Efetov}(1996)}]{Efetov_book}%
  \BibitemOpen
  \bibfield  {author} {\bibinfo {author} {\bibfnamefont {K.B.}\ \bibnamefont
  {Efetov}},\ }\href@noop {} {\emph {\bibinfo {title} {{Supersymmetry in
  disorder and chaos}}}}\ (\bibinfo  {publisher} {Cambridge University Press},\
  \bibinfo {year} {1996})\BibitemShut {NoStop}%
\bibitem [{\citenamefont {Mirlin}\ and\ \citenamefont
  {Fyodorov}(1991)}]{MF1991}%
  \BibitemOpen
  \bibfield  {author} {\bibinfo {author} {\bibfnamefont {A.~D.}\ \bibnamefont
  {Mirlin}}\ and\ \bibinfo {author} {\bibfnamefont {Y.~V.}\ \bibnamefont
  {Fyodorov}},\ }\bibfield  {title} {\enquote {\bibinfo {title} {{Localization
  transition in the Anderson model on the Bethe lattice: spontaneous symmetry
  breaking and correlation functions}},}\ }\href@noop {} {\bibfield  {journal}
  {\bibinfo  {journal} {Nucl. Phys. B}\ }\textbf {\bibinfo {volume} {336}},\
  \bibinfo {pages} {507} (\bibinfo {year} {1991})}\BibitemShut {NoStop}%
\bibitem [{\citenamefont {Mirlin}\ and\ \citenamefont
  {Fyodorov}(1994)}]{MF1994}%
  \BibitemOpen
  \bibfield  {author} {\bibinfo {author} {\bibfnamefont {A.~D.}\ \bibnamefont
  {Mirlin}}\ and\ \bibinfo {author} {\bibfnamefont {Y.~V.}\ \bibnamefont
  {Fyodorov}},\ }\bibfield  {title} {\enquote {\bibinfo {title} {{Statistical
  properties of one-point Green functions in disordered systems and critical
  behavior near the Anderson transition}},}\ }\href@noop {} {\bibfield
  {journal} {\bibinfo  {journal} {J. Phys. France}\ }\textbf {\bibinfo {volume}
  {4}},\ \bibinfo {pages} {655} (\bibinfo {year} {1994})}\BibitemShut {NoStop}%
\bibitem [{\citenamefont {Aizenman}(2013)}]{Aizenman2013}%
  \BibitemOpen
  \bibfield  {author} {\bibinfo {author} {\bibfnamefont {M.~Warzel~S.}\
  \bibnamefont {Aizenman}},\ }\bibfield  {title} {\enquote {\bibinfo {title}
  {{Resonant delocalization for random Schr{\"o}dinger operators on tree
  graphs}},}\ }\href {\doibase 10.4171/JEMS/389} {\bibfield  {journal}
  {\bibinfo  {journal} {Journal of the European Mathematical Society}\ }\textbf
  {\bibinfo {volume} {15}},\ \bibinfo {pages} {1167--1222} (\bibinfo {year}
  {2013})}\BibitemShut {NoStop}%
\bibitem [{\citenamefont {Mel'nikov}(1981)}]{Melnikov1981}%
  \BibitemOpen
  \bibfield  {author} {\bibinfo {author} {\bibfnamefont {V.~I.}\ \bibnamefont
  {Mel'nikov}},\ }\bibfield  {title} {\enquote {\bibinfo {title} {Fluctuation
  of resistance of finie disordered system},}\ }\href@noop {} {\bibfield
  {journal} {\bibinfo  {journal} {Sov. Phys. Solid State}\ }\textbf {\bibinfo
  {volume} {23}},\ \bibinfo {pages} {782--786} (\bibinfo {year}
  {1981})}\BibitemShut {NoStop}%
\bibitem [{\citenamefont {Khaymovich}\ \emph {et~al.}(2019)\citenamefont
  {Khaymovich}, \citenamefont {Haque},\ and\ \citenamefont
  {McClarty}}]{Behemoths2019}%
  \BibitemOpen
  \bibfield  {author} {\bibinfo {author} {\bibfnamefont {I.~M.}\ \bibnamefont
  {Khaymovich}}, \bibinfo {author} {\bibfnamefont {M.}~\bibnamefont {Haque}}, \
  and\ \bibinfo {author} {\bibfnamefont {P.~A.}\ \bibnamefont {McClarty}},\
  }\bibfield  {title} {\enquote {\bibinfo {title} {Eigenstate thermalization,
  random matrix theory, and behemoths},}\ }\href {\doibase
  10.1103/PhysRevLett.122.070601} {\bibfield  {journal} {\bibinfo  {journal}
  {Phys. Rev. Lett.}\ }\textbf {\bibinfo {volume} {122}},\ \bibinfo {pages}
  {070601} (\bibinfo {year} {2019})}\BibitemShut {NoStop}%
\bibitem [{\citenamefont {B{\"{a}}cker}\ \emph {et~al.}(2019)\citenamefont
  {B{\"{a}}cker}, \citenamefont {Haque},\ and\ \citenamefont
  {Khaymovich}}]{Baecker2019}%
  \BibitemOpen
  \bibfield  {author} {\bibinfo {author} {\bibfnamefont {Arnd}\ \bibnamefont
  {B{\"{a}}cker}}, \bibinfo {author} {\bibfnamefont {Masudul}\ \bibnamefont
  {Haque}}, \ and\ \bibinfo {author} {\bibfnamefont {Ivan~M}\ \bibnamefont
  {Khaymovich}},\ }\bibfield  {title} {\enquote {\bibinfo {title}
  {{Multifractal dimensions for random matrices, chaotic quantum maps, and
  many-body systems}},}\ }\href {\doibase 10.1103/PhysRevE.100.032117}
  {\bibfield  {journal} {\bibinfo  {journal} {Phys. Rev. E}\ }\textbf {\bibinfo
  {volume} {100}},\ \bibinfo {pages} {032117} (\bibinfo {year}
  {2019})}\BibitemShut {NoStop}%
\bibitem [{\citenamefont {Luitz}\ \emph {et~al.}(2019)\citenamefont {Luitz},
  \citenamefont {Khaymovich},\ and\ \citenamefont
  {Lev}}]{luitz2019multifractality}%
  \BibitemOpen
  \bibfield  {author} {\bibinfo {author} {\bibfnamefont {David~J}\ \bibnamefont
  {Luitz}}, \bibinfo {author} {\bibfnamefont {Ivan}\ \bibnamefont
  {Khaymovich}}, \ and\ \bibinfo {author} {\bibfnamefont {Yevgeny~Bar}\
  \bibnamefont {Lev}},\ }\href@noop {} {\enquote {\bibinfo {title}
  {Multifractality and its role in anomalous transport in the disordered xxz
  spin-chain},}\ } (\bibinfo {year} {2019}),\ \Eprint
  {http://arxiv.org/abs/1909.06380} {arXiv:1909.06380} \BibitemShut {NoStop}%
\end{thebibliography}%

\end{document}